%Paper: chao-dyn/9304012
%From: giovanni@cptsu2.univ-mrs.fr
%Date: Thu, 29 Apr 93 15:57:25 +0200
%Date (revised): Fri, 4 Jun 93 19:44:03 +0200
%Date (revised): Sun, 6 Jun 93 12:45:25 +0200

%%%%%%%%%%%%%%%%%%%%%%%%%%%%%%%%%%%%%%%%%% PLAIN TEX
%
%TO PRINT THE POSTCRIPT FIGURES THE DRIVER NUMBER MIGHT HAVE TO BE
%ADJUSTED. IF the 4 choices 0,1,2,3 do not work set in the following line
%the \driver variable to =5. Setting it =0 works with dvilaser setting it
%=1 works with dvips, =2 with psprint, =3 with dvitps, (hopefully).
%Using =5 prints incomplete figures (but sill understandable from the
%text). The value MUST be set =5 if the printer is not a postscript one

\newcount\driver \driver=1          %%%this is the value to set!!!

%%% the value =0,1 has been tested. The figures are automatically
%%% generated (do not worry). The figures are in p.18 and 19.
%%%%%%%%%%%%%%%%%%%%%%%%%%%%%%%%%%%%%%%%%%
\magnification=\magstep0\hoffset=0.cm
\voffset=-0.5truecm\hsize=16.5truecm\vsize=24.truecm
\parindent=4.pt
%\magnification=\magstep1\hoffset=0.cm
%\baselineskip=14pt plus0.1pt minus0.1pt \parindent=12pt
%\lineskip=4pt\lineskiplimit=0.1pt      \parskip=0.1pt plus1pt
%%%%%%%%%%%%%%%%%%%%%%%%%%%%%%%%%%%%%%%%%%
%\overfullrule=10pt
%
%%%%%GRECO%%%%%%%%%
%
\let\a=\alpha \let\b=\beta  \let\g=\gamma    \let\d=\delta \let\e=\varepsilon
\let\z=\zeta  \let\h=\eta   \let\th=\vartheta\let\k=\kappa \let\l=\lambda
\let\m=\mu    \let\n=\nu    \let\x=\xi       \let\p=\pi    \let\r=\rho
\let\s=\sigma \let\t=\tau    \let\f=\varphi
\let\ps=\psi  \let\o=\omega 
  \let\D=\Delta  \let\Th=\Theta  
     \let\F=\Phi       \let\O=\Omega

%%%%%%%%%%%%%%%%%%%%%%%%%%%%%%%%%%%%%%%%%%%%%%%%%%%%%%%%%%%%%%
%%%%%%%%%%%%%%%%%%%%%  Numerazione pagine
%%%%%%%%%%%%%%%%%%%%%  NUMERAZIONE PAGINE

{\count255=\time\divide\count255 by 60 \xdef\oramin{\number\count255}
        \multiply\count255 by-60\advance\count255 by\time
   \xdef\oramin{\oramin:\ifnum\count255<10 0\fi\the\count255}}

\def\ora{\oramin }

\def\data{\number\day/\ifcase\month\or gennaio \or febbraio \or marzo \or
aprile \or maggio \or giugno \or luglio \or agosto \or settembre
\or ottobre \or novembre \or dicembre \fi/\number\year;\ \ora}

\setbox200\hbox{$\scriptscriptstyle \data $}

\newcount\pgn \pgn=1
\def\foglio{\number\numsec:\number\pgn
\global\advance\pgn by 1}
\def\foglioa{A\number\numsec:\number\pgn
\global\advance\pgn by 1}

%\footline={\rlap{\hbox{\copy200}\ $\st[\number\pageno]$}\hss\tenrm
%\foglio\hss}
%\footline={\rlap{\hbox{\copy200}\ $\st[\number\pageno]$}\hss\tenrm
%\foglioa\hss}
%

%%%%%%%%%%%%%%%%% EQUAZIONI CON NOMI SIMBOLICI
%%%
%%% per assegnare un nome simbolico ad una equazione basta
%%% scrivere \Eq(...) o, in \eqalignno, \eq(...) o,
%%% nelle appendici, \Eqa(...) o \eqa(...):
%%% dentro le parentesi e al posto dei ...
%%% si puo' scrivere qualsiasi commento;
%%% per assegnare un nome simbolico ad una figura, basta scrivere
%%% \geq(...); per avere i nomi
%%% simbolici segnati a sinistra delle formule e delle figure si deve
%%% dichiarare il documento come bozza, iniziando il testo con
%%% \BOZZA. Sinonimi \Eq,\EQ,\EQS; \eq,\eqs; \Eqa,\Eqas;\eqa,\eqas.
%%% All' inizio di ogni paragrafo si devono definire il
%%% numero del paragrafo e della prima formula dichiarando
%%% \numsec=... \numfor=...  (brevetto Eckmannn); all'inizio del lavoro
%%% bisogna porre \numfig=1 (il numero delle figure non contiene la sezione.
%%% Si possono citare formule o figure seguenti; le corrispondenze fra nomi
%%% simbolici e numeri effettivi sono memorizzate nel file \jobname.aux, che
%%% viene letto all'inizio, se gia' presente. E' possibile citare anche
%%% formule o figure che appaiono in altri file, purche' sia presente il
%%% corrispondente file .aux; basta includere all'inizio l'istruzione
%%%           \include{nomefile}
%%%
%%%%%%%%%%%%%%%%%%%%%%%%%%%%%%%%%%%%%%%%%%%%%%%%%%%%%%%%%%%%%%%

\global\newcount\numsec\global\newcount\numfor
\global\newcount\numfig
\gdef\profonditastruttura{\dp\strutbox}
\def\senondefinito#1{\expandafter\ifx\csname#1\endcsname\relax}
\def\SIA #1,#2,#3 {\senondefinito{#1#2}
\expandafter\xdef\csname #1#2\endcsname{#3} \else
\write16{???? ma #1,#2 e' gia' stato definito !!!!} \fi}
\def\etichetta(#1){(\veroparagrafo.\veraformula)
\SIA e,#1,(\veroparagrafo.\veraformula)
 \global\advance\numfor by 1
 \write15{\string\FU (#1){\equ(#1)}}
 \write16{ EQ \equ(#1) == #1  }}
\def \FU(#1)#2{\SIA fu,#1,#2 }
\def\etichettaa(#1){(A\veroparagrafo.\veraformula)
 \SIA e,#1,(A\veroparagrafo.\veraformula)
 \global\advance\numfor by 1
 \write15{\string\FU (#1){\equ(#1)}}
 \write16{ EQ \equ(#1) == #1  }}
\def\getichetta(#1){Fig. \verafigura
 \SIA e,#1,{\verafigura}
 \global\advance\numfig by 1
 \write15{\string\FU (#1){\equ(#1)}}
 \write16{ Fig. \equ(#1) ha simbolo  #1  }}
\newdimen\gwidth
\def\BOZZA{
\def\alato(##1){
 {\vtop to \profonditastruttura{\baselineskip
 \profonditastruttura\vss
 \rlap{\kern-\hsize\kern-1.2truecm{$\scriptstyle##1$}}}}}
\def\galato(##1){ \gwidth=\hsize \divide\gwidth by 2
 {\vtop to \profonditastruttura{\baselineskip
 \profonditastruttura\vss
 \rlap{\kern-\gwidth\kern-1.2truecm{$\scriptstyle##1$}}}}}
}
\def\alato(#1){}
\def\galato(#1){}
\def\veroparagrafo{\number\numsec}\def\veraformula{\number\numfor}
\def\verafigura{\number\numfig}
\def\geq(#1){\getichetta(#1)\galato(#1)}
\def\Eq(#1){\eqno{\etichetta(#1)\alato(#1)}}
\def\eq(#1){\etichetta(#1)\alato(#1)}
\def\Eqa(#1){\eqno{\etichettaa(#1)\alato(#1)}}
\def\eqa(#1){\etichettaa(#1)\alato(#1)}
\def\eqv(#1){\senondefinito{fu#1}$\clubsuit$#1\write16{No translation for #1}%
\else\csname fu#1\endcsname\fi}
\def\equ(#1){\senondefinito{e#1}\eqv(#1)\else\csname e#1\endcsname\fi}

\def\include#1{
\openin13=#1.aux \ifeof13 \relax \else
\input #1.aux \closein13 \fi}
\openin14=\jobname.aux \ifeof14 \relax \else
\input \jobname.aux \closein14 \fi
\openout15=\jobname.aux %\write15
%
%%%%%%%%%%% GRAFICA %%%%%%%%%
%
% Inizializza le macro postscript e il tipo di driver di stampa.
% Attualmente le istruzioni postscript vengono utilizzate solo se il driver
% e' DVILASER ( \driver=0 ), DVIPS ( \driver=1) o PSPRINT ( \driver=2);
% o  DVITPS (\driver=3)
% qualunque altro valore di \driver produce un output in cui le figure
% contengono solo i caratteri inseriti con istruzioni TEX (vedi avanti).
%
%\newcount\driver \driver=1
%\ifnum\driver=0 \special{ps: plotfile ini.pst global} \fi
%\ifnum\driver=1 \special{header=ini.pst} \fi
\newdimen\xshift \newdimen\xwidth
%
% inserisce una scatola contenente #3 in modo che l'angolo superiore sinistro
% occupi la posizione (#1,#2)
%
\def\ins#1#2#3{\vbox to0pt{\kern-#2 \hbox{\kern#1 #3}\vss}\nointerlineskip}
%
% Crea una scatola di dimensioni #1x#2 contenente il disegno descritto in
% #4.pst; in questo disegno si possono introdurre delle stringhe usando \ins
% e mettendo le istruzioni relative nel file #4.tex (che puo' anche mancare);
% al disotto del disegno, al centro, e' inserito il numero della figura
% calcolato tramite \geq(#3).
% Il file #4.pst contiene le istruzioni postscript, che devono essere scritte
% presupponendo che l'origine sia nell'angolo inferiore sinistro della
% scatola, mentre per il resto l'ambiente grafico e' quello standard.
% Se \driver=2, e' necessario dilatare la figura in accordo al valore di
% \magnification, correggendo i parametri P1 e P2 nell'istruzione
%         \special{#4.pst P1 P2 scale}
%
\def\insertplot#1#2#3#4{
\par \xwidth=#1 \xshift=\hsize \advance\xshift
by-\xwidth \divide\xshift by 2 \vbox{
 \line{} \hbox{ \hskip\xshift  \vbox to #2
  {\vfil
 \ifnum\driver=0 #3%
  \special{ps: plotfile #4.ps}
 \fi
 \ifnum\driver=1  #3    \includegraphics{#4.ps} \fi
 \ifnum\driver=2  #3    \special{#4.ps 1.2 1.2 scale} %\special{inips.pst}
\fi
\ifnum\driver=5 #3 \fi
 \ifnum\driver=3
\psfig{figure=#4.ps,height=#2,width=#1,scale=1.2%,prolog=ini.pst
}
\kern-\baselineskip          #3 \fi }\hfil }
%\line{} \centerline{\geq(#3)} \line{}
}}

%%%%%%%%%%%%%%%%%%%%%%%%%%%%%%%%%%%%%%%%%%%%%%%%%%%%%%%%%%%%%%%%%%%%%%
%\input jphead92
%\let\octo=\eightrm
%%%%%%%%%%%%%%%%%%%%%%%%%
%%%%preloaded fonts%%%%%%
%%%%%%%%%%%%%%%%%%%%%%%%%
\newskip\ttglue

\newfam\msbfam

%\font\ninei=cmmi9

%\font\ninebf=cmbx9
%\font\ninett=cmtt9
%\font\ninesl=cmsl9
%\font\nineit=cmti9
\font\eightrm=cmr8
\font\eighti=cmmi8
\font\eightsy=cmsy8
\font\eightbf=cmbx8
\font\eighttt=cmtt8
\font\eightsl=cmsl8
\font\eightit=cmti8
\font\sixrm=cmr6
\font\sixbf=cmbx6
\font\sixi=cmmi6
\font\sixsy=cmsy6
%%%%%%%%%%%%%%%%%%%%%%%%%%%%%%%%%%%%%%%
%%% the following fonts force true %%%%
%%% computer modern behaviour      %%%%
%%% they are used to override      %%%%
%%% the ps-fonts for math symbols  %%%%
%%% like \dot \ne ....             %%%%
%%%%%%%%%%%%%%%%%%%%%%%%%%%%%%%%%%%%%%%
\newfam\truecmr
\newfam\truecmsy

\font\tentruecmr=cmr10
\font\tentruecmsy=cmsy10
\font\eighttruecmr=cmr8
\font\eighttruecmsy=cmsy8
\font\seventruecmr=cmr7
\font\seventruecmsy=cmsy7
\font\sixtruecmr=cmr6
\font\sixtruecmsy=cmsy6
\font\fivetruecmr=cmr5
\font\fivetruecmsy=cmsy5
%%%% add the definitions for 10pt %%%%%%%%
\textfont\truecmr=\tentruecmr
\scriptfont\truecmr=\seventruecmr
\scriptscriptfont\truecmr=\fivetruecmr
\textfont\truecmsy=\tentruecmsy
\scriptfont\truecmsy=\seventruecmsy
\scriptscriptfont\truecmr=\fivetruecmr
\scriptscriptfont\truecmsy=\fivetruecmsy
%%%%% size changes%%%%%%
\def \eightpoint{\def\rm{\fam0\eightrm}% switch to 8-point type
\textfont0=\eightrm \scriptfont0=\sixrm \scriptscriptfont0=\fiverm
\textfont1=\eighti \scriptfont1=\sixi   \scriptscriptfont1=\fivei
\textfont2=\eightsy \scriptfont2=\sixsy   \scriptscriptfont2=\fivesy
\textfont3=\tenex \scriptfont3=\tenex   \scriptscriptfont3=\tenex
\textfont\itfam=\eightit  \def\it{\fam\itfam\eightit}%
\textfont\slfam=\eightsl  \def\sl{\fam\slfam\eightsl}%
\textfont\ttfam=\eighttt  \def\tt{\fam\ttfam\eighttt}%
\textfont\bffam=\eightbf  \scriptfont\bffam=\sixbf
\scriptscriptfont\bffam=\fivebf  \def\bf{\fam\bffam\eightbf}%
\tt \ttglue=.5em plus.25em minus.15em
\setbox\strutbox=\hbox{\vrule height7pt depth2pt width0pt}%
\normalbaselineskip=9pt
\let\sc=\sixrm  \let\big=\eightbig  \normalbaselines\rm
\textfont\truecmr=\eighttruecmr
\scriptfont\truecmr=\sixtruecmr
\scriptscriptfont\truecmr=\fivetruecmr
\textfont\truecmsy=\eighttruecmsy
\scriptfont\truecmsy=\sixtruecmsy
}
\font\sette=cmr7\let\0=\noindent
\def\didascalia#1{\kern-0.4truecm\vbox{
\sette\0\it#1\hfill}\vskip0.3truecm}

\let\nota=\octo
%%%%%%%%%%%%%%%%%%%%%%%%

%%%%%%%%%%%%%%%%%%%%%%%%%%%%%%%%%%%%%%%%%% DEFINIZIONI VARIE
%
\def\V#1{\vec#1}\let\dpr=\partial\let\ciao=\bye
\let\io=\infty
\let\ii=\int\let\ig=\int
\def\media#1{\langle{#1}\rangle}

\def\guida{\leaders\hbox to 1em{\hss.\hss}\hfill}
\def\tende#1{\vtop{\ialign{##\crcr\rightarrowfill\crcr
              \noalign{\kern-1pt\nointerlineskip}
              \hglue3.pt${\scriptstyle #1}$\hglue3.pt\crcr}}}
\def\otto{{\kern-1.truept\leftarrow\kern-5.truept\to\kern-1.truept}}
\def\tto{{\Rightarrow}}
\def\acapo{\hfill\break}

%%%%%%%%%%%%%%%%%%%%%%%%%%%%%%%%%%%%%%%%%%LATINORUM

\def\etc{\hbox{\it etc}}\def\eg{\hbox{\it e.g.\ }}

\def\ie{\hbox{\it i.e.\ }}

\def\fiat{{}}

%%%%%%%%%%%%%%%%%%%%%%%%%%%%%%%%%%%%%%%%%%%

%%%%%%%%%%%%%%%%%%%%%%%%%%%%%%%%%%%%%%%%%%%DEFINIZIONI LOCALI

\def\AA{{\V A}}\def\aa{{\V\a}}
\def\nn{{\V\n}}\def\oo{{\V\o}}
\def\FF{{\V F}}
 \let\lis=\overline
\def\mm{{\V m}}

\def\II{{\cal I}}
\def\EE{{\cal E}}\def\MM{{\cal M}}
\def\TT{{\cal T}}\def\RR{{\cal R}}
\def\sign{{\rm sign\,}}

\def\={{ \; \equiv \; }}\def\su{{\uparrow}}\def\giu{{\downarrow}}
\let\ch=\chi

\def\Im{{\rm\,Im\,}}\def\Re{{\rm\,Re\,}}
\def\nn{{\V\n}}\def\lis#1{{\overline #1}}

\def\pps{{\V\ps{\,}}}
\let\dt=\displaystyle

\def\NN{{\cal N}}
\def\DD{{\cal D}} \def\2{{1\over2}}
\def\OO{{\cal O}}
\def\FF{{\cal F}}
\def\igb{{\ii \kern-9pt\raise4pt\hbox to7pt{\hrulefill}}}
\def\MM{{\cal M}}\def\mm{{\V\m}}

\def\acapo{\hfill\break}
\def\tst{\textstyle}

\def\st{\scriptscriptstyle}\def\fra#1#2{{#1\over#2}}
\let\\=\noindent

\def\*{\vskip0.3truecm}

%%%%%%%%%%%%%%%%%%%%%%%%%%%%%%%%%%%%%%%%%%%%%%%%%%%%%%%%%%%%%%%%%%%%%%
\catcode`\%=12\catcode`\}=12\catcode`\{=12
              \catcode`\<=1\catcode`\>=2

\openout13=f1.ps
\write13<%%BoundingBox: 0 0 240 130>
\write13<% 5.4>
\write13</tlinea { gsave moveto [4 4] 2 setdash lineto stroke grestore} def>
\write13</punto { gsave 2 0 360 newpath arc fill stroke grestore} def>
\write13</puntone { gsave 3 0 360 newpath arc fill stroke grestore} def>
\write13< 120 60 punto    >
\write13< 200 110 puntone   >
\write13< 200 90 puntone   >
\write13< 200 70 puntone   >
\write13< 200 10 puntone   >
\write13< 120 60 moveto 200 110 lineto>
\write13< 120 60 moveto 200 90 lineto>
\write13< 120 60 200 70 tlinea>
\write13< 120 60 moveto 200 10 lineto>
\write13< stroke>
\closeout13

\openout13=f2.ps
\write13<%%BoundingBox: 0 0 240 120>
\write13<% 5.7>
\write13</tlinea { gsave moveto [4 4] 2 setdash lineto stroke grestore} def>
\write13</punto { gsave  2 0 360 newpath arc fill stroke grestore} def>
\write13</puntone { gsave 3 0 360 newpath arc fill stroke grestore} def>
\write13<120 60 punto  >
\write13<200 110 puntone>
\write13<200 90 puntone >
\write13<200 70 puntone >
\write13<200 10 puntone >
\write13<120 60 moveto 200 110 lineto>
\write13<120 60 moveto 200 90 lineto>
\write13<120 60 200 70 tlinea>
\write13<120 60 moveto 200 10 lineto>
\write13<stroke>
\closeout13

\openout13=f3.ps
\write13<%%BoundingBox: 0 0 240 120>
\write13<% 5.8>
\write13</tlinea { gsave moveto [4 4] 2 setdash lineto stroke grestore} def>
\write13</punto { gsave 2 0 360 newpath arc fill stroke grestore} def>
\write13</puntone { gsave 3 0 360 newpath arc fill stroke grestore} def>
\write13<120 60 punto  >
\write13<200 110 puntone>
\write13<200 90 puntone >
\write13<200 70 puntone >
\write13<200 10 puntone >
\write13<120 60 moveto 200 110 lineto>
\write13<120 60 moveto 200 90 lineto>
\write13<120 60 200 70 tlinea>
\write13<120 60 moveto 200 10 lineto>
\write13<stroke>
\closeout13

\openout13=f4.ps
\write13<%%BoundingBox: 0 0 240 170>
\write13<% fig.pst>
\write13</punto { gsave 2 0 360 newpath arc fill stroke grestore} def>
\write13<0 90 punto     >
\write13<70 90 punto    >
\write13<120 60 punto   >
\write13<160 130 punto  >
\write13<200 110 punto  >
\write13<240 170 punto  >
\write13<240 130 punto  >
\write13<240 90 punto   >
\write13<240 0 punto    >
\write13<240 30 punto   >
\write13<210 70 punto   >
\write13<240 70 punto   >
\write13<240 50 punto   >
\write13<0 90 moveto 70 90 lineto>
\write13<70 90 moveto 120 60 lineto>
\write13<70 90 moveto 160 130 lineto>
\write13<160 130 moveto 200 110 lineto>
\write13<160 130 moveto 240 170 lineto>
\write13<200 110 moveto 240 130 lineto>
\write13<200 110 moveto 240 90 lineto>
\write13<120 60 moveto 240 0 lineto>
\write13<120 60 moveto 240 30 lineto>
\write13<120 60 moveto 210 70 lineto>
\write13<210 70 moveto 240 70 lineto>
\write13<210 70 moveto 240 50 lineto>
\write13<stroke>
\closeout13
\catcode`\%=14\catcode`\{=1
\catcode`\}=2\catcode`\<=12\catcode`\>=12

\def\equ{}

%\input ffiat

%\BOZZA

%\footline={\rlap{\hbox{\copy200}\ $\st[\number\pageno]$}\hss\tenrm
%\foglio\hss}

\vglue1.truecm

\0{\bf Twistless KAM tori, quasi flat homoclinic intersections, and other
cancellations in the perturbation series of certain completely
integrable hamiltonian systems.  A review.}\footnote{${}^*$}{\nota This
paper is deposited in the archive {\tt mp\_arc@math.utexas.edu}: to get
a TeX version of it, send an empty E-mail message to this address and
instructions will be sent back.}

\vskip1.truecm
\0{\bf Giovanni Gallavotti}\footnote{${}^1$}{\nota
E-mail: {\tt gallavotti\%40221.hepnet@lbl.gov}: Dipartimento di Fisica,
Universit\`a di Roma, ``La Sa\-pi\-en\-za", P. Moro 5, 00185 Roma, Italia.}

\vskip.2truecm
\0{\bf Abstract:} {\sl Rotators interacting with a pendulum via small,
velocity independent, potentials are considered.  If the interaction
potential does not depend on the pendulum position then the pendulum
and the rotators are decoupled and we study the invariant tori of the
rotators system at fixed rotation numbers: we exhibit cancellations, to
all orders of perturbation theory, that allow proving the stability and
analyticity of the dipohantine tori.  We find in this way a proof of
the KAM theorem by direct bounds of the $k$--th order coefficient of
the perturbation expansion of the parametric equations of the tori in
terms of their average anomalies: this extends Siegel's approach, from
the linearization of analytic maps to the KAM theory; the convergence
radius does not depend, in this case, on the twist strength, which
could even vanish ({\it "twistless KAM tori"}).  The same ideas apply
to the case in which the potential couples the pendulum and the
rotators: in this case the invariant tori with diophantine rotation
numbers are unstable and have stable and unstable manifolds ({\it
"whiskers"}): instead of studying the perturbation theory of the
invariant tori we look for the cancellations that must be present
because the homoclinic intersections of the whiskers are {\it "quasi
flat"}, if the rotation velocity of the quasi periodic motion on the
tori is large.  We rederive in this way the result that, under suitable
conditions, the homoclinic splitting is smaller than any power in the
period of the forcing and find the exact asymptotics in the two
dimensional cases ({\it e.g.} in the case of a periodically forced
pendulum).  The technique can be applied to study other quantities: we
mention, as another example, the {\it homoclinic scattering phase
shifts}.}
\*

\vskip1.truecm

{\bf\S1 Introduction}\pgn=1\numfig=1\numsec=1\numfor=1

\vskip0.5truecm

We discuss the invariant tori and the splitting of their homoclinic
stable and unstable manifolds for a special class of quasi integrable
hamiltonian systems.  We apply and extend, in the considered class, the
important ideas of Melnikov and Eliasson respectively on the theory of
low dimensional invariant tori and their manifolds, [Me], and on the
cancellations behind the convergence of the formal perturbation series
for the invariant tori of maximal dimension, [E], \ie for the KAM tori
([K],[A2],[M]).  We also point out the analogy of the methods with those
used in quantum field theory, particularly in the renormalization group
approaches, [G2]. %

The ideas of Melnikov and Eliasson have been around since quite a
while: but it seems that few realized their importance; a
possible explanation is that the original papers are plagued by
excessive generality.  Assuming that this is the reason it is desirable
to show what they imply in a "simple" case.%

There are, however, no really simple cases (as it is well known).

In his book on mathematical physics Thirring made an attempt to find a
"simple" model to explain the KAM theory, [T].  Although his models are
essentially as difficult to treat as the most general hamiltonian system
with twist, some aspects of the proofs are somewhat simpler and leave us
with the feeling that something more can still be done.  The class of
models consists of a family of rotators (\ie points on circles or, if
one deems this too abstract, cylinders with fixed axis), say
$l-1$ in number, interacting with a pendulum via a conservative force.
The inertia moments $J_j$, $j=1,\ldots,l-1$, of the rotators form a
matrix $J$ which is diagonal.  And they are supposed to be $J_j\ge
J_0>0$, if $J_0$ is the inertia of the pendulum.  Actually Thirring has
no pendulum, just a system of rotators: this can be viewed as a special
case in which the pendulum and the rotators do not interact. Also he
considers a somewhat more general model than model 1), by allowing the
matrix $J$ to depend on $\aa$.

More formally: we shall consider the $l$ degrees of freedom hamiltonian
$H_\m\=H_0+\m f$ given by one of the following two expressions:
$$\eqalign{
1)\quad&\oo\cdot\V A+{1\over2}J^{-1}\AA\cdot\AA+\m
\sum_{0<|\nn|\le N}f_\nn\cos\aa\cdot\nn\cr
2)\quad&\oo\cdot\V A+{1\over2}J^{-1}\AA\cdot\AA+{I^2\over2J_0}+g^2J_0
(\cos\f-1)+\m \sum_{{|\n|\le N}\atop{\nn\neq \V 0}}
f_\n\cos(\aa\cdot\nn+n\f)\cr}\Eq(1.1)$$
where $(I,\f)\in R^2,(\AA,\aa)\in R^{2(l-1)}$ are canonically
conjugated variables, $\oo\in R^{l-1}$, $\n\=(n,\nn)\in Z^l$,
$|\n|=|n|+|\nn|=|n|+\sum_{i=1}^l |\n_i|$ and $J,J_0>0$ (respectively,
``rotator's inertia moment'' and ``pendulum inertia moment''), $g>0$
($g^2$ is the ``gravity''), $f_\n$ are fixed constants.  And $\oo,\m$
are parameters.

To avoid trivial, or lengthy, comments here and there we suppose that
$J\ge J_0$ with $J_0>0$ fixed forever, setting a scale for the size
of the inertia moments, (we do not take it $1$ as we are convinced that
it is better not to fix units of measure), and that $f_{\V0}=0$ or
$f_{n,\V0}\=0$, for all $n$, (for what concerns the trivial comments),
and $|f_\n|\le J_0\oo^2$ (for the lengthy ones): this will be clearly
not restrictive.

Model 2) will be called {\it rotator--pendulum model}, or {\it simple
resonance} model (see below for motivation) or {\it Arnold model}, and
model 1) will be called {\it rotator model} or {\it Thirring model}:
they deserve a name, because they are a very interesting class with
many properties that cannot be discussed here, the analysis of a few
(non trivial) of which was begun in [A] for the first and in [T] for
the second.

Model 1) has many applications in plasma physics, and celestial
mechanics.  Model 2) has great relevance for the analysis of the
breakdown of invariant tori and the corresponding universal behaviour.

In this paper we suppose a priori that:
\*
\penalty-500
{\bf Hypothesis H: \it the parameters $\oo,\m$ verify, in general:
$$\oo=\fra{\oo_0}{\sqrt\h},\qquad |\m|\le b\h^Q, \qquad \h\le1\Eq(1.2)$$
with $Q$ and $b^{-1}$ which will be restricted to be large enough in
the course of the analysis.  In the case of model 1) we shall also fix
$\h=1,Q=0$.}
\*%
\penalty10000
\0and:
\*%
{\bf Hypothesis H': \it $\oo_0$ is a {\it diophantine vector}, \ie:
$$\tst |\oo_0\cdot\nn|\ge {1\over C_0|\nn|^\t}\ ,\qquad \hbox{for all}
\ \V0\ne\nn\in Z^{l-1} \Eq(1.3)$$
for some {\it diophantine constant} $C_0$ and some {\it diophantine
exponent} $\t>0$.}
\*

A natural {\it energy scale} for the model 1) will be $E=J_0\oo^2$ and
for model 2) one can take $E=J_0g^2$.

We prove the following theorem for model 1):
\*
{\bf Theorem 1} {\it (Twistless KAM):
Consider the solution flow of the hamiltonian equations of the Thirring
model 1), under the assumptions H,H' above.  Then there exists a
constant $b>0$ and a function (dimensionless)
$\V x(\pps,\m)$ of $\pps\in T^{l-1},\m\in
R$, holomorphic in $\m$ and $\pps$, in the complex
domains $|\m|<e^{-N\x} b$ and $|\Im
\psi_j|<\x$, parameterized by $\x>0$, and bounded by $1$, such that the:
$$\aa=\pps-\m\fra{E}J
\sum_{\nn\ne\V0}\V x_\nn(\m)\fra{\sin\nn\cdot\pps}{(\oo\cdot\nn)^2},
\qquad\AA=-\m E\sum_{\nn\ne\V0}\V x_\nn(\m)
\fra{\cos\nn\cdot\pps}{(\oo\cdot\nn)}\Eq(1.4)$$
with $\pps\in T^{l-1},\ E\=J_0\oo^2$, are the parametric equations of an
invariant torus (for the flow) on which the flow is quasi periodic with
spectrum $\oo$ and the angles $\pps$ are its average anomalies: \ie the
flow is $\pps\to \pps+\oo t$, so that the angles $\ps_j$ rotate at
constant velocity $\o_j$. The constant $b$ depends only on
$J_0,\oo^2,C_0,\t,N,l$; and $\V x_\nn(\m)=\V x_{-\nn}(\m)$
denotes the Fourier transform
of $\V x(\pps,\m)$.}
\*
{\it Remarks:}
\item{1) } The fact that the theorem is {\it uniform} in the size of
$t_w=\det J^{-1}$, provided $J\ge J_0$, means that it holds even if the
twist $t_w=0$. This is the reason we call the above theorem "twistless"
KAM theorem.  We prove it, under a slightly stronger diophantine
hypothesis (see \equ(7.1)), by finding a bound on the $k$--th order of the
$\m$--expansion of the function $\V x$: see \S7 and Appendices A3,A4.

\item{2) } If some or all the inertia moments $J_j$ are $+\io$ the above
theeorem is an easy consequence of the classical KAM theorem (or better
of its proof); see comments around \equ(4.16) for the case $J_j\=+\io$.
However the bound $b$ obtained via the classical proof depends on the
twist rate, \ie on the maximum among the $J_j$ {\it which are not
$+\io$}, and diverges as the rate approaches $0$. This non uniformity is
quite surprising, but it is an artifact of the classical proof (as a
direct careful analysis of the latter also shows).
\*
\0Our second class of results concerns the simple resonance model,
2) in \equ(1.1).\acapo
\0The $l=2$ and $J=+\io$ case will {\it not} be excluded and corresponds
to the "pendulum in a periodic force field"; if $l>2$ we take $J<+\io$,
for simplicity, to be a constant but our results can be extended,
essentially unchanged, to the case in which $J$ is a diagonal matrix
with the first element zero and the others positive (in fact one can
take $J$ to be a $(l-1)\times(l-1)$ symmetric matrix with the first row
and columns zero and with rank $l-2$).  This is also implicitly shown,
in the framework of the classical proofs, by the proof of KAM theorem
for model 1) developed in [T], if particularized to model 1): it yields
results uniform in the twist rate.  This is probably the case, as well,
for the original proof of Kolmogorov, [K], if particularized to model
1).
\*
{\bf Remark:} To motivate the name of "simple resonance model" and
hypothesis H one should recognize that model 2) presents the ``basic"
structure (after a few simple canonical changes of variables) of a
perturbation of a completely integrable system, near a ``simple
resonance" (see [BG], [CG]).  Suppose that the original (\ie before the
above mentioned change of variables) unperturbed hamiltonian is non
degenerate, and $\h$, $0<\h\le1$, is a measure of the size of the
original perturbation. Then, provided $\h$ is small enough, one finds
that the parameters $\oo$ and $\m$ verify: $\oo\={\oo_0\,\h^{-1/2}}$,
$|\m|\le \h^{Q_0}$, where $Q_0>1$ can be prefixed at the beginning of
the construction of the change of coordinates, .  This is, perhaps, the
strongest motivation to study the hamiltonians \equ(1.1), model 2), with
$\oo$ given proportional to $1/\sqrt\h$.%
\*

For $\m=0$, the hamiltonian equations generated by \equ(1.1), (\ie
$\dot I=-\dpr_\f H_\m$, $\dot \f=\dpr_I H_\m$, $\dot \AA=-\dpr_\aa
H_\m$, $\dot \aa=\dpr_\AA H_\m$), admit $(l-1)$--dimensional invariant
tori:
$$\TT_0\=\{I=0=\f\}\times \{\AA\=\AA^0\ ,\aa\in T^{l-1}\} \Eq(1.5)$$
possessing homoclinic stable and unstable manifolds, {\it called
``whiskers"}. The manifolds equations are:
$$\tst W_0^{\pm}\=W_0\=\{
{I^2\over2J_0}+g^2J_0(\cos\f-1)=0 \} \times \{\AA\=\AA^0\ ,\aa\in
T^{l-1}\} \Eq(1.6)$$
The integrability is reflected also by the
degeneracy property that $W_0^+\=W_0^-$.

Then, it follows ``from KAM theory'', [Me],[E],[CG], that "many"
unperturbed tori around the torus $\V A^0=\V0$ (including the one $\V
A^0=\V0$ itself) can be {\it  continued analytically} (in $\m$), togheter
with their whiskers, into invariant tori with the same $\oo$,
for all $|\m|<b \h^Q$ (if $b$ is a suitable constant,
explicitly computable in terms of a few parameters associated with
$H_0,f$ in \equ(1.1), see for instance[CG]) and for $Q$ large enough; we
shall call such tori {\it persistent}. The determination of $b,Q$
reqires going through an analysis very similar to that of the classical
KAM theorem: hence we say that such tori and whiskers are ``obtained by
KAM analytic continuation''.\footnote{${}^2$}{\nota
If one wants to get an idea of the kind of numbers that might be
involved here (which is {\it not logically necessary} for reading the
present paper) one can look at [CG].  It follows from [CG] that $Q$ can
be taken to be $10$: see (5.76) of [CG] where $E_0= O(\h^{-1/2})$ and
all other parameters are $\h$--independent; including $C_0$, which is
not related to the $C_0$ of the present paper because in [CG] the quoted
inequality was obtained under the hypothesis that the constant bearing
there the same name, besides being a bound on the diophantine constant
for $\oo$, was also larger than $g^{-1}$.  In the present case the
$g^{-1}$ is of order $O(1)$, while the diophantine constant for $\oo$
is, by \equ(1.2), \equ(1.3), $C_0\h^{1/2}$ so that in applying the
quoted result to our case we are forced to take the constant $C_0$ in
(5.76) of [CG] of order $O(1)$.
\acapo
Note that in applying the results of [CG] to find an estimate for $Q$ we
apply the (5.76) and not the final result (5.90), which would give a
more stringent condition ($Q=71$), because we are using here only an
intermediary result whose proof was completed under the condition
(5.76). Such result is the persistence of a single invariant torus with
given rotation vector $\oo$: the paper [CG] was concerned with the
persistence of a whole family of invariant tori, and subject to the
condition of lying on a surface of prefixed energy (which led to add to
(5.76) two further conditions, (5.81), (5.85) in [CG]).
\acapo
In any event the above bounds should not be taken too seriously from a
quantitative viewpoint as they are very likely far from optimal (as no
effort at all was devoted to obtaining numerically good bounds).
}.

{\it Remark:} If $l=2$, however, one does not need an elaborated method
and a direct elementary check of the persistence of the invariant tori,
which in this case are periodic orbits, is possible, togheter with a
reasonably straightforward estimate of $b,Q$. One finds, for
instance, that for all $J\le+\io$ the value $Q=1$ is sufficient, (hint:
write the persistence condition as a fixed point equation for the
Poincar\'e map of the periodic orbit).

In some applications the \equ(1.2) is not a very strong condition
because of the remark following \equ(1.1) whereby the $Q_0$ mentioned
there can be taken as large as wished, by finding suitable coordinates.
In particular, if $Q_0=Q+1$ then the analyticity domain in $\m$
contains, well inside, the value $\h^{Q_0}$ for $\h$ small
enough.\footnote{${}^{3}$}{\nota \eg $Q_0=11$ according to the just
mentioned estimates in [CG].}

We shall always suppose that $\m$ is taken small enough so that the
invariant torus that we are considering is persistent: \ie we shall
always suppose that $|\m|<b\h^{Q}$ where $Q,b$ are the constants
mentioned in the discussion following the hypothesis \equ(1.2), so that the
above persistence properties hold (the values of the constants can be
determined in terms of properties of $H_0,f$ in \equ(1.1), see for
instance [CG] where such estimates are obtained).

We shall denote by $W^\pm_\m$ and $\TT_\m$ the stable and unstable
whiskers and, respectively, the whiskered tori obtained by the KAM
analytic continuation (in $\m$). The stable and unstable whiskers
$W^\pm_\m$ are characterized by the fact that distance$(S^t_\m
X^\pm_\m,\TT_\m) \to 0$ exponentially fast as $t\to\pm \io$; here
$S^t_\m$ is the hamiltonian flow generated by \equ(1.1). The flow on the
persistent whiskered tori can be described, in suitable coordinates and
for all $|\m|$ small (\ie $|\m|\le b\h^Q$), by $\V\ps\to\V\ps+\oo' t$,
for suitable diophantine vectors $\oo'$ {\it independent of $\m$} (\eg
$\oo'=\oo$ in the continuation of the unperturbed torus $\AA=\V 0$,
$I=0=\f$).

A legacy of the unperturbed situation (see \equ(1.6)) is that the
persistent whiskers $W^\pm_\m$ arising from \equ(1.6) are, for $|\m|$
small enough, {\it graphs over the angles} $\aa\in T^{l-1}$ and $\f$, at
least if $|\f|<2\p-\d$ for any prefixed $\d>0$; hence they can be
written as:
$$W^\pm_\m=\{(I,\AA,\f,\aa)=(I^\pm_\m(\aa,\f),\AA^\pm_\m(\aa,\f),\f,\aa):
\aa\in T^{l-1}, |\f|<2\p-\d\}\Eq(1.7)$$
for suitable real--analytic (in $(\aa,\f)$ {\it and} $\m$) functions
$\AA^\pm,I^\pm$.  It is also not difficult to check that the {\it
parity} (in $(\aa,\f)$ of \equ(1.1) implies that $(\aa,\f)=(\V 0,\p)$ is
a homoclinic point, \ie $(\V 0,\p)\in W^+_\m\cap W^-_\m$ (see \equ(4.18)
below), {\it for all $\m$} small enough (so small that the above tori
and whiskers can be proved to exist).

In this context, it is natural to {\it measure the splitting between
$W^+_\m$ and $W^-_\m$} at $\f=\p$ and $\aa=\V 0$ by the quantity:
$$\d(\aa)\=\det \dpr_\aa(\AA^+_\m-\AA^-_\m)|_{\f=\p}\Eq(1.8)$$
and its $\aa$--derivatives at $\aa=\V 0$.

Using the theory normal forms ([N], see also [BG]), one can show, see
[Nei], that if $\m\le b\h^{Q}$ then $\d$ is {\it smaller than any power
in $\h$ as $\h\to 0$}.

Here we use an algorithm derived in [CG], (see also \S2 below), for
the computation of the $\m$--expansion coefficients of the functions
$\AA^\pm,I^\pm$. And we derive the above smallness result by
explicitly checking several interesting {\it cancellation mechanisms},
operating to all orders of perturbation theory, and which are behind the
smallness of the splitting $\d$ when $\m\le b\h^{Q}$, see also [E],
[ACKR].

As a byproduct we obtain, and extend, the results of [Nei]. We obtain
also the exact asymptotics in the cases $l=2$, (with $0<J\le+\io$), see
\S8:
\*

{\bf Theorem 2} {\it (Quasi flat homoclinic intersections):
Let $l=2$, $J>J_0$, $J\le +\io$; let $N_0$ be the
degree in $\f$ of the trigonometric polynomial $f$ in \equ(1.1); then,
provided $\h$ is small enough:
$$\d(0)= \fra\m{\h^{N_0-1/2}}\, A_*\, e^{-\fra{\p\o_0}{2g\sqrt\h}}
(1+O(\sqrt\h))\Eq(1.9)$$
if $|\m|\le \h^q$ with $q>3N_0+5$, if $A_*=-g^{-1}(\fra{\o_0}g
)^{2N_0-1}(f_{N_0,1}+f_{N_0,-1})\fra{4\p(-1)^{N_0}}{(2N_0-1)!}$,
provided $A_*\ne0$.  And in fact $\d(\a)$ is a holomorphic function of
$\a$ for $|\Im \a|<\z$, for all $\z>0$, if $\h$ is small enough
(depending on $\z$). And:
$$|\d(\a)|\le D_\z\fra{|\m|}{\h^{N_0-1/2}}
e^{-\fra{\p\o_0}{2g\sqrt\h}},\qquad |\Im \a|<\z\Eq(1.10)$$
for a suitable $D_\z>0$ and for $\h$ small enough (depending on $q,\z$).}
\*

{\bf Remarks:}
\*
\item{1) }The optimal result is probably $q>N_0-\fra12$. In Appendix A1
we sketch how to get easily a better result: $q>N_0+\fra92$ instead
of the pessimistic one $q>3N_0+5$ given above.
\item{2) }For completeness we quickly rederive, in \S2 and \S4,
the recursive
formulae for the whiskers (see [CG], appendix 13).
\item{3) }It will appear that our technique consists in representing
the coefficient of the $\m^k$-th order contribution to $\d(\aa)$ as a
time integral from $t=-\io$ to $t=+\io$. The estimates should follow if
analyticity properties of the integrands, allowing
the shift of the $t$ integration to a region where the integrand is
small, could be checked.
\acapo
Such an idea has been present in the literature since a long time, and
sometimes it led to errors. It is also behind the available studies of
the asymptotic behaviour of $\d(\aa)$ as $\h\to0$ (and $\m=\h^Q$ with
$Q$ large enough).\acapo
In our representation of $\d(\aa)$ the above approach is not so
straightforward: {\it we cannot prove the holomorphy of the integrand in
the integral representation of the $k$-th order coefficient of the $\m$
expansion of $\d(\aa)$}, which in fact {\it is not holomorphic} in any
useful region for the above idea to work literally, for $k\ge2$.  Rather
we show that the integrand can be written as a sum of a holomorphic term
which is small, because of an analyticity argument like the one
mentioned above, and of a non holomorphic term which, however, is ``as
small as needed'' if the bounds on the coefficients of the lower orders
are ``as small as needed''.  Since the first order trivially has all the
necessary properties, the result is obtained by induction.
\item{4) }It is worth mentioning a related property, discovered in
[CG], that will not be needed nor discussed further in the present
work but that is closely related to theorem 2.\acapo
In many cases in which a hamiltonian like model 2) arises in the
analysis of a resonant motion of a slightly pertubed completely
integrable system, it happens that the original unperturbed hamiltonian
is strongly degenerate, \ie it does not depend on all the action
variables (a feature common in celestial mechanics).  In such cases,
that we shall call "degenerate", the vector $\oo$ in \equ(1.1) has (at
least) one component which is of $O(\sqrt{\h})$.  \acapo
More physically one can say that the models 2) verifying
\equ(1.2),\equ(1.3)
contain two basic time scales, of order $O(\h^{1/2})$ (``fast'') and
$O(1)$ (``slow'') corresponding to the frequences $\oo$ and $g$. But in
the degenerate cases there are (at least) three time scales of order
$O(\h^{1/2})$, $O(1)$ and $O(\h^{-1/2})$ (``secular''), as the very slow
secular scale reflects that the perturbation of the original system
removes its degeneracy to order $O(\h)$, relative to the $\oo$'s size.
\acapo
It is remarkable that a second order computation along the lines
discussed below, shows that {\it generically the splitting $\d$, with
$\m\=\h^{Q}$, is of $O({\h^M})$ for some $M>0$}: {\it hence it is not
smaller than any power in $\h$}. The latter fact can be used to show, at
least in some cases, the existence of {\it Arnold's drift and diffusion}
for any $\h>0$ small enough for systems which are small perturbations of
degenerate completely integrable systems, see [CG].\acapo
{\it The latter problem is still completely open in the non degenerate
case and probably requires some really new ideas for its solution}.

\item{5) }The proof that we give in the case $l=2$ and in the general
case have different nature: in both cases we use explicitly some general
results on the existence of $(l-1)$--dimensional tori, [M], in the form
derived in [CG].  However we shall show that the $l=2$ case is not
really relying on such results.  We think that also the $l>2$ cases can
be freed from its dependence on [CG]: the reason is the validity of
theorem 1 and its proof: it should be possible to carry out a similar
proof also for theorem 2, if $l>2$.

\vskip1.truecm

%\ciao

\penalty-200

{\bf\S2 Recursive formulae}\pgn=1
\vskip0.5truecm\numsec=2\numfor=1

\penalty10000

In this section we recall some basic facts from the KAM theory mentioned
in the introduction and we derive from such facts simple recursive
formulae for the functions $I^\pm_\m$, $\AA^\pm_\m$ in \equ(1.7)
and their time evolution.

Let us consider, for concreteness, the $\m$ dependence of the above
mentioned (see comments following \equ(1.6)) analytic continuation in
the parameter $\m$ of the unperturbed whiskers having $\AA^0=\V 0$.
The unperturbed motion is simply:
$$X^0(t)\=(I^0(t),\V 0,\f^0(t), \aa+\oo t)\Eq(2.1)$$
where $(I^0(t),\f^0(t))$ is the separatrix motion, generated by the
pendulum in \equ(1.1) starting at, say, $\f=\p$.  If we call $P(I,\f)$
the unperturbed pendulum energy in \equ(1.1), it is $P(I^0,\f^0)\=0$ (see
\equ(1.6)).

As mentioned in the introduction, under the hypotheses H,H' (see \S1)
the unperturbed whiskers $W^\pm_0$ \- $\=$ \- $W_0$ persist for $|\m|\le
\m_0\=b \h^{Q}$ and can be analytically continued into whiskers $W^\pm_\m$.

Let $X^\s_\m(t;\a)$, $\s=\pm$, be the evolution, under the flow
generated by \equ(1.1), of the point on $W^\s_\m$ given by
$(I^\s_\m(\aa,\p),\AA^\s_\m(\aa,\p),\p,\aa)$ (see \equ(1.7); from now on
we shall fix $\f=\p$, which amounts to studying the whiskers at the
``Poincar\'e section" $\{\f=\p\}$).

The mentioned analyticity in $\m$ (see [CG], for instance, \S5,\S6)
allows us to consider the Taylor series expansions of the whiskers
equations; let:
$$X^\s_\m(t)\=X^\s_\m(t;\aa)\equiv \sum_{k\ge 0} X^{k\s}(t;\aa) \m^k=
\sum_{k\ge 0} X^{k\s}(t) \m^k,\qquad \s=\pm\Eq(2.2)$$
be the power series in $\m$ of $X^\s_\m$, (convergent for $\m$ small);
note that $X^{0\s}\=X^0$ is the unperturbed whisker. We shall often not
write explicitly the $\aa$ variable among the arguments of various $\aa$
dependent functions, to simplify the notations.

{}From KAM theory, it follows that the $t$--dependence of $X^\s_\m(t)$
has the form:
$$X^\s_\m(t)=X^\s_\m(\oo t,t) \=X^\s_\m(\oo t,t;\aa)\Eq(2.3)$$
where $X^\s_\m(\V\ps,t;\aa)$ is a real analytic function, of {\it all}
its arguments ($\m$ included), which is periodic in $\V\ps$ and
$\aa$; furthermore it has a holomorphy domain:
$$\tst \DD\=\DD_{\x,\x_0,K,\m_0}\=\{|\Im \ps_i|<\x,\quad|\Im\a_i|<\x_0
,\quad |\Im t|<Kg^{-1},\quad|\m|<b\h^Q\}\Eq(2.4)$$
where $\x,\x_0,K,b,Q$ are suitable positive parameters. And on $\DD$
the following bound holds (see (6.28) in [CG], and [CG] for a
complete proof):
$$\tst\sup_\DD |X^\s_\m(\V\ps,t;\aa)-X^\s_\m(\V\ps,\s\io;\aa)|
\le D e^{-|\Re g t| \k},\qquad
\sup_\DD |X^\s_\m(\V\ps,t,\aa)|\le D\Eq(2.5)$$
where $\s={\,\rm sign\,}(\Re t)$ and $|X|=|X_-|+|X_\giu|+
(J_0g)^{-1}\big( |X_+|+|X_\su|\big)$.

In particular, if $X^{k\s}(\V\ps,t;\aa)$ is the $k^{\rm th}$ Taylor
coefficient (in the $\m$ expansion) of the function
$X_\m^\s(\V\ps,t;\aa)$ and if $\m_0=b\h^Q$, one has for all $k\ge0$:
$$\eqalign{
&\sup_\DD |X^{k\s}(\V\ps,t;\aa)-X^{k\s}(\V\ps,\s\io;\aa)|
\le D \ \m_0^{-k}\e^{-|\Re g t| \k},\qquad \m_0=b\h^Q\cr
&\sup_\DD |X^{k\s}(\V\ps,t,\aa)|< D\ \m_0^{-k}\cr}\Eq(2.6)$$
and $X^{k\s}(t)$ in \equ(2.2) is recognized to coincide with
$X^{k\s}(\oo t,t;\aa)$.

{\it We number the components of $X$ with a label $j$,
$j=0,\ldots,2l-1$, with the convention that:
$$X_0=X_-,\quad (X_j)_{j=1,\ldots,l-1}=\V X_\giu,\quad
X_l=X_+,\quad (X_j)_{j=l,\ldots,2l-1}=\V X_\su\Eq(2.7)$$
\ie we write first the angle and then the action components; first the
pendulum and then the rotators.}

Inserting \equ(2.2) into the Hamilton equation associated with
\equ(1.1), we see that the coefficients $X^{k\s}(t)\=$\- $X^{k\s}(\oo
t,t;\aa)$ satisfy the hierarchy of equations:
$${d\over dt} X^{k\s}\= \dot {X}^{k\s}=L X^{k\s}+F^{k\s}\Eq(2.8)$$
where:
$$\tst
L\=L(t)=\pmatrix{
0                    &\V 0        & J_0^{-1}       &\V 0     \cr
\V 0                 &0           &\V 0            &J^{-1}   \cr
 g^2J_0 \sin\f^0(t)  &\V 0        &0               &\V 0     \cr
\V 0                 &0           &\V 0            &0        \cr},\quad
F^1(t)=\pmatrix{0\cr0\cr
-\dpr_\f f(\f^0(t),\aa+\oo t)\cr
-\dpr_\aa f(\f^0(t),\aa+\oo t)\cr}\Eq(2.9)$$
and where $F^{k\s}$ depends upon $X^0,...,X^{k-1\s}$ but not on
$X^{k\s}$; here (as everywhere else) the arrows denote $(l-1)$--vectors.
Note that the entries of the $(2l\times 2l)$ matrix $L$ have different
meaning according to their position: the $\V 0$'s in the first and third
row are $(l-1)$ (row) vectors, the $\V 0$'s in the first and third
column are $(l-1)$ (column) vectors, and (even more confusing) the $0$'s
and $J^{-1}$ in the second and fourth column are $(l-1)\times (l-1)$
matrices, while the $0$'s in the first and third columns are scalars.

More explicitly, and more generally, if we write the Hamilton
equations for $X=X_\m^{k\s}(t)$, with hamiltonian \equ(1.1), as:
$$\dot{X}= G_0(X)+\m  G(X)\Eq(2.10)$$
(\ie $G_0=E\dpr H_0$, $G=E\dpr f$, $E\=$ standard symplectic matrix,
$\dpr\=(\dpr_I,\dpr_\AA,\dpr_\f,\dpr_\aa)$) then, by Taylor expansion,
we can rewrite \equ(2.8) for $k\ge1$ as:
$$\eqalign{
\dot X_r^{k\s}=&\sum_{j=1}^{2l}(\dpr_j {G_0}_r)(X^0(t)) X^{k\s}_j
+\sum_{|\V m|>1}({G_0}_r)_{\V m}(X^0(t))\sum_{(k^i_j)_{\V m,k}}
\prod_{i=0}^{2l-1}\prod_{j=1}^{m_i}X^{k^i_j\s}_i+\cr
&+\sum_{|\V m|\ge0}
(G_r)_{\V m}(X^0(t))\sum_{(k^i_j)_{\V m,k-1}}
\prod_{i=0}^{2l-1}\prod_{j=1}^{m_i}X^{k^i_j}_i\cr}\Eq(2.11)$$
where we have used the notation ($k^i_j\ge 1, m_i>0$):
$$\eqalign{
(G)_{\V m}(\cdot)\=&\Bigl(
{\dpr^{m_1}_I\dpr^{m_2}_{A_1}
\ldots\dpr^{m_{l}}_{A_{l-1}}\dpr^{m_{l+1}}_\f\dpr^{m_{l+2}}_{\a_1}
\ldots\dpr^{m_{2l}}_{\a_{l-1}}\,G\over
m_1!\,m_2!\,m_3!\,\ldots m_{l+2}!\,\ldots m_{2l}!}\Bigr)(\cdot)\cr
(k^i_j)_{\V m,p}\=&(k^1_1,\ldots,k^1_{m_1},k^2_1,\ldots,k^2_{m_2},
\ldots,k^{2l}_1,\ldots,k^{2l}_{m_{2l}})\qquad {\rm
s.t.\ }\sum k^i_j=p\cr}\Eq(2.12)$$
and $r=1,...,2l$, and the case $k=1$ requires a convenient interpretation
(that the reader can easily work out, as this is the very easiest case,
and as the result must be consistent with \equ(2.8)).

Expression \equ(2.11) gives in particular a formula for
$F^{k\s}$ in terms of the coefficients $X^0,...,X^{k-1\s}$ and of the
derivatives of $H_0$ and $f$.  Therefore, denoting, as above, by
indices $+,\su,-,\giu$ the components $I,\AA,\f,\aa$, we see that, in
our special case, \equ(2.8) takes the particularly simple form:
$$\eqalign{
& {d\over dt} X_+^{k\s}= (g^2 J_0 \sin\f^0) X_-^{k\s}+ F_+^{k\s}
\ ,\quad
{d\over dt} X^{k\s}_\su=\V F^{k\s}_\su\cr
&{d\over dt} X_-^{k\s} = J_0^{-1} X_+^{k\s}\ ,\quad\kern3.cm
{d\over dt} \V X^{k\s}_\giu=J^{-1} \V X^{k\s}_\su\cr}
\Eq(2.13)$$
as $F^{k\s}_-$, $\V F^{k\s}_\giu$ vanish identically, for $k\ge 1$.
And if $X_-\=X_0$, for all $k\ge1$ it is:
$$\eqalignno{
& F_-^{k\s}\=0\ ,\  \V F_\giu^{k\s}\=\V 0\ ,\
\V F_\su^{k\s} = - \sum_{|\V m|\ge0} (\dpr_\aa f)_{\V m}(\f^0,\oo t)
\sum_{(k^i_j)_{\V m,k-1}} \prod_{i=0}^{l-1}\prod_{j=1}^{m_i}
X^{k^i_j\s}_i&\eq(2.14)\cr
& F_+^{k\s} \= \sum_{|\V m|\ge 2} (g^2 J_0 \sin \f)_{\V m} (\f^0)
\sum_{(k_j)_{k}} \prod_{j=1}^{m}X^{k_j\s}_- -
\sum_{|\V m|\ge0} (\dpr_\f f)_{\V m}(\f^0,\oo t)
\sum_{(k^i_j)_{\V m,k-1}} \prod_{i=0}^{l-1}\prod_{j=1}^{m_i}
X^{k^i_j\s}_i\cr}$$
where $(k^i_j)_{\V m,k},(k^i_j)_{\V m,k-1}$ (are defined in \equ(2.12)).
Note that the first sum in the expression for $\V F^h_+$ can only
involve vectors $\V m$ with $m_j=0$ if $j\ge1$, because the function $J_0
g^2\cos\f$ depends only on $\f$ and not on $\aa$, (hence also $k^i_j=0$
if $i>0$). We use here the above notation to uniformize the notations.

The evolution of $X^k$ is determined by integrating
\equ(2.13), if the initial data are known.  The $k=1$ case requires a
suitable interpretation of the symbols, which we leave to the reader
(and the result has to be \equ(2.9),\equ(2.8)).

We recall that the {\it wronskian matrix} $W(t)$ of a solution $t\to
x(t)$ of a differential equation $\dot x= f(x)$ in $R^n$ is a $n\times
n$ matrix whose columns are formed by $n$ linearly independent
solutions of the linear differential equation obtained by linearizing
$f$ around the solution $x$ and assuming $W(0)=$ identity.

The solubility by elementary quadrature of the free pendulum equations
on the separatrix (see for instance \equ(3.3) below), leads after a
well known classical calculation to the following expression for the
wronskian $\lis W(t)$ of the separatrix motion of the pendulum
appearing in \equ(1.1), with initial data at $t=0$ given by $\f=\p,I=2g
J_0$:
$$\lis W(t)=\pmatrix{
{1\over\cosh gt}&{{\bar w}\over4J_0g}\cr
-J_0g{\sinh gt\over\cosh^2 gt}&
(1-{{\bar w}\over4}{\sinh gt\over\cosh^2gt})\cosh gt\cr},
\qquad{\bar w}\={2gt+\sinh 2gt\over\cosh gt}\Eq(2.15)$$
And the evolution of the $\pm$ (\ie $I,\f$) components can be determined by
using the above wronskian:
$$\pmatrix{X^{k\s}_-\cr X^{k\s}_+\cr}=\lis W(t)
\pmatrix{0\cr X^{k\s}_+(0)\cr} +
\lis W(t)\ii_0^t{\lis W\,}^{-1}(\t)\pmatrix{0\cr F^{k\s}_+(\t)\cr}\ d\t
\Eq(2.16)$$
Thus, denoting by $w_{ij}$ ($i,j=0,l$) the entries of $\lis W$
we see immediately that:
$$\eqalign{ & X^{k\s}_+(t)=w_{ll}(t)X^{k\s}_+(0)+w_{ll}(t)
\ig^t_0 w_{00}(\t) F^{k\s}_+(\t) d\t-w_{l0}(t)\ig^t_0 \bar
w_{0l}(\t) F^{k\s}_+(\t)\,d\t\cr & X^{k\s}_-(t)=\bar
w_{0l}(t)X^{k\s}_+(0)+w_{0l}(t) \ig^t_0w_{00}(\t)
F^{k\s}_+(\t) d\t-w_{00}(t)\ig^t_0w_{0l}(\t)
F^{k\s}_+(\t)\,d\t\cr} \Eq(2.17)$$
The integration of the equations
\equ(2.13) for the $\su,\giu$ components is ``easier'' yielding:
$$\eqalign{ \V X_\su^{k\s}(t)=&\V X_\su^{k\s}(0)+\ig_0^t\V
F^{k\s}_\su(\t)d\t\cr \V X_\giu^{k\s}(t)=&J^{-1}\Big(t \V
X_\su^{k\s}(0)+ \ig_0^td\t \,(t-\t)F^{k\s}_\su(\t)\Big)\cr}\Eq(2.18)$$
having used that the $\V X^{k\s}_\giu\=\V 0$ because the initial datum
is fixed and $\m$ independent; and \equ(2.17), \equ(2.18) can be used
to find a reasonably simple algorithm to represent the whiskers
equations to all orders $k\ge1$ of the perturbation expansion.

We shall regard the two functions $X^{k\s}(t)$, as forming a
single function $X^k(t)$:
$$X^k(t)=\cases{X^{k+}(t)& if $\s={\rm sign}\,t=+$\cr
X^{k-}(t)& if $\s={\rm sign}\,t =-$\cr}\Eq(2.19)$$
and at this point it is useful to open a parenthesis to define some
integration operations that can be performed on such functions; such
operations arise as soon as one tries to determine the initial data
(still unknown) in \equ(2.17),\equ(2.18).
\vskip1.truecm
\penalty-200

{\bf\S3 The improper integration $\II$.}\pgn=1

\penalty10000

\vskip0.5truecm\numsec=3\numfor=1

\penalty10000

The operation is simply the integration over $t$ from $\s\io$ to $t$,
$\s=\sign t$. In general such operation cannot be defined as an ordinary
integral of a summable function, because the functions on which it has
to operate (typically the integrands in \equ(2.17),\equ(2.18)) do not,
in general, tend to $0$ as $t\to\io$.

But the simplicity of the initial hamiltonian has the consequence that
the functions $X^k(t)$, and the matrix elements $w_{ij}$ in \equ(2.15),
belong to a very special class of analytic functions on which the
integration operations that we need can be given a meaning.

To describe such class we introduce various spaces of functions; all of
them are subspaces of the space $\hat \MM$ of the functions of $t$
defined as follows.
\*%

\0{\bf Definition 1:} {\it Let $\hat\MM$ be the space of the functions
of $t$ which can be represented, for some $k\ge 0$, as:
$$M(t)=\sum_{j=0}^k{(\s t g)^j\over j!} M_j^\s(x,\oo t)\ ,\quad
x\=e^{-\s gt}\ ,\quad \s={\rm sign}\, t\Eq(3.1)$$
with $M_j^\s(x,\V\psi)$ a trigonometric polynomial in $\V\ps$ with
coefficients holomorphic in the $x$-plane in the annulus $0<|x|<1$,
with: 1) possible singularities, outside the open unit disk, in a closed
cone centered at the origin, with axis of symmetry on the imaginary axis
and half opening $<\fra\p2$; 2) possible polar singularities at $x=0$;
3) $M_k\ne0$.  The number $k$ will be called the {\it $t$--degree} of
$M$. The smallest cone containing the singularities will be called the
{\it singularity cone} of $M$.}
\*%

It is not difficult to see that if a function admits a representation
like \equ(3.1), with the above properties, then such a representation is
unique, (see [CG], eq. (10.15), note that in [CG] $\hat\MM$ is called
$\MM$).

It is convenient to consider the functions of $t$ defined by the
monomials:
$$\tst
\s^\ch\,\fra{(\s t g)^j}{j!} x^h e^{i\oo\cdot\nn \, t}\Eq(3.2)$$
where $\ch=0,1$ and we use the notations in \equ(3.1).  We remark that
the functions in $\hat\MM$ can be expanded in the above monomials (\ie
the \equ(3.2) "span" the space $\hat\MM$).  The functions $M\in\hat\MM$
such that the residuum at $x=0$ of $x^{-1}\media{M_j^\s(x,\cdot)}$ is
zero, (here the average is over $\pps$, \ie it is an "angle average"),
form a subspace $\hat\MM_0$ of $\hat\MM$.

The functions in $\hat\MM,\hat\MM_0$ are not bounded near $x=0$, in
general.  We denote $\MM$ and $\MM_0$ the respective subspaces of the
functions bounded near $x=0$: which means that the $M_j$ have no pole at
$x=0$ and, furthermore, that $M_j(0,\V\psi)=0$ if $j>0$.

The symbols $\hat\MM^k,\hat\MM_0^k,\MM^k,\MM_0^k$ will denote the
subspaces of $\hat\MM,\hat\MM_0,\MM,\MM_0$, respectively, containing the
functions of $t$--degree $\le k$.

Note that $M\in \hat \MM$ can be written as $M=P+M'$ with $P$ being a
polynomial in $\s t$ (with $\s$ dependent coefficients) and with
$M'\in\hat\MM_0$: this can be done in only one way and we call $P$ the
``polynomial component'' of $M$.  Likewise $M\in \MM$ can be written as
$M=p+M'$ with $p$ being a constant function (with constant value
depending on $\s$) and $M'\in\MM_0$: $p$ will be called the ``constant
component'' of $M$.  In both cases $M'$ will be the "non singular"
component of $M$.

The coefficients of the above mentioned expansions and polynomials
depend on $\s=\pm$, \ie each $M\in\hat\MM$ is, in general, a pair of
functions $M^\s$ defined and holomorphic for $t>0$ and $t<0$,
respectively (and, more specifically, in a domain including
$\{\s \Re t>0$, $|\Im gt|<\p/2\}$).

The functions $M^\s(t)$ might sometimes (as in our cases below) be
continued analytically in $t$ but in general $M^+(-t)\ne M^-(-t)$ even
when it makes sense (by analytic continuation) to ask whether equality
holds. For the purpose of comparison with [CG] we note that the only
spaces introduced there are $\hat\MM$ and $\hat\MM_0$, and in [CG] they
are called $\MM$ and $\MM_0$ respectively.

Note however that if $M\in \MM$ the points with $\Re t=0$ and $|\Im
t|<\p/(2g)$ ($gt=\pm i\p/2$ corresponds to $x=\mp i$) are, (by our
hypothesis on the location of the singularities of the $M_j$ functions),
regularity points so that the values at $t^\pm$, "to the right" and "to
the left" of $t$, will be regarded as well defined and given by
$M(t^\pm)\=\lim_{t'\to t,\,\Re t'\to \Re t^\pm} M(t')$, in particular
$M^\pm(0^\pm)\= M_0^\pm(1^-,\V 0)$.

It must be remarked (as this will be essential later) that, since
$\f^0(t)=4 \arctan e^{-gt}$:
$$\tst
\cos \f^0=1- {2\over (\cosh gt)^2}= 1- 8 {x^2\over (1+x^2)^2}\ ,
\quad \sin \f^0= 2 {\sinh gt\over (\cosh gt)^2} = 4\s x {1-x^2\over
(1+x^2)^2}\Eq(3.3)$$
and since $f$ in \equ(1.1) is a trigonometric polynomial, the function
$F^1$, see \equ(2.9), belongs to $\MM$ and, in fact, the component $\V
F_\su^1$ belongs to $\MM_0$ (as accidentally does $F^1_+$ as well).

In general if a function $M\in\hat\MM$ is {\it holomorphic in} $t$ in
the strip $|\Im gt|<\fra\p2$ and it has a given parity, then it follows
that: $\tst M^+_j(x)=\pm(-1)^jM^-_j(x^{-1})$,
where the sign is $+$ if $M$ is even, and $-$ if it is odd. This means
that $\s^jM_j$ have the same parity for $x\otto x^{-1}$ as $M$ for
$t\otto-t$. Interesting examples are the functions in \equ(3.3).

It will be checked, by induction, that the functions $X^k$ and $F^k$ are
in $\MM$ and their representation \equ(3.1) is such that the sum over
$j$ runs up to $2k$.  We say that $F^k,X^k\in \MM^{2k}$, see definition
1.  More precisely:
$$X^k\in\MM^{2k-1},\qquad F^k\in\MM^{2(k-1)},\qquad \V F^{k}_\su\in
\MM^{2(k-1)}_0\Eq(3.4)$$
and, furthermore, the singularity cone consists of just the imaginary
axis (\ie the singularities of the functions defining $X^k,F^k$ are
on the segments on the imaginary axis $(-i\io,-i]$ and $[+i,+i\io)$).
See below for details.

On the class $\hat\MM$ we can define the following operation.  If
$M\in\hat\MM$, and $t=\t+i\th$, with $\t,\th$ real, {\it and $\t=\Re
t\ne0$, $\s=\sign \Re t$}, the function:
$$\II_R M(t)\=\ig_{\s\io+i\th}^te^{-Rg\s z} M^\s(z)\,dz\Eq(3.5)$$
is defined for $\Re R>0$ and large enough, the integral being on an
axis parallel to the real axis.

If $M\in\hat\MM$ then the function of $R$ in \equ(3.5) admits an
analytic continuation to $\Re R<0$ with possible poles at the integer
values of $R$ and at the values $i\oo\cdot\nn$ with $|\nn|<$
(trigonometric degree of $M$ in the angles $\V\psi$); and we can then
set:
$$\II M(t)\=\oint\fra{d R}{2\p i R} \,\II_R M(t)\Eq(3.6)$$
where the integral is over a small circle of radius $r<1$ and
$r<\min |\oo\cdot\nn|$, with the minimum being taken over the
$\nn\ne\V0$ which appear in the Fourier expansion of $M$  (which is
finite by definition 1).

{}From the above definition it follows immediately that if $M(t)=t^j$ then:
$$\tst\II M(t)= \fra1{j+1}\, t^{j+1}\Eq(3.7)$$
and more generally, if $j,h$ are integers and $\chi=0,1$, the
$\II$ acts on the monomial \equ(3.2) as:
$$\II M(t)=\cases{- g^{-1} \s^{\chi +1} x^h e^{i\oo\cdot\nn t}
\sum_{p=0}^j{
(g\s t)^{j-p} \over(j-p)!} {1 \over(h- i \s g^{-1} \oo\cdot\nn)^{p+1}}
& if $|h|+|\nn|>0$\cr
g^{-1}\s^{\ch+1}\fra{(\s gt)^{j+1}}{(j+1)!}& otherwise\cr}\Eq(3.8)$$
showing, in particular, that the radius of convergence in $x$ of $\II
M$, for a general $M$, is the same of that of $M$. But in general the
singularities at $\pm i$ will no longer be polar, even if those of the
$M_j$'s were such.

In fact one might want to check that $\II M\in \hat\MM$: this certainly
happens if no singularity can appear in the $(\II M)_j$'s outside the
singularity cone of $M$.  Note that $M\in\hat \MM$ can be written, for
fixed $\s=\sign \Re t$, as a sum of finitely many monomials $t^j x^{-r}
e^{i\O t}$ with $r\ge0$ plus a function $M'(t)$ of the form \equ(3.1)
with $M_j(0,\V\psi)=0$.

If the $\II$ operates on the monomials with $r\ge0$ (\ie on monomials
not vanishing at $x=0$ or with a polar singulariy at the origin) then
it can be explicitly computed by using \equ(3.8), and the check of the
claimed property is immediate.

The $\II M$ for $M$ such that $M_j(0,\V\psi)\=0$ can be computed as an
ordinary integral: it is obvious from \equ(3.5) that in such
case the $\II_R M(t)$ is holomorphic in $R$ for $R=0$ (as the integral
in \equ(3.5) is convergent for $R=0$).  Consider first
the case in which $M(t)= e^{i\oo t}M^\s(x)$ with
$M^\s(x)$ in $\hat\MM$ and $M^\s(0)=0$, one gets for
all $\O$ real:
$$\tst\ig_{\s\io}^t e^{i\O \t} M(e^{-\s g\t}) \,d\t= \ig_0^x e^{-i\O\s
g^{-1}\log y} M(y) \fra{dy}y= e^{i\O t}\ig_0^1 e^{-i\O\s g^{-1}\log y}
M(yx) \fra{dy}y\Eq(3.9)$$
which is clearly holomorphic in $x$ in the above considered $x$ plane
deprived, outside the unit disk, of the singularity cone of $M$.  The
more general case, in which we consider $t^j e^{i\O t}M(x)$ with
$M(0)=0$, is derived from \equ(3.9) by differentiation with respect to
$\O$. Any other case is a finite linear combination of the considered
cases.

Furthermore, if $M(t)\in\hat\MM$ and if $\sign \Re t= \sign \Re t_0$, it
follows from \equ(3.8) (as declared at the beginning of \S3)
that the $\II M$ function is a primitive of $M$:
$$\ig_{t_0}^tM(\t)d\t=\II M(t)-\II M(t_0)\Eq(3.10)$$
In general, $\II:\hat\MM^k\to\hat\MM^{k+1}$, because of \equ(3.7).
Furthermore, because of the similarities of the $\II$ operation with a
definite integral, we shall often use the notation:
$$\igb_{(\s)}^tM(\t)d\t\=\II M(t)\ ,\qquad M\in\hat\MM,\
\s=\hbox{sign}\,\Re t\Eq(3.11)$$
In fact many standard properties of integration are, in such a way,
extended to the space $\hat \MM$; for instance:
$$
\igb_{(\s)}^t\ = \igb_{(\s)}^{0^\s}\ + \ig_{0}^t\ \ \ ,\qquad
\II^2\=\igb_{(\s)}^t\ \igb_{(\s)}^\t \ =\igb_{(\s)}^t (t-\t)
\Eq(3.12)$$
where $\s=\sign t$ and $\igb_{(\s)}^{0^\s}$ means $\lim_{\e\to0^\s}
\igb_{(\s)}^{\e}$, of course.

This leads to a few more natural definitions and properties.

If $\s_t=\sign\Re t$, if $M\in\hat\MM$ and if $R$ is large enough, we
define $\II_{\pm,R}M$ via:
$$\II_{\pm,R}M(t)\=\ig_{\pm\io}^tM(\t)e^{-R\s_\t\t g}d\t=
\II_R M(0^\pm) + \ig_0^t M(\t) e^{-R\s_t\t g}d\t\Eq(3.13)$$
where we use the definition \equ(3.5), and in the r.h.s. the integral is
on the straight line joining $0$ to $t$.  This allows us to consider the
analytic continuation in $R$ of $R^{-1}\II_{\pm,R}M(t)$ and its residue
$\II_\pm M(t)$ at $R=0$. The latter is linked to the operation $\II$
already defined in \equ(3.6) so that the following two definitions (of
$\II_{\pm}M(t)$ and the consequent one of $\igb_{-\io}^{+\io} M(\t)d\t$)
are natural:
$$\eqalign{ \II_\pm M(t)\=&\igb_{\pm\io}^t M(\t)\,d\t=\II
M(0^\pm)+\ig_0^tM(\t) d\t\cr \igb_{-\io}^{+\io} M(\t) d\t=&\II_-
M(t)-\II_+ M(t)\cr}\Eq(3.14)$$
where the integral from $0$ to $t$ is over a straight path joining $0$
with $t$.  The \equ(3.14) will be a quite useful extension of the
operation $\II$ introduced in \equ(3.11).

Given the definition \equ(3.14), a natural question arises at this
point: is there a class of functions $M\in \MM$ such that the following
{\it shift of contour} formula:
$$\igb_{-\io}^{+\io} M(\t) d\t=\igb_{-\io}^{+\io} M(\t+ig^{-1}\x)d\t
\Eq(3.15)$$
holds for all $\x$ smaller than the complement to $\fra\p2$ of the half
opening of the singularity cone of $M$ (\ie for all $\x$ for which
\equ(3.15) makes sense)? A very simple answer follows immediately from
the definitions: \equ(3.15) holds if $M$ is holomorphic as a function of
$t$ in a strip around the real axis, wider than $g^{-1}\x$.  Note that
this might be at first surprising as the operation $\igb$ is an improper
integral operating on generally non summable functions.

The \equ(3.15) can be proved by remarking that for $R$ large (and
positive):
$$\II_{-,R}M(0^-)-\II_{+,R}M(0^+)=
\ig_{-\io}^0 e^{R\t g} M(\t)d\t+\ig_0^{+\io}e^{-R\t g} M(\t)
d\t\Eq(3.16)$$
which, by the assumed analyticity of $M$, differs from:
$$\ig_{-\io}^0 e^{R\t g} M(\t+ig^{-1}\x)d\t
+\ig_0^{+\io} e^{-R\t g}M(\t+ig^{-1}\x)d\t\Eq(3.17)$$
precisely by:
$$\eqalignno{\tst
&(e^{i R\x}-1)\ig_{-\io}^0d\t e^{R\t g} M(\t+ig^{-1}\x)+
(e^{-i R\x}-1)\ig_{0}^{+\io} d\t e^{-R\t g} M(\t+ig^{-1}\x)+\cr
-&i\ig_0^{g^{-1}\x} (e^{-R i\t g}-e^{+Ri\t g})M(i\t)\,d\t&\eq(3.18)\cr}$$
\ie \equ(3.16) is the sum of \equ(3.17) and \equ(3.18), by the
analyticity properties of $M$.  This implies \equ(3.15) by taking the
residues at $R=0$, as the analytic continuation of \equ(3.18) vanishes
if $R\to0$, when $M\in \hat\MM_0$, \ie it has no polynomial component;
while if $M$ is a polynomial in $t$ one can check by direct calculation
(using \equ(3.7)) that both sides of \equ(3.15) give $0$.

It is also useful for the purposes of a better understanding, to realize
that, if $M$ is holomorphic as a function of $t$ in a strip wider than
$g^{-1}\x$, then for $t$ real it is:
$$\fra12\sum_\r\igb_{\r\io}^{t+ig^{-1}\x} M(\t)d\t=\oint\fra{d R}{2\p i
R}\fra12\sum_\r\ig_{\r\io}^t e^{-R\s_\t g (\t+ig^{-1}\x)}
M(\t+ig^{-1}\x)d\t\Eq(3.19)$$
where the integrals in the r.h.s.  have to be considered to be the
analytic continuation on $R$ from $R>0$ and large.  This is so because
the two sides differ by the residue at $R=0$ of
$-iR^{-1}\ig_0^{g^{-1}\x} d\t (e^{-R i \x}$ - $e^{R i \x})M(i\t)$ which
vanishes.

Note that the r.h.s. is {\it different} from $\fra12\sum_\r
\igb_{\r\io}^t M(\t+i g^{-1}\x)d\t$, which would be the
residue ar $R=0$ of $R^{-1}\sum_\r\ig_{\r \io}^t$\- $e^{-R\s_\t g\t}$\-
$M(\t+ig^{-1}\x)\,$\- $d\t$. The two quantities coincide under the
mentioned analyticity assumption, however, if $M$ is in $\MM_0$; and one
could verify that this remains true if $M$ has no polynomial component,
\ie $M\in \hat\MM_0$.

This completes the discussion of the operations $\II$.
\vskip0.5truecm
%%%

\vskip1.truecm
\penalty-200

{\bf\S4 Analytic expressions of the expansion coefficients for the
whiskers and the KAM tori. Parity properties.}\pgn=1

\penalty10000

\vskip0.5truecm\numsec=4\numfor=1

\penalty10000
{\bf A): \it Whiskers.}
\*
We shall show that $X^{k}$ admits rather simple expressions in terms of
the operation $\II$,(and other related operations introduced below).
Recall that in \S2 we have fixed $\aa\in T^{l-1}$ and $\f=\p$, and we
are looking for the motions, on the stable ($\s=+$) or unstable ($\s=-$)
whisker, which start with the given $\aa$ and $\f=\p$ at $t=0$; in the
following $\aa$ is kept constant and usually notationally omitted.

We suppose inductively that $F^{h}\in\MM^{2(h-1)},\,F^h_\su\in
\MM_0^{2(h-1)}$ for $h\le k$, and
that $X^{h\s}\in\MM^{2h-1}$, for $h<k$: see definition 1, \S3.
This means, in particular, that $F^{h\s},X^{h\s}$ can be represented as:
$$\eqalign{
F^{h\s}(x,\V\psi,t)=&\sum_{j=0}^{2(h-1)}\fra{t^j}{j!}
F^{h\s}_j(x,\V\psi),\qquad h=1,\ldots,k\cr
X^{h\s}(x,\V\psi,t)=&\sum_{j=0}^{2h-1}\fra{t^j}{j!}
X^{h\s}_j(x,\V\psi),\qquad h=1,\ldots,k-1\cr}\Eq(4.1)$$
by setting $\pps=\oo t$, $\s=\sign t$, $x=e^{-g\s t}$; with
$F^{k\s}_j,X^{k\s}_j$ holomorphic at $x=0$ and vanishing at $x=0$ if
$j>0$. Hnece if $x=e^{- g\s t}$ and $\pps$ is kept fixed
the $F^h_j,X^h_j$ tend exponentially to
zero as $t\to\io$, if $j>0$; while if $j=0$ they tend exponentially
fast to a limit as $t\to\s\io$ (\ie as $x\to0$), which we denote
$F^h(\pps,\s\io)$ dropping the subscript $0$ as there is no ambiguity.

Furthermore the inductive hypothesis is enriched by:
$$\V F^{k\s}_{\su\V0}(\s\io)=\V0, \qquad {\rm for\  all}\  h\le
k\Eq(4.2)$$
recalling that, in general, a subscript $\nn$ affixed to a function
denotes the Fourier component of order $\nn\in Z^{l-1}$ of the
considered function.

We denote $X^{h\s}_{j\nn}(t)$ and $F^{h\s}_{j\nn}(t)$ the Fourier
transforms in $\V\psi$ of $X^{h\s}_j(t,\V\psi)$ and
$F^{h\s}_j(t,\V\psi)$.  {\it It follows from the KAM theory mentioned
in \equ(2.6), that $X^{h\s}(t)$ and, from \equ(2.14), hence also
$F^{h\s}(t)$ are bounded as $t\to\s\io$ for all $h$}, so that
$X^{h\s}_j(0,\V\psi)=0$ if $j\ge1$. We show that the latter information
is very strong and permits us to determine $X^k$.

We note that, since $F^{k\s}\in\MM^{2(k-1)}$ and $\V
F^{k\s}_{\su\V0}=\V0$ hold, the function $\V X^{k\s}_\su(t)$, given by
the first of \equ(2.18), is in fact in $\MM^{2(k-1)}$ (by integration).
But of course we do not know (yet) the initial data $X^{k\s}(0)$.

To find expressions for $X^k_\su$ we start from the
equations \equ(2.13) with initial time at some instant $T$.
And we use that $\II F(t)$ is a primitive of the function $F(t)$, see
\equ(3.10), so that:
$$\V X^{k\s}_\su(t)=\V X^{k\s}_{\su}(T)+\II F^{k\s}(t)-\II
F^{k\s}(T)\Eq(4.3)$$
where $\s={\rm\,sign\,}t,$ and $T$ has the same sign of $t$.

The function $\V X_\su^{k\s}(T)$ tends to become quasi periodic with
exponential speed as $T\to\s\io$: in fact it becomes asymptotic to the
$j=0$ component, see \equ(4.1), at $x=0$: $X^{k\s}_{0\su}(0,\oo T)$,
(in the sense that the difference tends to $0$, bounded by
$(g|T|)^{2k-1}e^{-g|T|}$).  The function $\II F^{k\s}_\su(T)$ also
becomes asymptotically quasi periodic with exponential speed {\it and
$\V0$ average}, because $\V F^{k\s}_\su\in \MM_0^{2(k-1)}$ and by the
definition of $\II$: therefore the two quasi periodic functions of $T$
must cancel modulo a constant equal to $\media{\V
X^{k\s}_{0\su}(0,\cdot)}\=\V X_{\su\V0}^{k\s}(\s\io)$.

Hence it follows that:
$$\V X^{k\s}_\su (t)=\V X_{\su \V 0}^{k\s} (\s\io)+ \II \V F^{k\s}_\su (t)
\Eq(4.4)$$
and, by inserting \equ(4.4) into the second of\equ(2.18), (considering
also that the time average of $\II \V F^{k\s}_\su$ vanishes, and
therefore $\ig_0^t\t\V F^{k\s}_\su(\t)\,d\t= t\II\V F^{k\s}_\su(t)+$ a
$t$-bounded function), we see that the $\V X_\giu^{k\s}(t)$ can be
bounded only if:
$$\V X_{\su\V 0}^{k\s}(\s\io)=\V 0,\kern 1.truecm\hbox{hence:}
\kern1.truecm \V X_\su^{k\s}(t)=\II \V F_\su^{k\s}(t)\Eq(4.5)$$
yielding, setting $t=0^\s$, the initial values of $X_\su^k$ {\it and}
a simple form for its time evolution.  Analogously, recalling that $\V
X_\giu^{k\s}(0)=\V 0$, essentially by definition, one finds:
$$
\V X_\giu^{k\s}(t)= J^{-1}\big( \II^2 \V F_\su^{k\s}(t)-
\II^2\V F_\su^{k\s}(0^\s)\big)\=J^{-1}\bar\II^2 \V F_\su^{k\s}(t)
\Eq(4.6)$$
which gives a simple form to the time evolution of the $\aa$ (\ie $\giu$)
component of $X^k$ in terms of the operator $\lis\II^2$ defined by the
r.h.s. of \equ(4.6).

Likewise considering the \equ(2.17) and the behaviour at $\s \io$ of
$\lis W$ in \equ(2.15) and {\it recalling that $X^{k\s}(t)$ has to be
bounded at $\s\io$ by \equ(2.6)}, we see from the second of \equ(2.17)
that:
$$X_+^{k\s}(0)=-\ii_0^{\s\io} w_{00}(\t) F_+^{k\s}(\t)\ d\t
\Eq(4.7)$$
Thus we get (defining at the same time also $\OO$ and $\OO_+$):
$$\eqalign{
& X^{k\s}_+(t)=w_{ll}(t)\igb_{(\s)}^t
w_{00}(\t)F^{k\s}_+(\t)d\t-w_{l0}(t)\ig^t_0w_{0l}(\t)
F^{k\s}_+(\t)d\t\=\OO_+ F^{k\s}_+(t)\cr
&
X^{k\s}_-(t)= w_{0l}(t)\igb_{(\s)}^t w_{00}(\t)
F^{k\s}_+(\t)d\t-w_{00}(t)\ig^t_0w_{0l}(\t)
F^{k\s}_+(\t)d\t\=\OO F_+^{k\s}(t)\cr}
\Eq(4.8)$$
The \equ(4.5),\equ(4.6),\equ(4.8), and \equ(2.5) imply \equ(2.2) for
$h=k+1$.  As already remarked before
\equ(4.3) we note again that, since $F^{h\s}_{\su\V0}(\s\io)=\V0$ for
$h\le k$, the $\V X_\su^{k+1},\V X^{k+1}_\giu$ functions are in fact in
$\MM^{2(k-1)}$, (as the $\II$ operation, on such $\V F^k_\su$ functions
does not increase the degree).  Also, if one looks carefully at the
$X^{h\s}_\pm$--evaluation in terms of $F^{h\s}_+$, one realizes that
the $\OO,\OO_+$ operations may increase the degree but by at most $1$.
Thus the inductive hypothesis made in connection with \equ(4.1) is
proved for $X^{k+1}$, and it remains to check it for $F^{k+1}$.

The latter check follows from the expression of $F^{k+1}$, see
\equ(2.14), in terms of the $X^h$ with $h\le k$: see \equ(2.14).  One
treats separately the sums in \equ(2.14) with $|\V m|\ge2$ and $|\V
m|\ge0$.  One just has to consider that in the first case, which might
look dangerous for the inductive hypothesis, the products of $X$'s
contains {\it at least two factors} (which therefore have order labels
smaller than $k$ and verify the inductive hypothesis); and,
furthermore, the coefficients $(\dpr_\aa f)_{\V m}(\f_0,\oo t)$ or $g^2
J_0\sin\f_0$ or $g^2 J_0\cos\f_0$, by \equ(3.3), do not contain
any terms that can possibly increase the degree. Hence $F^{k+1}\in
\MM^{2k}$.

To see that $\V F^{(k+1)\s}_\su\in \MM^{2k}_0$, \ie
$\V F^{(k+1)\s}_{\su\V0}=\V0$, we simply remark that otherwise the
second of \equ(2.18) could not be bounded in $t$ as $t\to\io$; but we
know that it is bounded by \equ(2.5).
\*

{\bf Remark: } The use of \equ(2.5) is clearly a spurious element: it
should not be necessary to invoke a rather involved analytic discussion
({\it e.g.} the "KAM" theorem yielding the \equ(2.5)) to prove an
algebraic fact, namely $\V F^{(k+1)\s}_{\su\V0}=\V0$, that if not valid
would prove the claims of the theorem leading to \equ(2.5) false.  At
least it is unpleasant to do so (although logically consistent) and a
direct check of the algebraic property is highly desirable: it would
show that
\equ(2.14), and the equations $\V F^{k\s}_{\su\V0}=\V0$, and
\equ(4.4),\equ(4.6) and \equ(4.8) yield recursively a formal power
series expression for the whiskers.
\acapo
{\it Such a check is possible, see [CG] appendix 14}.  This means that
the property $\V F^{k\s}_{\su\V0}=\V0$ and the \equ(4.4),\equ(4.6), and
\equ(4.8), coupled with the initial condition \equ(2.9), {\it always}
have a solution verifying \equ(3.4). It gives us a formal power series
solution to the problem of finding the whiskers equations (and those of
their tori as well, which are easily related to the $t\to\pm\io$
behaviour of the $X^k$ functions).
\acapo
The convergence of the series, however, does not follow from what said
so far. In the case $l=2$ it will be checked in \S8, so that the $l=2$
case can be made fully independent on the KAM--type results expressed by
the \equ(2.5) (and imply them, of course). If $l>2$ we still rely on the
KAM results for the convergence, although it will be clear that with
some extra work it should be possible to obtain convergence estimates
along the same lines as in the $l=2$ case.
\*

We can summarize the above analytic considerations as:
$$\tst
\V F_{\su\V 0}^{k\s}(\s\io)\=\ii_{T^{l-1}}
\V F^{k\s}_\su(\V\ps,\s\io){d\V\ps\over (2\p)^{l-1}}\=
\langle \V F_\su^{k\s}(\cdot,\s \io)\rangle=\V 0\Eq(4.9)$$
for all $k\ge1$, and, still for all $k\ge1$, by:
$$\eqalignno{\tst
X^h_-(t)=&w_{0l}(t)\II(w_{00}F^h_+)(t)-w_{00}(t)\big(\II(w_{0l}F^h_+)(t
)-\II(w_{0l}F^h_+)(0^\s)\Big)\=\OO(F^h_+)(t)\cr
\V X^h_\giu(t)=&J^{-1}\,\Big(\II^2(\V F^h_\su)(t)-\II^2(\V
F^h_\su)(0^\s)\Big)\=J^{-1}\lis\II(\V F^h_\su(t)&\eq(4.10)\cr
X^h_+(t)=&w_{ll}(t)\II(w_{00}F^h_+)(t)-w_{l0}(t)\big(\II(w_{0l}F^h_+)(t
)-\II(w_{0l}F^h_+)(0^\s)\Big)\=\OO_+(F^h_+)(t)\cr
\V X^h_\su(t)=&\II(\V F^h_\su)(t)\cr}$$
where $\OO,\OO_+,\lis\II^2,\II$ are defined here (and in \S3); and
$X^h\=(X_-,\V X_\giu,X_+,\V X_\su)=(X^h_j)$, $j=0,\ldots 2l-1$,
$F^h=(0,\V0,F_+^h,\V F^h_\su)$ so that $-,+$ are synonims of $0,l$
respectively and $\giu,\su$ denote collectively the labels
$j=1,\ldots,l-1$ and $l+1,\ldots,2l-1$ respectively (see also \S3). Note
that while $X^h$ has non zero components over both the {\it angle}
($j=0,\ldots,l-1$) components and over the {\it action}
($j=l,\ldots,2l-1$) the $F^h$ has only the action compnents non zero.
{\it Furthermore if $\s t>0$ the above functions describe a motion on
the whisker $W^\s$ with initial data at some $\aa$ and $\f=\p$.}
\*
{\bf B): \it Tori:}
\*

A case of special interest is the case in which $f$ in \equ(1.1)
is $\f$--independent. In such case the pendulum and the rotators
decouple and we are really studying the perturbation theory of a
completely integrable $(l-1)$--dimensional system, of rather special
form, namely that of model 1) in \equ(1.1).

The whiskers will, in this case, be degenerate and, at $\f=\p$,
have the form:
$$\aa\,\to\,(\V X_\su(\V0;\aa),I_0(\p))\Eq(4.11)$$
and we can deduce the geometric locus of the torus, by letting $\pps$
vary in $T^{l-1}$, via:
$$\tst\V A=\V X_\su(\V\ps,\s\io;\V0)=\sum_{k=1}^\io\m^k
\V X_\su^{k\s}(\V\ps,\s\io;\V0),\qquad
\aa=\V\ps+\sum_{k=1}^\io\m^k\V X^{k\s}_\giu(\V\ps,\s\io;\V0)\Eq(4.12)$$
and, in fact, in this case $X^\s(\V\ps,t;\V0)$ is $t$--independent, and
also $\s$--independent.

If $0<J_0<J<+\io$ we can use the general KAM theory to say that the
above invariant torus does exist for $|\m|<b^{-1}$ small enough.  The
analysis in A) of the whiskered tori then yields, in this particular
case, that the motion $t\to X(t)=(\V X_\giu(t),\V X_\su(t))$, given by
the following \equ(4.13),\equ(4.14), is a quasi periodic motion on the
invariant torus:
$$\V X^h_\giu(t)=J^{-1}\big(\II^2\V F^h_\su(t)-\II^2\V
F^h_\su(0)\big),\qquad \V X^h_\su(t)=\II \V F_\su^h(t)\Eq(4.13)$$
where we have dropped the $\s$ labels as the functions $\V F,\V X$ no
longer depend on them. And, by \equ(2.14), if $\V
m=(m_1,\ldots,m_{l-1})$, $k^i_j\ge1$:
$$\V F^h_\su=-\sum_{|\V m|>0}
(\dpr_\aa f)_{\V m}(\oo t)
\sum_{(k^i_j)_{\V m,k-1}} \prod_{i=1}^{l-1}\prod_{j=1}^{m_i}
X^{k^i_j\s}_i\Eq(4.14)$$
because the pendulum and the whiskers really disappear as a consequence
of the decoupling.

The analysis in A) of the whiskered tori can be repeated to show that
$\V X^h_i,\V F^h_i$, for $i=1,\ldots,l-1$ ($\giu$ components) and for
$i=l+1,\ldots,2l-1$ ($\su$ components), are in $\MM^0$ for all $h\ge1$.
This is a stronger statement than \equ(3.3), valid only for the $\V X^k$
components and in the present special case: note that in our notations
$\MM^0$, with the label $0$ as superscript, means that not only the
functions in $\MM^0$ are bounded (\ie in $\MM$), but also that they have
$t$--degree $0$: see definition 1, \S3.

Furthermore:
$$\V F^{h\s}_{\su\V0}=\V0\Eq(4.15)$$
As in A) the property that $\V F^h_{\su\V0}=\V0$ is derived from the KAM
theorem, (as done in \equ(4.9)).  But also in this case one can check it
directly.  In this way we free ourselves from the KAM theorem and we are
in a position to study it independently.  This means that the property
$\V F^{k\s}_{\su\V0}=\V0$ and the \equ(4.13),\equ(4.14), coupled with
the initial condition $\V F^1_\su(t)=\dpr_\aa f(\oo t)$, (see
\equ(2.9)), {\it always} have a solution with $\V X^h$ quasi periodic
(with spectrum $\oo$).  Such solution gives us a formal power series
solution to the problem of finding the invariant tori equations.  In
this case we shall discuss and prove the convergence of such series.
The first proof ("non KAM") of \equ(4.15) is due to [CZ]: in \S6 we
shall find a more direct proof of \equ(4.15), see remark 3) after
\equ(6.18): which is very simple although far less general than [CZ].
But for clarity of exposition we prefer to appeal to [CZ] and continue,
without stopping to prove \equ(4.15).  \*

The above statements do dot require that $J$ be a constant: it can be,
for instance, a diagonal matrix (of dimension $l-1$), provided we
interpret conveniently the multiplications by $J^{-1}$. It is then
interesting to note that all the above statements remain true if some,
or all, the elements of the diagonal matrix $J$ become infinite. This is
not difficult to check by going through the classical proof of KAM
theorem, (it is, in fact, not easy to find it explicitly stated except
in the case in which either only one of the $J_i$ is infinite
(``periodically forced systems'') or all of them are infinite).  However
one finds a result with a analyticity radius (in $\m$) which, as
mentioned in the introduction, {\it is not bounded away from $0$
uniformly in the size of the largest among the non infinite $J_i$'s}.
Although this is nevertheless sufficient to prove that $\V
F^{k\s}_{\su\V0}=\V0$, it is reassuring that the algebraic check, that we
refer to above, is independently possible.

{\it If we let $J$ be a diagonal matrix and we allow for some, or all,
its elements to be $+\io$ and if we set $\h\=1$ in \equ(1.2),
and ask whether the torus $\V A=\V0$ can be analytically continued for
$\m$ complex, $|\m|<b^{-1}$, with $b$ being $J$--{\sl independent},
then we say that we are considering a "twistless KAM" problem.}

The convergence of the formal series for the tori equations studied in
\S7 will yield a radius of convergence, $b^{-1}$, independent on the
$J_j$ (a result stronger than the usual KAM theorem relying on the twist
property). Hence we show, by direct bounds, that the just posed
twistless KAM problem has a solution (a fact that could be checked also
by a careful exam of some of the classical proof of KAM theore, as
mentioned in the introduction).

This is less surprising if one studies the $J=+\io$ case, see [G1]:
problems 1,16,17 \S5.10, showing that the system is completely
integrable.  In the language of the present work the quoted reference
[G1] can be easily worked out: one finds that $\V X_\giu^h\=\V 0$ if
$h\ge1$ and $\V F^h_\su\=\V0,\V X^h_\su\=\V0$ if $h\ge2$, while $\V
X^1_\su(t)=\II(\dpr_\aa f)(t)$, so that:
$$\V A=\V A(\aa)\=-\m \sum_{\nn\ne\V0} i\nn\,f_\nn
\fra{e^{i\oo\cdot\nn t}e^{i\aa\cdot\nn}}{i\oo\cdot\nn},\qquad
\aa\=\aa\in T^{l-1}\Eq(4.16)$$
are the equations of the invariant, twistless, torus arising from the
unperturbed $\V A=\V0$ torus.
\*

{\bf C)} {\it Parity properties. (I).}
\*
\0We close this section by pointing out a few {\it parity properties}
of the operator $\II$, which will be useful below (see [CG], \S10).

Suppose that $M$ depends on other $l-1$ dimensional angles $\aa$ as a
linear combination (with $\s$--independent coefficients) of monomials:
$${(g\s t)^p\over p!} x^k\s^\chi\cos_{\chi'}(\oo\cdot\nn
t+\aa\cdot\mm)\= {(g\s t)^p\over p!} x^k\s^\chi \fra1{2i^{\ch'}}
\sum_{\r=\pm1}\r^{\chi'} e^{i \r (\oo\cdot \nn t +\mm\cdot
\aa)}\Eq(4.17)$$
with $\chi,\chi'=0,1$ and $\cos_{\chi'} y=\cos y$ if $\chi'=0$ and
$\cos_{\chi'} y=\sin y$ if $\chi'=1$, then $\II M$ has the same form.

We shall say that $M$ is {\it time-angle even} if $\chi+\chi'=$ even for
all monomials of $M$. If, instead, $\chi+\chi'=$ odd for all monomials
we say that $M$ is {\it time-angle odd}.  It then follows that the time
angle parities of $M$ and $\II M$ are opposite (when either is well
defined).

After the above remarks we make the inductive assumption that
$F^{h\s}(t;\aa)$ has action components $(+,\su)$, denoted symbolically
$d$, of odd time-angle parity in the above sense and angle components
$(-,\giu)$, denoted $p$, of even time angle parity. Opposite parity
assumptions will be made for $X^{h\s}(t;\aa)$. We shall write:
$$F^h=\pmatrix {p\cr d\cr},\quad X^h=\pmatrix{d\cr p\cr}\Eq(4.18)$$
dropping the label $\s$ from $F$ and $X$. In fact the main goal of the
above formalism is to treat simultaneously the stable and the unstable
whiskers: for $t>0$ it is $\s=1$ and $F^h,X^h$ represent $F^{h+},X^{h+}$
while for $t<0$, $\s=-1$ and $F^h,X^h$ represent $F^{h-},X^{h-}$. Hence
we can symbolically write:
$$\eqalign{
F^h=&\sum\,\d\,x^k\,(g\s t)^{k'}\s^\chi\cos_{\chi'}(\oo\cdot\nn
t+\aa\cdot\mm)\cr
X^h=&\sum\,\x\,x^k\,(g\s t)^{k'}\s^\chi\cos_{\chi'}(\oo\cdot\nn
t+\aa\cdot\mm)\cr}\Eq(4.19)$$
with suitable $\s$-independent coefficients $\d,\x$ and $\chi+\chi'=$
even for the $(+,\su)$ components and odd for the $(-,\giu)$ components
in the case of $X$, and with reversed parities in the case of $F$ (the
symbol $\cos_\chi y$ being defined after \equ(4.17)).

By remarking that $X(t)$ can be expressed via the wronskian:
$${X}(t)=W(t)\Bigl(X(0)+\ig^t_0W(\t)^{-1}F(\t)d\t\Bigr)\Eq(4.20)$$
where the wronskian matrices $W(t)$ and $W(t)^{-1}$ are respectively:
$$\tst(w_{jq})\=\pmatrix{
w_{00}(t)&0&w_{0l}(t)&0\cr
0&1&0&J^{-1}t\cr
w_{l0}(t)&0&w_{ll}(t)&0\cr
0&0&0&1\cr}\ ,\quad
(w_{jq}^{-1})\=
\pmatrix{w_{ll}(t)&0&-w_{0l}(t)&0\cr
0&1&0&-J^{-1}t\cr
-w_{l0}(t)&0&w_{00}(t)&0\cr
0&0&0&1\cr}\Eq(4.21)$$
($w_{jq},\,j,q=0,l$, being the matrix in \equ(2.15), with $w_{jj}$ even
in $t$ and $w_{l0},w_{0l}$ odd in $t$), one deduces immediately from the
above property of $\II$ that $X^h$ will have the {\it opposite}
structure to $F^h$ (\ie if $F^h=\pmatrix{p\cr d\cr}$ then
$X^h=\pmatrix{d\cr p}$).

The above remark and \equ(2.11) imply that if $X^{h'}$ has the structure
$\pmatrix{d\cr p}$ for $h'<h$ then $F^h$ has $\pmatrix{p\cr d}$
structure. And since it is obvious that $F^1$ has $\pmatrix{p\cr d\cr}$
structure, the \equ(4.18) follows by induction.

An important consequence of the parity of $X^k_+,X^k_\su$ is that if
$\aa=\V0$ they are even functions of $t$ so that, see \equ(4.17), if
$\chi=0$ it is $\chi'=1$ and viceversa. Hence
$X^{k+}_j(0,\V0)=X^{k-}_j(0,\V0)$ and we see that $\aa=\V0,\f=\p$ is a
homoclinic point (and more precisely $\a_j=0,\p,$ and $\f=\p$ ar
$2^{l-1}$ homoclinic points. In the following we study the homoclinic
point $\V\a=\V0,\f=\p$.
\*
{\bf D) \it Parity properties. (II).}
\*
It is important to study the following operators:
$$\eqalign{
U\,f(t)=&\fra1{2}\sum_{\r=\pm}\igb_{\r\io}^t J^{-1}(t-\t) f(\t)\,d\t\cr
V\,f(t)=&\fra12\sum_{\r=\pm}\igb_{\r\io}^t
(w_{0l}(t)w_{00}(\t)-w_{0l}(\t)w_{00}(t))f(\t)\,d\t\cr}\Eq(4.22)$$
Then one checks that $U,V$ {\it preserve the time angle parity of $f$ as
well as the holomorphy}. This is not the case of the operation $\II$:
the latter has {\it floating } integration axtremes (depending on the
siign of $t$): therefore it maps analytic functions of $t$ into functions
with a possible non analyticity at $\Re t=0$, unless the function is odd
in $t$.

Hence if $f$ is analytic in $t$ in a strip $\Im t<\x g^{-1}$ and time
angle even it will be:
$$U\,f(0^\s)=\s\sum_\nn u^1_\nn\cos a\,\aa\cdot\nn+
\sum_\nn u_\nn\,\sin\aa\cdot\nn\Eq(4.23)$$
for suitable coefficents $u^1,u$; and similarly for $V$. But \equ(4.23)
has to be analytic; hence:
$$U\,f(0)=\sum_\nn u_\nn\sin\aa\cdot\nn,\qquad V\, f(0)=\sum_\nn
v_\nn\sin\aa\cdot\nn\Eq(4.24)$$
for suitable coefficients $u_\nn,v_\nn$. If $f$ is analytic and odd we
exchange the role of sines and cosines.

Note, as implicit in the discussion of C) and D) above,
that if $f$ time angle even and analytic then $\II f(0^\s)$ has the form
$\sum_\nn (f^1_\nn\sin \aa\cdot\nn+\,f^2_\nn \s\cos \aa\cdot\nn)$ ,
\ie the generally present (in the
non analytic cases) part $\sum_\nn\,f^2_\nn \s\cos \aa\cdot\nn$ is not
necessarily $0$; and if
$f$ is time--angle odd and analytic then $\II f(0^\s)=
\sum_\nn (f^1_\nn\,\s\,\sin \aa\cdot\nn+f^2\nn\cos\aa)$:
\ie "no simplification" occurs because of analyticity.

We see that the parity properties, although very simple, can become
quite intricate to visualize.

\vskip1.truecm

\penalty-200

%\ciao
\fiat

{\bf\S5 Trees, roots, nodes, branches and fruits: the formalism.}\pgn=1
\vskip0.5truecm\numsec=5\numfor=1\numfig=1

\penalty10000

\vglue0.5truecm
\let\tto=\Rightarrow
We develop a graphical formalism to represent, via \equ(4.10) and
\equ(2.14), the generic $h$-th order contribution to various quantities
related to the invariant tori and their whiskers.

We shall consider, for instance, the whiskers splitting in model 2),
\equ(1.1), at the point with coordinates $\f=0$ and $\aa$, which is
$\D_{j}^k(\aa)=X^{k+}_{j}(0;\aa)- X^{k-}_{j}(0;\aa)$ if $j$ is an
action component subscript (hence a homoclinic point (at $\f=0$)
corresponds to the values $\AA$ such that $\D^k_j(\aa)=0$).  We shall
also consider, as a second example, the {\it homoclinic scattering
phase shifts} (a notion introduced in [CG]), and as a third example the
parametric equations of the invariant tori in model 1).

The case of model 1) can be regarded as a special case of model 2),
with $f$ independent on the pendulum position $\f$, hence the two
problems can be treated with the same formalism: although the latter is
simpler and, if independently formulated, would require a slightly
easier analysis.

Recall our label convention, in \equ(2.7), for the action angle
variables: we label with $j=0$ the $\f$ angle, with $j=1,\ldots,l-1$
the angles $\a_1,\ldots,\a_{l-1}$ and we label with $j=l$ the pendulum
action $I$ and with $j=l+1,\ldots,2l-1$ the rotators actions $\AA$.
The $F^k$ are given by \equ(2.14) and the $X^k$ are related to the
$F^k$ by \equ(4.10).

Note that the action angle labels are enumerated, between $0$ and
$2l-1$, in the order in which they appear in \equ(4.10): because in
\equ(2.14) {\it a subscript $-$ or $+$ is synonimous of a subscript $0$
or $l$; and $\giu$ or $\su$ denote, collectively, the subscripts
$j=1,\ldots,l-1$ or $j>l$}.  Furthermore $F^k$ depends only on $X^h_j$,
$h<k$, with $j=0,\ldots,l-1$, (\ie it depends only on the the lower
order $-,\giu$ components of $X^k$; this is a consequence of the $f$ in
\equ(1.1) depending only on the angles $\f,\aa$).

We imagine to use \equ(2.14) recursively to express everything in terms
of $F^1$ only. The structure of \equ(2.14), (or, more generally,
\equ(2.11); see also [G1] Ch.  5, \S11), leads us naturally to a
graphical representation, quite familiar in perturbation theories, see
[G2].  This fact was noted explicitly in the context of KAM theory by
[E], (see also [V]).

We can see a tree grow out, of \equ(2.14), \equ(4.10) as follows.

First we represent $F^h_j(\t)$ as a fat point $v$:
$$F^h_j(\t)=_{\quad v}\bullet^{\t,h,j}\Eq(5.1)$$
Then we consider the improper integration operations with upper limit
$t$, denoted $\OO,$ $J^{-1}\lis\II^2$, $\OO_+$, $\II$ in \equ(4.10).  We
represent them with a line segment of unit length, called {\it branch},
joining two points $r,v$, and calling $r$ the {\it branch root}, and $v$
the {\it branch node}.  The branch will bear a {\it branch label} $j=0$
when representing $\OO$, or a label $j=1,\ldots,l-1$ for $J^{-1}\lis
\II^2$, or $j=l$ for $\OO_+$, or $j=l+1,\ldots,2l-1$ for $\II$, see
\equ(4.10).  A label $t$ will be attached to the root $r$:
$$\hbox{\hglue0.2cm}_r^t\,
\raise6pt\hbox{$\st j\atop{\st \bullet}
\kern-2pt\raise1pt\hbox{\vrule height0.2pt width3.cm}$}
\kern-3pt\raise3pt\hbox{${\st \bullet}_v$}\Eq(5.2)
$$
Then we can represent the whole formula \equ(4.10) as:
$$X^h_j(t)= \hbox{}^t_r\,\raise6pt\hbox{$\st j\atop{\st \bullet}
\kern-2pt\raise1.5pt\hbox{\vrule height0.2pt width3.cm}
\kern-0.9pt{\displaystyle\bullet}_{\displaystyle v}^{\displaystyle
\t,h,\tilde\jmath}$}\Eq(5.3)$$
and there is an implicit constraint as $j,\tilde\jmath$ cannot be
independent, as one sees from \equ(4.10): in fact
$j=0\tto\tilde\jmath=l$ (first or third line in \equ(4.10));
and $j=1,\ldots,l-1\tto\tilde\jmath=j+l$ (second line in \equ(4.10));
and $j\ge l\tto\tilde\jmath=j$ (fourth line in \equ(4.10).

Going back to the representation of $F^h_j(\t)$ as a fat point,
\equ(5.1), we can use \equ(2.14) to express it in terms of
$X^{h^i_p}_i(\t)$ with $\sum_{i=0}^l\sum_{p=1}^{m_i}
h^i_p=$ $h$ or $h-1$.

Using the representation \equ(5.1) for $X_s^{h^s_p}(\t)$ we see that
\equ(2.14) can be written, ($j\ge l$):

\insertplot{240pt}{130pt}{
%(5.4)
\def\nn{{\V \n}}
\ins{-55pt}{70pt}{$\dt F^h_j(\t)=_{\quad v} \bullet^{\t,h,j}\ =\
\sum\fra1{\prod_{i=0}^{l-1} m_i!}
\quad{}_v\,$}
\ins{205pt}{120pt}{${}^{\t,h^0_1,l}$}
\ins{205pt}{100pt}{${}^{\t,h^0_2,l}$}
\ins{205pt}{80pt}{${}$}
\ins{205pt}{20pt}{${}^{\t,h^{l-1}_{m_{l-1}},2l-1}$}
\ins{295pt}{70pt}{$\eq(5.4)$}
}{f1}

\0where the first $m_0$ branches are labeled $0$, the next $m_1$ are
labeled $1$, etc.; $\d=0,1$ and $\sum h^i_p=h-\d$, because of the
meaning of the symbols $(k_p^i)_{\V m,k}, (k_p^i)_{\V m,k-1}$ in
\equ(2.14).

The vertex $v$, that we call a {\it node}, corresponds to the factors $\V
m!\, (g^2 J_0 \sin\f)_{\V m}(\f^0)$ and $\V m!\,(\dpr f)_{\V m}$
appearing in \equ(2.14) and deprived of the combinatorial factor $\V
m!=\prod_{i=0}^{l-1} m_i!$, see \equ(2.14), \equ(2.12).  And the label
$\n$ is introduced to split such factors as sums of their Fourier
components.  Namely, see \equ(1.1), let:
$$\eqalign{ f^{\d}(\aa,\f)\=& \sum_{\n=(n,\nn)} f^\d_\n \, e^{i
(n\f+\nn\cdot\aa)} ,\qquad \d=0,1\cr
f^0(\f,\aa)\=&J_0
g^2\cos\f=\sum_{\n,\,\nn=\V0\atop n=\pm1} \fra{f^0_\n}2 \,e^{i n\f}
,\qquad f^1(\f,\aa)\=\sum_\n\fra{f_\n^1}2\, e^{i
(n\f+\nn\cdot\aa)}\cr}\Eq(5.5)$$
then the factor represented by the node $v$, with its labels $\t,\n,\d,j$
is:
$$\fra{i \n_{j-l}}2\cdot\Big(\prod_{s=1}^{l-1}(i\n_s)^{m_s}\Big)\,
f^\d_\n e^{i (n\f^0(\t)+\nn\cdot(\aa+\oo \t))}\Eq(5.6)$$
where $j-l$ is the branch label of the branch leading to $v$,
and the introduction of the above Fourier representation is convenient
as it eliminates the derivatives with respect to $\f,\aa$ in the
coefficients of \equ(2.14).

We call $\t$ a {\it time label}, $\n$ a mode label, $\d$ an {\it order}
label and $\tilde\jmath$ an {\it action} label associated with the node
$v$.

One can further simplify the \equ(5.4) by imagining that the
$m=\sum_{i=0}^{l-1} m_i$ branches coming out of $v$ are distinguished by
a number counting them from $1$ to $m$ and appended to them, which
however is not explicitly written and is thought as associated with the
branch. Then we drop the condition that the branch labels
$j=0,\ldots,l-1$ are in non decreasing order and rewrite \equ(5.4) as:
\insertplot{240pt}{120pt}{
%(5.7)
\ins{-25pt}{85pt}{$\dt F^h_j(\t)=_{\quad v} \bullet^{\t,h,j}\ =\
\sum\fra1{m!}
\quad{}_v\,{\st\bullet}
\raise14pt\hbox{\kern-8pt\hbox{$\eqalign{\st \t&\st \n\cr
\noalign{\vskip-10pt}\st \d&\st j\cr}$}}$}
\ins{205pt}{120pt}{${}^{\t,h_1,\tilde\jmath}$}
\ins{205pt}{100pt}{${}^{\t,h_2,\tilde\jmath}$}
\ins{205pt}{80pt}{${}$}
\ins{205pt}{20pt}{${}^{\t,h_m,\tilde\jmath_m}$}
\ins{295pt}{70pt}{$\eq(5.7)$}
}{f2}
\0so that we can put freely the labels without worrying about the order
(if we did not replace the combinatorial factor in \equ(5.4) with that
of \equ(5.7) we would count the same term $m!\prod_{i=0}^{l-1}m_i!^{-1}$
times and this explains the change in the combinatorial factor).

It is now easy to write \equ(4.10) with the $F^h$ in it expressed in
terms of lower order $X^{h'}$, $h'<h$, via \equ(2.14): thinking $X^h$
as written in \equ(5.3) and \equ(2.14) as written in \equ(5.7), we
get:
\insertplot{240pt}{120pt}{
%(5.8)
\ins{-25pt}{88pt}{$\dt X^h_j(t)=\sum\fra1{m_{v_0}!}
\ \lower6pt\hbox{$\tst r $}
{\st\bullet}
\kern-2pt\hbox{\raise4pt\hbox{$\st j\atop
\vrule height0.2pt width2.cm depth.1pt$}}
\kern-3pt{\st\bullet_{
\raise-6pt\hbox{$\tst\kern-5pt
v$}}}\kern-20pt\raise14pt\hbox{$\eqalign{\st \t_{v_0}&
\st \n_{v_0}\cr\noalign{\vskip-4pt}
\st \d_{v_0}&\st j_{v_0}\cr}$}$}
\ins{205pt}{120pt}{${}^{\t,h_1,\tilde\jmath}$}
\ins{205pt}{100pt}{${}^{\t,h_2,\tilde\jmath}$}
\ins{205pt}{80pt}{${}$}
\ins{205pt}{20pt}{${}^{\t,h_{m_{v_0}},\tilde\jmath_{m_{v_0}}}$}
\ins{295pt}{70pt}{$\eq(5.8)$}
}{f3}
where the sum is over all the possible choices of the labels, subject to
the conditions described in their definitions ({\it e.g.}
$j_{v_0}=\tilde\jmath$).

We can now iterate the representation \equ(5.8) by replacing the fat
points by nodes out of which new branches emerge, until we reach fat
points with order labels $h=1$, where the expansion ends and we call the
last fat points {\it top nodes}, drawing them as slim as the other
nodes.  The root $r$ of the first branch drawn {will not} be regarded as
a node, but it will be called {\it root}.

We see that the above expansion leads to a sum over {\it trees} bearing
"decorations": such trees have branches emerging from the same node
which are regarded as distinguished by a number, never marked in the
drawings, see fig. 1 below. On such trees a natural group of
transformations acts: it is generated by the following operations; fix a
node $v$ and permute the subtrees emerging from it. Two trees that can
be transformed into each other by a transformation of the group just
defined will be regarded as identical.

\insertplot{240pt}{170pt}{%fig.tex
\def\nn{{\V \n}}
\ins{-35pt}{90pt}{\it root}
\ins{-10pt}{100pt}{$0^\s$}
\ins{25pt}{110pt}{$j_\l$}
%\ins{15pt}{80pt}{$h_{\l_0},\nn_{\l_0}$}
\ins{60pt}{85pt}{$v_0$}
\ins{50pt}{125pt}{$\matrix{\t_{v_0}\,\n_{v_0}\cr\d_{v_0}\,j_{v_0}\cr}$}
%\ins{115pt}{106pt}{$h_{\l_1},\nn_{\l_1}$}
\ins{115pt}{132pt}{$j_{\l_1}$}
\ins{152pt}{120pt}{$v_1$}
\ins{140pt}{165pt}{$\matrix{\t_{v_1}\,\n_{v_1}\cr\d_{v_1}\,j_{v_1}\cr}$}
\ins{110pt}{50pt}{$v_2$}
\ins{190pt}{100pt}{$v_3$}
\ins{230pt}{160pt}{$v_5$}
\ins{230pt}{120pt}{$v_6$}
\ins{230pt}{85pt}{$v_7$}
\ins{230pt}{-10pt}{$v_{11}$}
\ins{230pt}{20pt}{$v_{10}$}
\ins{200pt}{65pt}{$v_4$}
\ins{230pt}{65pt}{$v_8$}
\ins{230pt}{45pt}{$v_9$}
}{f4}
\kern1.3truecm
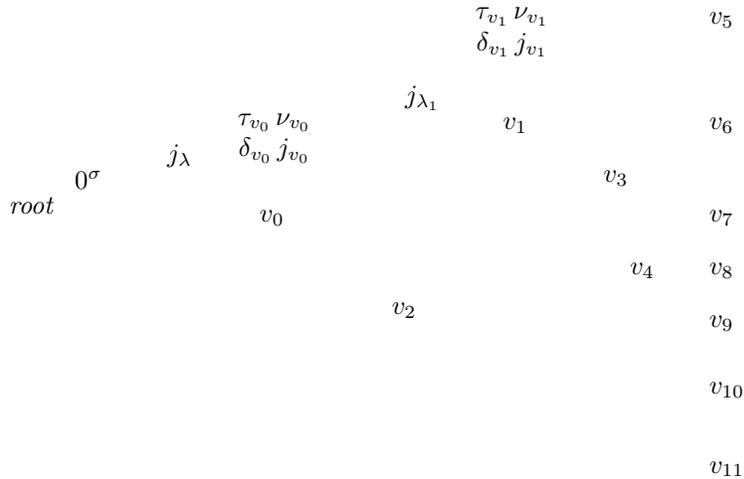
\didascalia{fig. 1: A tree $\th$ with
$m_{v_0}=2,m_{v_1}=2,m_{v_2}=3,m_{v_3}=2,m_{v_4}=2$ and $m=12$,
$n(\th)$=$n(\th_1)n(\th_2)$ $=(1\cdot1^2)(2!2!)=4,$ $\prod
m_v!=2^4\cdot6$, and some decorations.}

We call the above trees {\it semitopological trees}: we call {\it
topological trees} the trees in which the branches emerging from the
same node are not numbered, and also such trees will be identified when
superposable modulo the action of the same group of transformations.

A third type of trees, that we call "numbered trees" or simply {\it
trees}, are obtained by imagining to have a deposit of $m$ branches
numbered from $1$ to $m$ and depositing them on the branches of a
topological tree with $m$ branches: numbered trees that can be
superposed  by transforming them with the above group of tranfsormations
will still be regarded as identical. The number of such trees asociated
with a given semitopological tree is $m!\,\prod_v m_v!^{-1}$ and using
such trees, which will be the only ones considered from now on, unless
otherwise stated, the \equ(5.8) becomes:
$$X^h_j(t)=\sum_{\th \in trees}\fra1{m(\th)!}\sum_{labels;\,
\sum_v\d_v=h} \,\Big[{\rm Fig.\ 1}\Big]\Eq(5.9)$$
where $m(\th)=$ number of branches of $\th$; and the drawing of Fig. 1
symbolizes, now, a well defined hierarchical chain of operations of
improper integrations.\footnote{$\hbox{}^4$}{\nota More formally a
topological tree is a partially ordered set with the property that any
two elements follow some common predecessor (hence there is a minimum
element, or {\it root}) and have no common follower if they are not
comparable.  This is visualized (see fig. 1 above) by representing its
elements by oriented unit segments.  We draw the first segment (in the
partial order) and attach to its endpoint the $m_{v_0}\ge0$ segments
representing its immediate followers, and so on.  The endpoints of the
segments become {\it nodes} of the tree: to each node $v$ is associated
a {\it branching number} $m_v\ge0$ as the segments will be called {\it
branches}.  The first branch will be called the {\it root branch} and
its first point will not be a node, but it will, nevertheless, be called
the {\it root}.  The endpoint $v_0$ of the root branch will be called
the {\it first node}.  Two topological trees giving rise to the same
partially ordered set will be considered identical.  Each node $v$ with
$m_v>0$ can be regarded as the root of a {\it subtree}: and the
operation of graphical permutation of two subtrees emerging from the
same node establishes an equivalence relation between trees which is a
notion coinciding with that of giving rise to the same partially ordered
set.
\acapo
It is clear that the set of topological trees with $m$ branches
contains at most as many elements as the random walks on the lattice of
the integers $\ge1$ with $2m-2$ steps: hence the number of topological
trees with $m$ branches is bounded by $2^{2m-2}$.}

The expression \equ(5.9) is very convenient, as we shall see, from a
combinatorial point of view. It is however clear tha the summation in
\equ(5.9) over the labels produces a number $w(\th)$ depending {\it
only} on the topological tree associated with $\th$, \ie it does not
depend on the branch identification numbers (hidden in the pictures)
of the branches. Therefore in real calculations the above sum over trees
can be replaced by a single sum over semitopological trees, or by an
even simpler sum over topological trees.\footnote{${}^{5}$}{\nota
For this purpose the following identity is used.  A natural
combinatorial factor $n(\th)$ that can be associated with a topological
tree $\th$ that bifurcates at the first node $v_0$ into $m_{v_0}$
subtrees among which there are only $q$ topologically different
subtrees $\th_1,\ldots,\th_q$ each of which is repeated
$p_1,\ldots,p_q$ times, is:
$$\tst n(\th)=\prod_{i=1}^q p_i!\,n(\th_i)^{p_i}\Eq(5.10)$$
Such combinatorial factors are useful when one has to consider sums
over trees $\th$ with $m$ branches (root branch included) of functions
$F(\th)$ whose values depend only on the topological tree.  In such
cases one finds:
$$\tst {1\over m!}\sum_{\th\in\,trees} F(\th)\=
\sum_{\th\in\2topological}(\prod_{v\in \th}{1\over m_v!})\,F(\th)\=
\sum_{\th\in topological}{1\over n(\th)}\,F(\th)\Eq(5.11)$$
The above identity is closely related to Cayley's formulae, see also
[G2], (6.1), (6.2) and (5.13)), and [FG].  It simply reflects that the
number $N(m;\{m_v\})$ of trees with $m$ branches and branching numbers
$m_v$, hence generating the same semitopological tree, is $m!\prod_v
m_v!^{-1}$ (which is (an adaptation of) Cayley's formula, (see [H],
\S1.7)), while the number of trees generating the same topological tree
$\th$ is $m!\,n(\th)^{-1}$.}

Let us fix some terminology and conventions: given a tree $\th$ and a
family of labels for it we denote $\Th$ their pair and call it a {\it
labeled tree}.
\item{(a) }the trees are dawn from left to right and are regarded as
partially ordered sets of nodes in the obvious sense (see footnote
$\hbox{}^4$ above), with the higher nodes to the right (\ie we draw them
as "fallen trees", unfortunately: the vertical notation would have
required too much space). The branches are naturally ordered as well;
all of them have two nodes at their extremes (possibly one of them is a
top node) except the lowest or {\it first} branch which has only one
node, the first node $v_0$ of the tree. The other extreme $r$ of the
first branch, (which is the root of the branch), will be called the {\it
root} of the tree and it {\it will not} be regarded as a node.\acapo If
$v_1$ and $v_2$ are two nodes we say that $v_1<v_2$ if $v_2$ follows
$v_1$ in the order established by the tree: \ie if on has to pass $v_1$
before reaching $v_2$, while climbing the tree.

\item{(b) }each node carries a {\it time label} $\t_v$,
a {\it mode label} $\n_v$, and {\it order label} $\d_v$ and an {\it
action label} $\tilde\jmath_v=l,\ldots,2l-1$.  Each branch carries and
{\it angle label}, $j_\l=0,\ldots,l-1$; but if $\l$ is the root branch
its label can also be an {\it action label} $j_\l\ge l$.  If $\l$ leads
to $v$ then $j_\l=\tilde
\jmath_v+l$ if $j_\l=0,\ldots,l-1$, while $\tilde\jmath_v=j_v$
if $j_\l=l,\ldots,2l-1$ (which
is allowed only if $\l$ is the root branch).

\item{(c) } the {\it order} $h(\Th)$ of the labeled tree $\Th$ is $h=
\sum_v \d_v$, \ie the sum of the order labels of the nodes. The number of
branches emerging from the node $v$ is $m=1+\sum_v m_v$ (as one has to
count also the root branch). Of course, as the order label $\d_v=0,1$
and as each node $v$ with $\d_v=0$ {\it must} have $m_v\ge2$, (see above
and \equ(2.14)), it is $h\le m<2h$.

\item{(d) }the {\it momentum} of a node $v$ or of the branch $\l_v$
leading to $v$ is $\nn(v)=\sum_{w\ge v} \nn_w$, if $\n_v=(n_v,\nn_v)$ is
the {\it mode label} of $v$. The {\it total momentum} is
$\nn(v_0)\=\sum_{v\ge v_0}\nn_v$.

\item{(e) }\0Given all the above decorations on a labeled tree $\Th$
{\it we define its
value} $V_j(t;\Th)$ via the following operations:
\acapo%
(1) We first lay down a set of parentheses $()$ ordered hierarchically
and reproducing the tree structure (in fact any ordered (topological)
tree can be represented as a set of matching parentheses representing
the tree nodes). Matching parentheses corresponding to a node $v$ will
be made easy to see by appending to them a label $v$.  The root will not
be represented by a (unnecessary) parenthesis.
\acapo
(2) Inside the parenthesis $(_v$ and next to it we write, setting
$\tilde\jmath_v\=j_v-l$, the {\it node function}, see \equ(5.6):
$$\tst\Big(_v-{1\over2}i(\n_v)_{\tilde\jmath_v}\,f^{\d_v}_{\n_v} e^{i
(n_v\f(\t_v)+(\aa+\oo \t_v)\cdot\nn)}\,\prod_{s=0}^{l-1}
(i\n_{vs})^{m^s_v}\Eq(5.12)$$
here $j_v$ is the branch label of the branch leading to $v$.  The last
product is missing if no nodes follow $v$.
(3) Furthermore out of $(_v$ and next to it we write a symbol
$\EE^T_{v}$ which we interpret differently, depending on the label
$j'=j_{\l_v}$ on $\l_v$ and on the action label $j^{\prime\prime}=
\tilde\jmath_v$ on $v$:
$$\tst
\EE_{v}^T \Big(_v \cdot \Big)_v \= \cases{\OO\Big(_v  \cdot\Big)_v
(\t_{v'}),\quad &if $v>v_0\ ,\quad j_{\l_v}=0$; \cr J^{-1} \lis \II^2
\Big(_v  \cdot \Big)_v (\t_{v'})\ ,\quad &if $v>v_0\ ,\quad 1\le
j_{\l_v}\le l-1$;\cr} \Eq(5.13)$$
for $v>v_0$, otherwise:
$$\tst\EE_{v}^T \Big(_v \cdot \ \Big)_v
\= \cases{\OO\Big(_v \cdot\Big)_v (t^\s)\ , & if $v=v_0\ ,
\quad j_{\l_v}= 0$\ ,\cr
J^{-1} \lis\II^2
\Big(_v  \cdot \ \Big)_v (t^\s) &if $v=v_0\ ,\quad 1\le j_{\l_v}\le
l-1$\ ,\cr \OO_+\Big(_v \cdot\ \Big)_v (t^\s)\ ,\quad &if $v=v_0\
,\quad j_{\l_{v}}=l$; \cr \II \Big(_v   \cdot\ \Big)_v (t^\s)\ ,\quad &if
$v=v_0\ ,\quad l+1\le j_{\l_v}\le 2l-1$\cr} \Eq(5.14)$$
$t$ being the root time label of the tree and the superscript $\s$
attached to $t$ is important only if $t=0$: in such case \equ(5.14) has
to be interpreted as the limit as $t\to0^\s$.
\*

{\bf Remarks:}

{\bf 1)} One realizes that the giant symbol thus constructed
has a perfectly defined meaning as a hierarchically ordered chain of
operations $\II$: it gives a ``single contribution" to the value $
X^{h}_{j\nn}(t;\aa)\= X^{h}_{j\nn}(t)e^{i\aa\cdot\nn}$.  For instance:
given $\s=\pm$ multiply $\s$ times all the above values of the trees
$\Th$ with order $h(\Th)=h$, momentum $\nn$, and $j>l,t=0^\s$.  {\it
Such sum will give the $h$-th order contribution
$\D^h_{j}=X^h_j(0^+;\aa)-X^h_j(0^-;\aa)$ to the homoclinic splitting}.

{\bf 2):} If we do not perform the operation $\EE^T$ relative to the time
$\t_{v_0}$ of the first node $v_0$ and set it to be equal to $t$,
setting also $j\=j_{v_0}$, we see
that the result is a representation of $F^h_{j_{v_0}}(t)$, if $j_{v_0}$
is the label of the node $v_0$.

{\bf 3):} Note that if $\aa=\V0$ then we are at a homoclinic point,
because the hamiltonian \equ(1.1) is even: so that the sum in remark 1)
above yields a value $0$ for all components $j=l,\ldots 2l-1$, see the
final comments in \S4, C), and all fixed tree shapes.

{\bf 4):} Suppose that one whishes to study model 1) in \equ(1.1).
For instance suppose one wants to determine the parametric equations for
the invariant torus run quasi periodically with angular velocity $\oo$,
analytically continuing the unperturbed torus $\V A=\V0$.
The \equ(4.10)
must be replaced by \equ(4.13) and \equ(2.14) by \equ(4.14), while
\equ(4.9) remains the same. But {\it very little} has to be changed in
the above graphical representations.%
\acapo
In fact it is sufficient to consider, in the sum $X^h_{j,\nn}(t)=
\sum_{\Th:\,h(\Th)=h} V_{j\nn}(t;\Th)e^{i\nn\cdot\aa}$, only trees
$\Th$ with {\it
no node or branch labels} $j$ equal to $0,l$ (\ie all the labels have to
be rotators labels); and $\d_v\=1$, so that $m(\Th)=h(\Th)=h$. This is
obvious because we can think that the pendulum is present but that it is
completely decoupled from the rotators.%
\acapo
The functions $\V X^h_\giu(t), \V X^h_\su(t)$ thus defined are, of
course, the solutions of the equations of motion of a point starting at
$t=0$ on the above invariant torus $\TT$ at a angular position $\aa$ and
at action position $\AA_\su(0)=\sum_{h=1}^\io \m^h \V X^h_\su(0)$, for
$\m$ small enough (this is a convergent series for $\m$ small, at fixed
$J_j$, by KAM, as already mentioned): if, in view of the aim of {\it
proving} the theorem by directly showing the convergence of the series,
one does not want to use the KAM theorem then the latter series will
only be a formal power series solution to the problem of finding a
motion on the invariant torus $\TT$.
\acapo
Therefore setting:
$$\aa(t)=\aa+\oo t+\sum_{h=1}^\io \m^h \V X_\giu(t),\qquad
\V A(t)=\sum_{h=1}^\io\m^h\V X_\su(t)\Eq(5.15)$$
and letting $t$ vary in $[0,+\io)$, or $(-\io,0]$ (or, for that matter,
in any infinite connected subinterval of the above two) we describe (at
least formally) a dense set on the torus; \ie the torus itself.

The formalism necessary to see the cancellations, and to make use of
them, is completed.
\vskip1.truecm
\penalty-200

\fiat
{\bf\S6 Cancellation mechanisms.}\pgn=1

\penalty10000

\vskip0.5truecm\numsec=6\numfor=1\numfig=1

\penalty10000

\vglue0.5truecm

\0{\bf A):} {\it Homoclinic cancellations.}

{\it 1) (representation of $\EE^T$):}

To study cancellation mechanisms for the whiskers splitting we consider
trees with root time $t=0$, (\ie $t=0^\s$), and we introduce a useful
{\it representation of the $\EE$ operations}.  Given a tree $\th$
contributing to order $h$ and a total momentum $\nn$, let $v$ be a
node.  We fix our attention on one such node $v$ and call $v'$ its
predecessor: the case in which $v$ is a top node and the predecessor
$v'$ of $v$ is the root is the simple first order case, so that this
case will not be considered as it can be studied easily by direct
evaluation and no cancellations occurr in it, in general, (hence $v'$
will not be the root and $v>v_0$).

Supposing that $v'$ is not the root, we realize that the ``integral"
describing the $\II$ operation (associated with the vertex $v$) can be
written, if $j=j_{\l_v}$ is a $\giu$ component (\ie if $0<j \le l-1$, see
\equ(4.6)):
$$\tst J^{-1}\Big(\igb_{(\s)}^{\t_{v'}}(\t_{v'}-\t_v)
S_{v,\th}(\t_v)d\t_v-\igb_{(\s)}^0(-\t_v)S_{v,\th}(\t_v)d\t_v\Big)
\Eq(6.1)$$
with $S_{v,\th}(\t_v)$ being the result of the operations performed to
evaluate the integrals and sums inside the parentheses $(_v)_v$.  Note
that the $S_{v,\th}(\t)$ depends only on the subtree $\th_v\subset\th$
rooted at $v'$ and consisting of the nodes following $v$ in $\th$ (and
bearing all the decorations of $\th$).

Calling $t\=\t_{v'},\,\t\=\t_v$, using \equ(3.13),\equ(3.14),
we can replace the above expression by:
$$\tst J^{-1}\Big(
\ch(\s<0)t\igb_{-\io}^{+\io} S_{v,\th}(\t)d\t+
\igb_{+\io}^t (t-\t)S_{v,\th}(\t)\,d\t-
\igb_{+\io}^0(-\t)S_{v,\th}(\t)\,d\t\Big)\Eq(6.2)$$
Note that a rearrangement has been necessary to simplify the term with
the characteristic function as well as the other; here we define
$\igb_{-\io}^{+\io}$ to be identical to
$\igb_{-\io}^{0^-}-\igb_{+\io}^{0^+}$, see \equ(3.14).

A more symmetric representation is obtained by noting that a ``mirror"
formula must hold with $-\io$ playing the role of $+\io$.  Averaging
over the two formulae for the same quantity \equ(6.1) we get:
$$\eqalignno{
&J^{-1}\Big(-{1\over2}
\s t \igb_{-\io}^{+\io} S_{v,\th}(\t)\,d\t+\2\Big[
\igb_{+\io}^t (t-\t)S_{v,\th}(\t)\,d\t+
\igb_{-\io}^t (t-\t)S_{v,\th}(\t)\,d\t\Big]+\cr
&+\2\big(\igb_{+\io}^{0}\t S_{v,\th}(\t) d\t+
\igb_{-\io}^{0}\t S_{v,\th}(\t) d\t)\Bigr)&\eq(6.3)\cr}$$
such symmetrization is not really needed, but it has some aestethic
value.

A similar representation can be achieved also for the case $j=j_{\l_v}=0$:
$$\eqalign{\tst
&\ch(\s<0)w_{0l}(t)\ig_{-\io}^{+\io} w_{00}(\t) S_{v,\th}(\t) d \t
+\Big(w_{0l}(t)\ig_{+\io}^tw_{00}(\t)S_{v,\th}(\t)d \t- \cr&-
w_{00}(t)\igb_{+\io}^t w_{0l}(\t) S_{v,\th}(\t)d \t\Big)+
w_{00}(t)\igb_{+\io}^0w_{0l}(\t)S_{v,\th}(\t)d \t\cr}\Eq(6.4)$$
see \equ(2.17) and symmetrization over the choice of $\pm\io$ yields
the more symmetric formula:
$$\eqalign{\tst
&-{1\over2}\s w_{0l}(t)\igb_{-\io}^{+\io} w_{00}(\t) S_{v,\th}(\t) d\t+\cr
&+{1\over2}\Big[\igb_{+\io}^t\Big(w_{0l}(t)w_{00}(\t)-w_{00}(t)w_{0l}(\t)
\Big)S_{v,\th}(\t)d\t+\igb_{-\io}^t ({\rm same})d\t\Big]+\cr
&+\2 w_{00}(t)\Big(
\igb_{+\io}^{0}w_{0l}(\t)S_{v,\th}(\t)d\t+
\igb_{-\io}^{0}w_{0l}(\t)S_{v,\th}(\t)d\t\Big)\cr}\Eq(6.5)$$

We describe the above representation of the contribution of the considered
tree branch by affixing a ``bubble" around the node $v$ and enclosing all
the branches following it: the bubble will have a wiggly boundary or a
smooth boundary to distinguish between the first and the third terms in
the above sums. If a node is not enclosed in a bubble that cuts the
branch linking it to the previous node then this means that in the
above sum we selected instead the intermediate term: in this case we
mark the node $v$ by a label $\RR$ (in analogy with renormalization
theory, see [G2]; here $\RR$ has nothing to do with the parameter $R$
appearing in the definition of $\II$).

We can repeat the same representation operation on the (improper)
integrations pertaining to all the other nodes, starting from the
highest on the tree and going down towards the root: we stop at the
first nodes following $v_0$ (the latter node is somewhat
different from the others as the corresponding $\EE^T$ integration has a
un upper limit $t$ fixed).

{\it Thus the most general contribution to $X^h_j(t)$ by trees with
order $h$ will be represented by a tree with $h$ nodes with index
$\d=1$ (and up to $2h-1$ branches, see c) preceding \equ(5.12)) and all
the labels introduced so far ($\RR$ labels included) plus an arbitrary
number of bubbles drawn around tree nodes above $v_0$, and drawn as
described (avoiding overlappings).} Of course we could replace the
bubbles with labels affixed on the nodes following the root: but we
prefer the ``bubbles notation" as it reminds us of the analogous
bubbles that can be used to describe the renormalization cancellations
in quantum field theory, (see [G2]).

We call {\it free} the branches that are not enclosed inside bubbles of
any type.  The total {\it free momentum} of the tree will be the sum of
the node modes associated with the free nodes: {\it note that it is
quite different from the total tree momentum previously defined (see (d))
preceding \equ(5.12)}.

Looking at \equ(6.3),\equ(6.5) we see that the first and the last
terms contain constant factors which can be taken out of the integration
operations associated with the other tree nodes.  Thus the bubbles
involve simpler integrals. Furthermore the three terms in the
\equ(6.3),\equ(6.5) individually preserve the time--angle parities,
as all the operations in \S4 did: this is essential as it will permit
us to use the parity arguments derived at the end of \S4 (see C) and D)
in \S4).

\vskip0.3truecm
{\it 2): (Resummations):}

Consider a bubble containing a subtree of
order $h$ linked to its root node $v'\ge v_0$ by a branch $v'v$
carrying a momentum $\nn=\sum_{w\ge v}\nn(w)$
and a label $j$ ($0\le j\le l-1$).  Fixed
$\nn,j,h$ we can sum over the possible choices (consistent with the
labels $\nn,j,h$) of the subtree and its decorations:
$$\eqalignno{\textstyle
\b^{h,1}_{\nn j} e^{i\nn\cdot\aa}=&
\cases{
(J g^2)^{-1}
\sum{1\over 2}\igb_{-\io}^\io F^h_{\nn j}(\t)\,d g\t\ , &$j=1,...,l-1$\cr
(J_0g^2)^{-1}
\sum\2\igb_{-\io}^\io w_{00}(\t) F^h_{\nn j}(\t)\,d g\t\ , &$j=0$\cr}
&\eq(6.6)\cr \b^{h,2}_{\nn j}e^{i\nn\cdot\aa}=&
\cases{-(J g^2)^{-1}
\sum{1\over
2}\igb_{-\io}^{+\io}\s\,g \t F^h_{\nn j}(\t) \,d g\t
\ , &$j=1,...,l-1$\cr
-(J_0g^2)^{-1}\sum\2\igb_{-\io}^{+\io}\s\,g \bar w_{0l}(\t)
F^h_{\nn j}(\t) \,d g\t
\ ,&$j=0$\cr} \cr}$$
where $\bar w_{0l}\= (J_0 g)^{-1} w_{0l}=\fra14\bar w$, see \equ(2.15),
is a convenient adimensional matrix element, and $F^h_{\nn j}(\t)$ is
the function resulting from the resummation.  The terms $\b^{h,1}_{\nn
j}$ correspond to wiggly bubbles, while the $\b^{h,2}_{\nn j}$
correspond to smooth bubbles.  The "value" of the subtrees on which we
are summing is given according to the rules described in 1)$\div$ 4)
above.  The $\b^{h,1}_{\nn j}$ have the interpretation, which we leave
to the reader to verify, of being proportional to the $j$-th component
of the Fourier transform of mode $\nn$ of the homoclinic splitting
$\D^h(\aa)$ (hint: just note that $-\igb_{-\io}^\io=\igb_{+\io}^0-
\igb_{-\io}^0$ and recall the expression of $X^{h\,+}(0^+)$ and
$X^{h-}(0^-)$ in terms of trees and the remark 2) following \equ(5.14)
which implies that $F^h_{\nn j}$ in \equ(6.6) is really the same
function $F^h$ considered in the previous sections, see \equ(2.14)).

Taking care also of the constants of proportionality it is:
$$\D^h_{\nn,l}= J_0 g\,\b^{h,1}_{\nn,0}\ ,\qquad
\D^h_{\nn,l+j}= J g\,\b^{h,1}_{\nn,j}\ ,\quad (1\le j\le l-1)
\Eq(6.7)$$
(note that the bubbles values are dimensionless).

The $\b^{h,2}_{\nn j}$ are new objects, {\it while the quantities
$\s^h_{\nn j}$ that one would obtain by eliminating the sign function
$\s$ from the integral $\igb$ in the definition of $\b^{h,2}$ would be
the $j$-th component of the scattering phase shift} (see [CG] for a
definition of the notion of phase shift): however such interpretation
will not play a role in our discussion; hence we do not pursue the
enterprise of checking the latter statement.

The parity properties, discussed at the end of \S4, allow us to
conclude immediately that:
$$\D^h_{\nn j}\=-\D^h_{-\nn j},\quad
\b^{h,2}_{\nn j}\=-\b^{h,2}_{-\nn j},
\quad\s^h_{\nn j}\=\s^h_{-\nn j}\Eq(6.8)$$
Hence: $\D^h_{\V0 j}\=0,\b^{h,2}_{\V0 j}\=0$.
\vskip0.3truecm
{\it 3): (Definition of resummation trees and of dry and
ripe fruits):}

The above remark suggests introducing a new type of trees, that we
shall call ``resummation trees", or $\RR$-trees.  A resummation tree,
with all its decorations ($\RR$ and bubbles included), is defined by
drawing a tree with all its decorations and by {\it deleting} the
contents of the bubbles, leaving only the branches connecting the top
free nodes to the nodes $v$ inside the outermost bubbles (if any).  We
shall leave the bubble around each $v$ and on the branch leading to $v$
we write the label $j_{\l_v}$ and the total momentum $\nn_{\l_v}$ of the
deleted subtree together with the total order $h_{\l_v}$ of the deleted
subtree.

By construction the resummation trees have bubbles which can contain
only one node $v$.  It is natural to call such resummation bubbles {\it
"fruits"} (``of the resummations'' or ``of the trees'', as preferred).  If
the bubble is wiggly we call it a "dry fruit" while if it is smooth we
call it a "ripe fruit"; the node of the fruit will be called the {\it
seed} of the fruit.  As we shall see the dry fruits values will be
estimated, in our inductive construction of bounds on $\D_\nn^h$,
easily in terms of the inductive hypothesis, while the ripe fruits
will, at each inductive step, be just "ripe to be bounded".

It is convenient to define the ``value" of a resummation tree by the
same prescription we used for the previously introduced trees, but
changing the symbol corresponding to the seed $v$ (preceded by $v'$)
from \equ(5.12) to (see \equ(6.3), \equ(6.5), \equ(6.6)):
$$x_j^1(\t_{v'})\b^{h,1}_{\nn j} e^{i\nn\cdot\aa},\qquad {\rm with}\qquad
x^1_j(t)=\cases{\s g t,\quad &if $j=1,\ldots,l-1$\cr
\s \bar w_{0l}(t),&if $j=0$\cr}\Eq(6.9)$$
if the fruit surrounding the node is dry
(the dimensionless matrix element $\bar w_{0l}$ is defined after
\equ(6.6)); or:
$$x^2_j(\t_{v'})\b^{h,2}_{\nn j}e^{i\nn\cdot\aa},\qquad {\rm with}\qquad
x^2_j(t) =\cases{1,\quad &if $j=1,\ldots,l-1$\cr w_{00}(t),&if
$j=0$\cr}\Eq(6.10)$$
if the fruit is ripe.

The name {\it seed} is fairly appropriate as it really represnts a sum
of appropiately evaluated tree values of trees having the seed has a
root.  "A seed can be magnified to reveal its content in trees" whose
values add up to the value, \equ(6.9),\equ(6.10) of the fruit
containing the seed (as it should be).

Finally we {\it renumber} such trees (as
described at the beginning of the section) according to their
topological structure (\ie resummation trees do not inherit the
numbering from the trees that generated them).

This is a good definition as we can now say that the sum of
the values of the decorated resummation trees of given order is equal to
the sum of the values of the previously considered trees of the same
order. The non trivial part of such claim is dealing with the
combinatorial factors; the correctness of the combinatorics is easily
verified if the trees are regarded (as we are doing) as having all the
branches distinct.

A simple cancellation can already be seen: consider the contribution to
$F_{\nn,j}^h(\t)$ coming from trees {\it with $\V0$ total free momentum
and even fruits number}: then given a tree we can consider another
``conjugated'' tree with the node momenta of the free nodes
simultaneously changed in sign; summing their two contributions we see
that the integrals to be performed are integrals of a time odd analytic
function over the whole line: such integral will vanish (by the parity
considerations of \S4).  Hence we have the mechanism:

\item{I) }Cancellation of the contribution to $\b^{h,1}_{j\nn}$ from
the $\RR$--trees with an even number of fruits, summing over the
choices of a global sign multiplying the free node modes.  {\it Hence
trees without fruits and with $\V0$ free momentum do not contribute to
the value of $\b^{h,1}_{j,\nn}$, \ie to the Fourier transform (in the
$\aa$ angles) of the splitting (note that $\nn$ is the total momentum,
hence in general it is $\nn\ne\V0$).} We call this a "parity
cancellation".

A deeper cancellation mechanism is:

\item{II) }Cancellation of the contribution to the splitting
$b^{h,1}_{j,\nn}$ from $\RR$--trees carrying at least one fruit with
label $j>0$, summing over all the possible ways of attaching the fruits
to the tree.  We call the latter the "KAM cancellations" as they extend
the cancellations discussed below, leading to I), II) at the end of
\S6, and are sufficient to carry to KAM tori the Siegel--Brjiuno
method (\S7).  The (easy) proof is exactly the same.  And it is
not a parity cancellation: it will be called a "KAM cancellation".

We see that there are still quite a few cases of trees with $\V0$ free
momentum that can contribute to the homoclinic splitting. A few more
cancellations can be spotted.\footnote{${}^6$}{
\nota\0for instance:\acapo
III) Cancellation of the contributions to $\b^{h,1}_{j,\nn}$ from
$\RR$--trees with $\V0$ free momentum, ripe fruits only, and with ripe
fruits carrying only the label $j=0$, {\it and} with all the free
branches carrying a $j>0$ label, by summing over all the possible ways
of attaching the fruits to the tree.  This cancellation is also
remarkable but its role is not very clear; its proof, not immediate,
will not be given here.}

We shall see that we do not have to worry
about trees containing at least one dry fruit; but the {\it
basic problem remains for the trees with only ripe fruits}.
The problem is solved, for the trees with a root label $j>l$ by the
following really remarkable mechanism, which is a non trivial extension
of the above mechanism II).

\item{IV) }Cancellation of the  $\RR$--tree values contributions
to the splitting from $\V0$--free momentum trees with root
carrying a label $j>l$, by summing over all the possible ways of shifting
the root branch location to other free nodes of a given {\it
rootless} tree.

This is another cancellation which is not a parity cancellation: it
implies as a particular case the KAM cancellation above, but the latter
is very elementary (see below), while the more general cancellation IV)
is much deeper, and it is essential for the homoclinic splitting theory
(for which the KAM cancellation is not sufficient).  With some thought
one realizes that without cancellation II) for the KAM theory, and IV)
for the homoclinic splitting the above "nice" formalism would be
essentially useless.  The proof of IV) is based on the "tree root
identity", due to L.  Chierchia.

\vskip0.3truecm
{\it 4): (Tree root identity),} (Chierchia):

Consider a resummation tree $\th$ without fruits and consider its
$R$--value.  Suppose that the total mode of $\th$ is $\nn(v_0)=\V0$,
and let $j$ be the root label, $\bar h$ the order of $\th$.  Fix
$\nn_1,\ldots,\nn_s$ and $j_1,\ldots,j_s$, where $s\ge0$ and
$j_i\in(0,\ldots,l-1)$.  Consider all trees obtained by adding to
$\th$, in all possible ways, $s$ branches with a ripe fruit carrying
the labels $h_i,\nn_i,j_i$ with $i=1,\ldots,s$, or no fruit at all
($s=0$).  Suppose that the root label is $j>l$.

We want to prove that such trees (with vanishing free mode) contribute
$0$ to the splitting $\D^h_{\nn\, j}$, by a cancellation mechanism based
on the symmetry property of the following quadratic form, setting
$F_R(t)=F(t)e^{-R\s_t t g}$:
$$\eqalignno{
Q G(t)=&\sum_{\r=\pm}\ig_{\r\io}^t w(t,\t) G(\t)\,d\t&\eq(6.11)\cr
(F_{R_1},Q G_{R_2})\=&\ig_{-\io}^{+\io} e^{-R_1\s_t t g}d t
\sum_{\r=\pm}\ig_{\r\io}^t w(t,\t)F(t)e^{-R_2\s_\t\t g}G(\t)\, d\t
=(QF_{R_1},G_{R_2}) \cr}$$
where $w(t,\t)$ is either $t-\t$ or $w_{00}(t)w_{0l}(\t)-w_{00}(\t)
w_{0l}(t)$, see \equ(6.3), \equ(6.5), and $F,G$ are {\it arbitrary}
elements of $\hat\MM$.  The \equ(6.11) is immediately checked, for any
$R_1,R_2$ large enough, from the definitions and uses only the
antisymmetry of $w(t,\t)$.

Integrating \equ(6.11) over $R_1,R_2$ on the appropriately
small contours, to get the residues at $R_1,R_2=0$, see appendix A2:
$$\igb_{-\io}^{+\io} dt\sum_{\r}\igb_{-\r\io}^t w(t,\t) F(t)G(\t)d\t=
\igb_{-\io}^{+\io} dt\sum_{\r}\igb_{-\r\io}^t w(t,\t)G(t) F(\t)d\t
\Eq(6.12)$$
This identity, and the remark that it is relevant for the cancellations,
is a key property and is due to L. Chierchia (private communication).

If we consider a tree $\th$ with root label $j>l$ and we take
$t=\t_{v_0}$, $\t=\t_{v_1}$, $v_1$ being any of the nodes immediately
following $v_0$, the above identity has a simple graphical
interpretation (formally discussed in appendix A2).  Let
$\th_0=\th,\th_1,\ldots,\th_n$ be all the trees that can be obtained
from $\th$ by detaching the root branch from the node $v_0$ and
attaching it to another node $v_j$ with $\d_{v_j}=1$ (\ie to a node to
which the above operation gives rise to another of our trees).  Then
the symmetry \equ(6.12) implies that the $\RR$ value of any of the
trees thus obtained has the form $V\cdot i(\nn_{v_p})_j$ with $V$ {\it
independent} of $\th_p$

{\it Hence we see the following basic cancellation mechanism}: if the
root free mode is $\V0$ then $\sum_{v,\,\d_v=1}\n_{vj}=0$ and the sum
of the contributions to the homoclinic splitting coming from the above
family of trees vanishes.

{\it We conclude, in particular, that the $\RR$--tree value of the
contribution to the action splitting of the resummation $\RR$--tree
values $\th$ with ripe fruits only and $\V0$ free mode is always $0$,
if the root label is $j>l$.} Hence we do not consider them in the
analysis of the splitting.
\*
{\it 5): (The energy conservation cancellation):}
\*
We cannot, however, conclude that also the general contribution to the
splitting $\D^h_{l}$ of the resummation $\RR$--tree values from trees
with $\V0$ free mode is always $0$.

This will not be so bad, after all, as the energy of the two whiskers
is the same so that (see \equ(1.1), \equ(2.2)) $H_\m(X_\m^\s(t;\aa)\=
E_\m\=\sum_{h\ge 0} E_h \m^h$ where, by the KAM results reported in
\S2, $E_\m$ is an analytic function of $\m$ near $\m=0$ and is
independent of $\s$ (and of course of $t$ and $\aa$).  Recalling that
$\V X_{\m\giu}(0;\aa)\=\aa$, $X_{\m-}(0;\aa)\=\p$ and that
$I^0(0)\=X^0_+(0)=2J_0 g$ we find for $h\ge 1$:
$$\eqalign{
&\oo\cdot\V X_\su^{h\s}(0)+2g X_+^{h\s}(0)+\2 \sum_{ {h_1+h_2=h}\atop
{1\le h_i\le h-1}}\left( J^{-1}
\V X_\su^{h_1\s}(0)\cdot \V X_\su^{h_2\s}(0)+\right.\cr
&+\left. J_0^{-1} X_+^{h_1\s}(0) X_+^{h_2\s}(0)\right)+
\d_{h1} \sum_{ {|\n|\le N}\atop{\nn\neq\V 0}} f_\n \cos (\aa\cdot\nn+
n\p)=E_h\cr}\Eq(6.13)$$
where $X^{h\s}(0)$ is short for $X^{h\s}(0;\aa)$ and $\d_{1h}$ is the
Kronecker symbol (and for $h=1$ the sum over $h_i$ is absent); recall
that a label $l$ has been equivalently denoted $+$ and that the labels
$j>l$ are collectively denoted $\su$. Taking the difference when $\s=+$
and $\s=-$ one obtains:
$$\eqalign{
\D_+^h(\aa)=&(-2g)^{-1} \Big(
\oo\cdot \V \D_\su^h(\aa) + \2 \sum_{k=1}^{h-1}J^{-1}
\V \D_\su^{h-k}(\aa) \cdot [ \V X_\su^{k+}(0;\aa) +
\V X_\su^{k-}(0;\aa)] +\cr
&+ J_0^{-1} \D_+^{h-k}(\aa) [ X_+^{k+}(0;\aa) +
X_+^{k-}(0;\aa)] \Big)\cr}\Eq(6.14)$$
Hence in some sense the $\D_l^h$ components of the splitting may be
"less important" to control. One may also think that the energy
conservation allows us to "transfer" the cancellations that we have
shown to exist in the expressions for the rotator components of the
homoclinic splitting to the pendulum components.
\*

\0{\bf B):} {\it KAM cancellations.}
\*
The above analysis can be applied also to the case in which the
pendulum in \equ(1.1) is decoupled from the rotators, to provide a
graphical representation of the invariant torus which continues the
unperturbed torus $\V A=\V0$.  The simplest representation  of the torus
is:
$$\aa=\aa\in T^{l-1},\qquad\V A=\V X_\su(0;\aa)\Eq(6.15)$$
which is obtained by simply considering the trees $\th$ with only
``rotators labels'', \ie with node and branch, angle or action, labels
$j$ taking values $j=1,\ldots,l-1$ and $j=l+1,\ldots,2l-1$ (hence
excluding the uninteresting $0,l$). The root time label is $t=0^+$ (or
$t=0^-$) as we are computing $X_\su$ at time $0$).
\acapo
The root time $t$ is again taken to be $t=0$, and:
$$\V X_{\su\nn}(\V0;\aa)=\igb^0_{+\io} F^h_{\su\nn}(\t)\,d\t,\qquad \V
X_{\giu\nn}(0;\aa)\=\V0\Eq(6.16)$$
where $F^h_\nn$ is the same function in \equ(6.6), resulting from the
resummations (hence, by the remark following \equ(6.6) it is the
$\nn$--Fourier component, in the $\aa$--angles, contribution to
\equ(4.14)). This time we only integrate up to $0$ because we are not
looking at the splitting (which, see below, vanishes) but at the value
of $\V X_\su$ at $t=0$.

However the \equ(6.16) does not seem to be the most convenient
representation, not only from a technical viewpoint but also because it
does not lead to a simple representation of the motion on the torus.

A better representation is obtained by remarking that $F^h(\t)$
corresponds to a solution a solution $t\to X(t)$, on the torus and
starting at $\aa$, and has the form $F^h(t;\aa)\in \MM$, see \equ(3.4).
We shall denote $\F^h_i(\pps,t;\aa)$ the functions in the representation
\equ(3.1) of $F^h(t;\aa)=\sum_{i\ge 0}\fra{t^i}{i!}\F_i(\oo t,t;\aa)$.
The notation is slightly different from that of \S2,\S3,\S4 where
argument $t$ in $\F_i(\pps,t;\aa)$ appears in the form $x=e^{-\s gt}$;
(this is irrelevant and we shall see that $\F_i(\pps,t;\aa)$ is in fact
$t$--independent).

The absence of coupling between rotators and pendulum produces the
following great  simplifications, with respect to the case of the
homiclinic splitting.

\item{0) } The $+\io$ in \equ(6.16) can be replaced with $-\io$; \ie the
splitting vanishes.
\item{1) } The functions $\F_i$ have rotator components vanishing if
$i>0$, so that the rotator components of $F(t;\aa)$ can be simply
written $\F_0(\oo t,t;\aa)$ (\ie no secular terms $\fra{t^i}{i!}$ are
really present);
\item{2) } The function $\F_0(\pps,t;\aa)$
{\it is $t$--independent and $\V\F_{0\su}(\pps,t;\aa)$ has $\V0$
average over $\pps$}.

The above properties are an immediate inductive consequence of the
\equ(6.15), and of \equ(4.9),\equ(4.10), see below, (or, {\it
alternatively}, of the KAM theorem).

One sees, in fact, that the splitting is zero, to all orders of
perturbation theory, simply by remarking that from
\equ(4.14),\equ(4.13) we see that there are no traces, in the present
case, of the pendulum separatrix motion.  Therefore $F^1$ is a analytic
quasi periodic function with $\V0$ average.  Hence $X^1$ is also quasi
periodic and analytic because the operators $\II$ and $J^{-1}\lis\II^2$
do not generate any non quasi periodic nor non analytic, nor secular
terms {\it when applied to $\V0$ average quasi periodic functions of
$t$}.  Therefore $F^2_\su$, see \equ(4.14), is still analytic, and also
quasi periodic in $t$ with $\V0$ average over $t$, by the cancellation
\equ(4.15), {\it etc}.  Therefore we never generate non analytic non
quasi periodic terms.  In particular $X^h(0)$ can be equivalently given
by the \equ(6.16) with $-\io$ replacing $\io$: hence the splitting at
$0$ vanishes (being $X^h(0^+)-X^h(0^-)$.  This simply means that if
there is no interaction between the pendulum and the rotators, the
whiskers remain degenerate to all orders of perturbation theory.

The above implies that we can write:
$$\AA(t)=\II \V\F_\su(\oo t;\aa),\qquad \aa(t)= \aa+\oo
t+J^{-1}\lis\II^2 \V \F_\su(\oo t;\aa)\Eq(6.17)$$
where $\F(\pps;\aa)$ is the function $F(\pps;\aa)$ regarded at fixed $\aa$
and as a function of $\pps$.
\*

{\bf Remarks:}
\*
{\bf 1):} We {\it have not} invoked here the KAM theorem in describing
the above algorithm to build recursively the $\F^h(\o t;\aa)$, \ie to
construct a formal power series describing via \equ(6.17) the motion on
the invariant torus (which, unless we use the KAM theorem we still have
to prove to exist by proving the convergence of the series).
The \equ(6.17), which coincides with \equ(5.15), is here a well
defined formal power series with the coefficients defined by the
recursive algorithm discussed in connection with \equ(5.15)
and recalled above.

{\bf 2):} in the present case, in the notations of \S4, see comments to
\equ(4.1), it is $F^{h\s}(\pps,\s\io)\=$ $\F^h(\pps;\aa)$ so that
\equ(4.9) implies $\F^h_{\V0}=\V0$.

{\bf 3):} Since the $\oo$ is a diophantine vector the above points
$(\AA(t),\aa(t))$ cover densely the torus as $t$ varies; hence calling
$\oo t=\pps$ and setting $\aa=\V0$ and $\F(\pps)\=\F(\pps;\V0)$, we
realize that a natural parametrization of the torus is:
$$\eqalign{
\AA=&\sum_{k=1}^\io\m^k \,\V a^k(\pps),\qquad
\aa=\pps+\sum_{k=1}^\io \m^k \V b^k(\pps),\qquad \pps\in T^{l-1}\cr
\V a^h(\pps)=&\sum_{\nn\ne\V0}\V \F^h_{\su\nn}\, \fra{
e^{i\nn\cdot\pps}}{i\oo\cdot\nn}, \qquad\V b^h=\pps+
J^{-1}\sum_{\nn\ne\V0}\V\F^h_{\su\nn}\, \fra{
e^{i\nn\cdot\pps}-1}{(i\oo\cdot\nn)^2}\cr}\Eq(6.18)$$
and in such representation $\pps$ has the interpretation of "average
anomaly", \ie the evolution is simply $\pps\to\pps+\oo t$.
By \equ(4.9) $\V\F^h_{\V0}=\V0$, and the $\nn=\V0$ terms are therefore
absent (care has to be exercised here in not confusing the Fourier
transorm with respect to $\pps$ and that with respect to $\aa$: we are
not talking about the latter as $\aa$ is set $\=\V0$).  Since we set
$\aa=\V0$ we could use the parity properties of $\V\F^h$ which, if $\aa$
were not $\V0$, would be jointly odd in $\aa,\pps$, reflecting that
$F^h_\su$ is time angle odd and implying that if $\aa=\V0$ then it is
odd in $\pps$ and hence $\V \F_{\V0}=\V0$.  But $\V\F^h_{\V0}$ is equal
to $\V0$ also if $\aa\ne\V0$, by the cancellation remarked in
\equ(4.2) (see \equ(4.9); see also remark 2) above), which is not a parity
cancellation.

{\bf 4):} To calculate $\V F^h_\su$ at $\aa=\V0$, hence to
calculate $\V\F$ at $\aa=\V0$, we can use the tree expansion.  The
$\V\F^h_{j\nn}$ can be identified to be given by the trees of
order $h$ with time label $\t_{v_0}$ of the first node fixed equal to
the root time label $t$, and with $j_{v_0}=j$ and total free momentum
$\nn$. We use the remark 2) after \equ(5.14) to establish the
connection between $F^h$ and the tree expansion.\acapo
In fact, if we remark that:
$$\sum_\r \ig_{\r\io}^\t e^{i\oo\cdot\nn
\t'}\,d\t'\=\fra{e^{i\oo\cdot\nn\t}}{i\oo\cdot\nn}\Eq(6.19)$$
we see that a resummation tree with free momentum $\nn$ has a
$t_{v_0}$--dependence proportional to $e^{i\oo\cdot\nn \,\t_{v_0}}$.
Since the trees are evaluated stopping short of computing the "last
integral" over the $\t_{v_0}$ which is, instead, set equal to $t$ we see
that the tree value for the evaluation of
$e^{i\nn\cdot\pps}\F^h_\nn$ is then obtained by
replacing such exponentials in $t$ by $e^{i\nn\cdot\pps}$: summing over the
trees yields $e^{i\nn\cdot\pps}\V \F^h_\nn$.
In the above calculations $\aa$ is set
$\=\V0$.

{\bf 5):} since the splitting is identically zero to all orders there
will be no resummation trees with dry fruits.

{\bf 6):} thus we se that we can use the formalism developed in the case
of the theory of the splitting, to study $X^h$ for $\aa$ arbitrary.
\*

But to take advantage of the easy evaluation of the integrals and of
the fact that we only want $\aa=\V0$ in the present case, it is useful
to quickly go through an essentially independent and more detailed
analysis.  We can start by remarking that the integrals associated with
the tree vertices are straightforward because:
$$\ig_{(\s)}^t (t-\t)^\z \, e^{i\oo\cdot\nn
\t}\,d\t-\ig_{(\s)}^0 (t-\t)^\z \, e^{i\oo\cdot\nn
\t}\,d\t=\fra{e^{i\oo\cdot\nn t}}{(i\oo\cdot\nn)^{1+\z}}
\,-\,\z\,\fra{1}{(i\oo\cdot\nn)^{1+\z}},\qquad \z=0,1\Eq(6.20)$$
which takes the place of \equ(6.3) and \equ(6.5).

We can mimick, therefore, the procedure followed in the resummation
scheme developed for the homoclinic splitting cancellations.

Given a tree $\th$ contributing to order $h$ we represent the $\EE^T$
operations associated with a node $v$ by affixing a ``bubble" around the
node $v$ and enclosing all the branches following it, or by writing a
label $\RR$ on the node $v$.  The bubble will mean that after evaluating
the $J^{-1}\lis\II^2$ or $\II$ operation via \equ(6.20) we chose the second
term in the r.h.s.  of \equ(6.20) (with $\z=1$).  If the node is not
enclosed in a bubble that cuts the branch linking it to the previous
node, then this means that it is marked by $\RR$ and that in the sum in
\equ(6.20) we selected instead the first term (like in the previous
case).

We can repeat the same representation operation on the (improper)
integrations pertaining to all the other nodes, starting from the
highest on the tree and going down towards the root: we stop at the
first nodes following $v_0$, (the latter node is somewhat
different from the others as the corresponding $\II$ operation
will not be performed, as we evaluate $\V F^h$ and not $X^h$).

Thus the most general contribution by trees of order $h$ to $\V
\F^h_\su$ will be represented by a tree with $h$ nodes with index $\d=1$
(and $h$ branches, because now there are no nodes of order $\d=0$) and all the
labels introduced so far ($\RR$ labels included) plus an arbitrary
number of bubbles drawn around tree nodes above $v_0$, and drawn as
described (avoiding overlappings).

We call {\it free} the branches that are not enclosed inside bubbles of
any type. The total {\it free momentum} of the tree will be the sum of
the node modes associated with the free nodes.
By construction the $t$ dependence of the
tree value appears {\it only} through $e^{i\oo\cdot\nn t}$, if $\nn$ is
the total free momentum of the tree. This is special for the present
case, and it does not hold for the homoclinic case.

{\it Hence the $\nn$--th Fourier coefficients, in $\pps$, of $\V
\F^h_\su(\pps)$ are simply equal to the sum of the tree values of the
trees with order $k$ and total free momentum $\nn$, evaluated at $t=0$.}
They will be denoted, as usual, $\V\F^h_{\su\nn}$. This also provides an
interesting interpretation of the free momenta. Note that in this way
the concept of total momentum (as opposed to that of total free
momentum) does not even arise.

Looking at \equ(6.20) we see that the second term is a constant factor
which can be taken out of the integration operations $J^{-1}\lis\II^2$
associated with the other tree nodes.  Thus the trees with bubbles
involve somewhat simpler integrals (although not dramatically simpler
as in the case of the homoclinic splitting, because in this case all
the integrals are essentially trivial).

Consider a bubble containing a subtree of order $h$ linked to its root
node $v'\ge v_0$ by a branch $v'v$ carrying a total free momentum
$\nn\=\nn_f(v)=\sum_{w\ge v}^* \nn_w$, where the $*$ means that the sum
is over the nodes which are inside the bubble but not inside inner
bubbles, \ie which are ``relatively free''), and carrying a label $j$
($1\le j\le l-1$), and a label $h$ equal to the order.  Fixed $\nn,j,h$
we can sum over the possible choices (consistent with the labels
$\nn,j,h$) of the subtree and its decorations obtaining the ``resummed
bubble values":
$$\b^{h}_{\nn j}=-\fra1{(i\oo\cdot\nn)^2} \F^h_{\nn j}, \qquad
\nn\ne\V0\Eq(6.21)$$
The cancellation remarked after \equ(4.14), see also \equ(6.18), implies
that $\b^{h}_{\V0 j}\=0$ and \equ(6.21) is interpreted as $\V0$ for
$\nn=\V0$. In fact it could be easily checked by parity considerations
the more general relation:
$$\b^{h}_{\nn j}\=-\b^{h}_{-\nn j},\qquad j=1,\ldots,l-1\Eq(6.22)$$

Therefore the above resummation suggests introducing again "resummation
trees".  A resummation tree, with all its decorations (bubbles
included), is defined by drawing a tree with all its decorations and by
{\it deleting} the contents of the bubbles, leaving only the branches
connecting the top free nodes to the nodes $v$ inside the outermost
bubbles if any: we call $\tilde V_f$ the set of such "non free" nodes.
We shall leave the bubble around each $v$, and on the branch leading to
$v$ we write the label $j_{\l_v}$, together with the total free
momentum and
the order $h_{\l_v}$ of the deleted subtree.

By construction the resummation trees have bubbles which can contain
only one node.  It is natural to call such resummation bubbles {\it
"fruits"}, as in the previous case. And the nodes $v\in\tilde V_f$ we
again shall rightly call {\it seeds}, see remark following \equ(6.10).
Note that also from the new viewpoint there is only one type of fruits.

Setting $f_{\nn,0}\=f_\nn$ and calling $\nn_f(v)=\sum_{v\le w<\tilde
V_f} \nn_w$ the free momentum of the node $v$, the "value" of a
resummation tree contribution to $\F^h_{\su\nn}$ will be simply:
$${\tst\Big(-\fra12(i\n_{v_0})_{j_{v_0}} f_{\nn_{v_0}}
\prod_{j=1}^{l-1}(i\n_{v_0})_j)^{m^j_{v_0}}\Big)}
\cdot\prod_{v_0<v\not\in\tilde V_f}
\fra{-\fra12(i\n_v)_{j_v}f_{\n_v}\prod_{j=1}^{l-1}
(i\n_v)_j^{m^j_v}}{J\,(i\oo\cdot\nn_f(v))^2}
\,\prod_{v\in \tilde V_f}\fra{\b_{\nn(v)j_v}^{h_v}}
{J\,(i\oo\cdot\nn f(v))^2}\Eq(6.23)$$
by \equ(5.12), as in this case $n_v\=0,\,\d_v\=1$.  The value of $\V
F^h_{\su\nn}$ is obtained by summing over all trees $\th$ with order $h$
and total free momentum $\nn$: naturally the order $h$ is the number of free
nodes plus the orders of the fruits, but the total momentum is
the sum of the node modes {\it excluding} the modes of the fruits.

Finally we {\it renumber} such trees (as described at the beginning of
the section) according to their topological structure (\ie resummation
trees do not inherit the numbering from the trees that generated them).

This is a good definition as we can now say that the sum of the values
of the decorated resummation trees of given order is equal to the sum
of the values of the previously considered trees of the same order.  As
in the homoclinic splitting case the correctness of the combinatorics
is not difficult to verify if the trees are regarded (as we are doing)
as having all the branches distinct.
\*

{\bf Remarks:} \*
{\it i):}
The collection of all the trees of a given order and total
free momentum $\V0$ must give a zero contribution to $\V\F^h$ because
$\V F^h_{\su}(\pps;\V0)$ has zero average by \equ(4.9), or
\equ(6.22).  Hence $\nn_f(v)\ne\V0$ can, {\it and will}, be supposed
without fear of errors and without ever having to divide by zero in the
trees evaluations (recall the diophantine property \equ(1.3), showing
that zeros can only occur if $\nn_f(v)=\V0$): the terms in \equ(6.23)
with some of the denominators equal to $0$ have to be regarded as
$0$.\acapo
{\it ii)} {\it Another essential cancellation is the following.} Given a
tree $\th$ suppose it to be such that $\nn_f(v)=\nn_f(v')$, ($\ne\V0$),
with $v',v$ being a pair of consecutive nodes in $\th$, for some
$v>v_0$.  Then the $v'$ must be a node with a bifurcation with at least
two branches (as, of course, $\nn_v\ne\V0$ for all $v$).  We can then
consider the subtree $\th_2\subset\th$, with the same root as $\th$,
obtained by deleting the branch $v' v$ and the following ones.  Since
$\nn_f(v')=\nn_f(v)$ by assumption, such tree will be such that
$\sum_{w\ge v'}\nn_w=\V0$.\acapo% Let $\th_{v'}$ be the deleted subtree
with root $v'$: imagine to attach it to the remaining tree by pinning it
to the nodes $w>v'$. From \equ(6.23) it is clear that in so doing we get
terms equal to some $w$--independent constant times $i\n_{w j'}$ (if
$j'$ is the branch label of $v'v$) provided we could neglect that,
in so doing,  some of the denominators of a few of the branches above
$v'$ change value, The change in value is simply the addition or
subtraction of $\e=\oo\dot\nn(v)$. Hence this can be hoped to be a good
approximation if $\e$ is very small. In fact summing also over a
simoultanous change in sign of all the modes $\nn_w$ to which the
pinning can be done it is clear that the sum of the values of the
considered family of trees vanishes {\it to second order in $\e$}.

{\it iii):} The same argument holds if $v$ and $v'$ are comparable in
the partial order established by the tree, \ie if $v'<v$ but $v'$ is not
necessarily the immediate predecessor of $v$. %
\*
{\it Therefore we can approximately compute $\V X^h_\su(\V0;\aa)$ by
summing only trees such that:
\item{I) } $\nn_f(v)\ne\V0$ \ if\ $v$ is any node (seeds included).
\item{II) } $\nn_f(v)\ne\nn_f(v')$\ for all pairs of comparable nodes
$v',v$, (not necessarily next to each other in the tree order, however),
with $v'\ge v_0$.} \*
\0The consequences of the above proved cancellations are analyzed in
the next sections (and in the relative appendices).

\vglue1.truecm

\penalty-200

%\ciao
\fiat

\vskip0.5truecm

{\bf\S7 Twistless tori: Siegel--Brjiuno--Eliasson
method for KAM tori.}\pgn=1

\penalty10000

\vskip0.5truecm\numsec=7\numfor=1

This section has heuristic nature; we shall suppose that $\V
X_\su^{(h)}(\V0;\aa)$ can be computed by summing over the trees
verifying I),II) at the end of \S6. The corrextions, \ie the
contribution from the other trees are accounted for in the Appendix A3,
because the natural continuation of the heuristic argument is \S8, while
the corrections to the approximation II) require some new ideas
(constituting Eliasson's main contribution). It will be clear that the
approximation II) is very good, particularly if one applies it to
momenta $\nn(v)$ which are very large.

To simplify further the proof we shall make a further assumption on
$\oo_0$ besides \equ(1.3); namely we shall suppose that:
$$\min_{0\ge p\ge n}\big|C_0|\oo_0\cdot\nn|-\g^{p}\big|>\g^{n+1}\qquad
{\rm if}\ n\le0,\ 0<|\nn|\le (\g^{n+c})^{-\t^{-1}}\Eq(7.1)$$
and it is easy to see that the {\it strongly diophantine vectors}, as we
shall call the $\oo_0$ verifying \equ(1.3) and \equ(7.1), have full
measure in $R^l$ if $\g>1$ and $c$ are fixed and if $\t$ is fixed
$\t>l-1$: we take $\g=2,c=3$ for simplicity; note that \equ(7.1) is
empty if $n>-3$ or $p<n+3$.

We proceed to prove theorem 1, \S1: \ie the persistence of the torus
which for $\m=0$ is $\AA=\V0$. We recall the hypothesis H: hence,
in this section, the parameter $\h\=1$ and therefore the system time
scales are set by $|\oo|, C_0^{-1}$ and by the perturbation; and $J$ is
an arbitrary diagonal matrix with matrix elements $J_j\ge J_0$ where
$J_0>0$ is a positive constant.

The basic idea of the following proof goes back to Siegel, [S]: his
somewhat difficult proof has been greatly clarified by Brjuno and a very
clear exposition and generalization of Brjiuno's work by P\"oschel is
in [P]. But the connection with the KAM theorem and the tree
cancellations, exposed in \S6, are due to Eliasson, [E].

Let $\NN$ be the set of the harmonics $\nn$ for which
$f_\nn\=f_{0,\nn}$ in \equ(1.1) does not vanish: the number of such
"perturbation harmonics" does not exceed $(2 N+1)^{l-1}$ and the length
$|\nn|=\sum|\n_j|$ is $\le N$.  But the long sought connection with the
KAM theorem is due entirely to Eliasson, [E], who also realized the
role of the tree expansions in establishing the connection just
mentioned.

Consider the functions $\V\F^h_{\su\nn}$ defined in \S6, B), see
\equ(6.18).  We shall prove that:
$$b^h_{j\nn}\=\fra1{J_0|\oo\cdot\nn|^2}|\F^h_{j\nn}|\le D B^{h-1},\qquad
j=1,\ldots,l-1\Eq(7.2)$$
and $D,B$ will be determined below, and depend only on
$J_0,N,l,C_0,\t,|\oo|$. {\it Hence they do not depend on a lower bound
on $\det J^{-1}$}. Furthermore $\V F_{\su\nn}^h\=\V0$ if $|\nn|>N h$,
because the initial perturbation has only harmonics with $|\nn|\le N$.

The discussion of \S6, B), shows that \equ(7.2) implies theorem 1, if
valid under the just stated conditions, and the function $\V x$
can be taken $\V x_{\nn}=\fra{i}{E}\V \F_{\nn}$.

Consider a tree $\Th$ with labels and fruits and regard it as
representing a contribution to the l.h.s. in \equ(7.2).  We imagine to
express each fruit value as a sum over the trees that can be grown on
its seeds, and so on. We can think that "each seed is magnified to show
its content in subtrees", see remark following \equ(6.10).  In other
words we undo, for the purposes of the estimates, the collection of
contributions to $\F^h_{\nn j}$ which gave rise to the resummations and
to the notion of fruits.

{\it But we take into account the assumed cancellations by constraining,
from now on, the labeled trees appearing inside each of the bubbles
(representing a magnified seed) to verify the properties
I) (which is exact), and II) (which is approximated)  at the end of \S6.}

Therefore the contribution from the considered $\Th$, and as a
consequence the full value of $\F^h_{\nn j}$, is split as a sum of
contributions coming from trees $\th$ carrying an arbitrary number of
bubbles, hierarchically ordered to avoid overlappings and each of which
encloses a node of $\th$ as well as all the subsequent ones. The number
of nodes is of course $h$.

Having undone all our painful collection of the trees contributing to
$\F^h_{\nn j}$ it is convenient to regard again all the branches of the
trees as distinct and therefore the correct combinatorial factor with
which to multiply the tree contribution to the $\F^h_{\nn j}$ is again
$h!^{-1}$, see \equ(5.11).

Fixed $h$ and an unlabeled tree $\th$ the number of ways one can put
bubbles around the nodes is bounded by $2^h$ (as there can be at most
one bubble per node by our decoration rules); the number of node modes
$\nn_v$ is bounded by $(2N+1)^{(l-1)h}$; the number of branch labels is
$(l-1)^h$ and the number of trees is, by Cayley's count, see \S5,
footnote ${}^{5}$. Therefore a bound on $b^h_{\nn j}$ is simply:
$$\tst 2^h(2N+1)^{(l-1)h}(l-1)^h 2^{2h}\, M_h\Eq(7.3)$$
where $M_h$ is just the maximum possible value that a single fully
decorated tree can contribute to $b^h_{\nn j}$.

Suppose that in one of the possible trees $\th$ there are $n$ bubbles,
counting among them also an extra bubble (drawn for convenience) and
enclosing the whole tree except the root, and suppose that the $i$-th
of them encloses $h^*_i$ branches.  Then $h=n+\sum_{i=1}^n h^*_i$:
because, (having added one bubble encircling the whole tree but the
root), for every bubble there is one branch not entirely contained
inside a bubble.

Suppose that we can prove that for a fruitless tree the contribution to
$b^k_{\nn j}$ is bounded by $D_0 B_0^{k-1}$, for all $k\ge1$. Then
it clear from \equ(6.23) that one can take:
$$\tst M_h= \max \prod_{i=1}^n  D_0 B_0^{h^*_i}\Eq(7.4)$$
where the maximum is over $n,h^*_i$ such that $h=n+\sum_{i=1}^n h^*_i$.
Supposing that $B_0\ge D_0$ it follows immediately that \equ(7.2) holds
and that one can take (see \equ(7.3)):
$$D= (2^3(2N+1)^{l-1}(l-1))\cdot D_0,\qquad B=B_0\Eq(7.5)$$

Therefore the problem is reduced to the analysis of a fruitless tree.
Hence we only must study:
$$\d_{h,\nn}={\max_\th}^*\prod_{v\ge v_0} \fra1{|C_0
\oo\cdot\nn_f(v)|^2} \Eq(7.6)$$
where the maximum is considered over all fruitless trees with total free
mode $\nn$ and the $*$ reminds us that only trees with the considered
constraints are admmissible to compete for the maximum. The constant
$C_0$ is introduced to make $\d_{h,\nn}$ adimensional.

Proceeding as in [P], we fix, given $\nn,h$, one tree $\th$ on which the
maximum is attained and we consider its mode labels and the subtrees
$\th_v$ with first node at $v$ and root at $v'$, for any $v>v_0$. It is
necessarily true that the subtree $\th_v$ with the given mode labels is
also a maximizer for $\d_{h_v,\nn(v)}$.

A first remark, that should be at this point really obvious, is that
we can suppose that $0<|\nn_f(v)|\le h_v N$ if $h_v$ is the order of the
subtree with root at $v'$ and first node at $v$. And therefore we can
always suppose:
$$|C_0\oo\cdot\nn_f(v)|\ge\fra1{N^\t h_v^\t}\Eq(7.7)$$
by the diophantine assumption on $\oo$.

Given, arbitrarily, $q\ge1$ integer we say that a harmonic $\nn\ne\V0$
{\it is $q$--resonant} if:
$$C_0|\oo\cdot\nn|<\fra13 \fra1{N^\t q^\t}\=\z(q)\Eq(7.8)$$
where $\z(q)$ is defined here.

Then we define $N(\nn,h;q)$ to be the number of $q$--resonant harmonics
among the harmonics found in the tree where the maximum in \equ(7.6) is
reached.

The following (extension) of Brjuno's lemma (see [P]) holds:
$$N(\nn,h;q)\le 2\fra{k}{q}\Eq(7.9)$$
for the trees verifying I),II) at the end of \S6. We do not repeat the
proof as the present analysis has only heuristic value.

Assuming \equ(7.7) we can easily conclude the discussion; in
fact choosing $q=2^p$:
$$\prod_{v\ge v_0}\fra1{(C_0|\oo\cdot\nn_f(v)|)^2}\le
\prod_{p=0}^\io (3N^\t 2^{\t p})^{4 h 2^{-p}}\le B_4^h\Eq(7.10)$$
with $B_4=\exp(\sum_{p=0}^\io\fra4{2^p}\log(3 N^\t 2^{\t p})$.

Therefore we can bound \equ(7.6) by $B_4^h$.  Recalling also that in
the notations of \S1 $J\oo^2$ is also called $E$ and it is supposed to
be an upper bound on $|f_\n|$, see comment after \equ(1.1) and
\equ(1.4), and if $B_5\=\Big(\fra{J_0\oo^2}{J C_0^{-2}}\Big)$, this
means that $D_0,B_0$ in \equ(7.4) holds with:
$$D_0=(C_0^2\oo^2) (B_4 B_5),\qquad B_0=B_4B_5\Eq(7.11)$$
And therefore \equ(7.2) holds with:
$$B_0=b_l\,\fra{J_0}{J_{min}}\,\oo^2 C_0^2\, N^{8\t+l-1}2^{8\t},\qquad
D_0= b'_l B_0\Eq(7.12)$$
where $b_l,b'_l$ are suitable constants depending only on $l$ and $b'_l$
can be supposed without loss of generality $\ge1$.  Hence we see that
$B$ in \equ(7.2) does not depend on $J_{\max}$ {\it but} on $J_{\min}$.

We see that if $J_{\min}=+\io$ then $B=0$: and the system is in fact
completely integrable as already remarked in \S4, around \equ(4.16),
(see also [G1], probles 1,16,17 of \S5.11).

Clearly the bound \equ(7.2) implies the convergence of our approximation
to the series for the invariant tori for
$|\m|<B^{-1}$: this gives an idea of the connection between Siegel's
theorem and the proof of the KAM theorem. This was the situation before
Eliasson's work. To complete the proof we must turn to the corrections
present because the cancellation II) at the end of \S6 is only
approximate.
A complete analysis is in Appendix A3 and A4.

\vskip1.truecm

\penalty-200

\vglue0.5truecm

{\bf\S8 Theory of the homoclinic splitting.}\pgn=1

\penalty10000

\vskip0.5truecm\numsec=8\numfor=1

As a consequence of the above analysis we get that, in general, the
angles of homoclinic splitting (or $\d$, see \equ(1.8)) are smaller
than any power in $\h$.  Let us denote $\D_\nn^h$ the coefficient
of order $h$ in the Taylor expansion in powers of $\m$ and of order
$\nn$ in the Fourier expansion in $\aa$ of the splitting $(\m,\aa)\to$
$\D(\aa)\=X^+_\m(0;\aa)-X^-_\m(0;\aa)$; then the property of smallness
is an immediate consequence of the following bounds.

Let $d\in(0,\fra\p2)$, and let:
$$\e_h\=\e_h(d)\=\sup_{0<|\nn_0|\le Nh}
e^{-|\oo\cdot\nn_0|g^{-1}(\fra\p2-d)},\qquad\b=4(N_0+1),\qquad
p=4\t\Eq(8.1)$$
where $N_0$ is the maximal $\f$--harmonic of the perturbation $f$ in
\equ(1.1). Then, for $j=0,\ldots,l-1$ and for all $J\in [J_0,+\io)$
and $h\ge1$:
$$|\D^h_{\nn,j}|\le J_0g\,D\, (b\h^Q)^{-h},\qquad\qquad|\D^h_{\nn,
j}|\le J_0g\,D\,
d^{-\b}(B d^{-\b })^{h-1} (h-1)!^p\e_h\Eq(8.2)$$
where $b<1,D,Q,B$ are suitable dimensionless constants depending on the
various parameters describing \equ(1.1), {\it but not on the
perturbation parameters $\h,\m$}; the proof in the appendix A1 shows
how to construct a bound on $B$; the latter is the only real problem as
the existence of the $b,Q,D$ is well known, (see \S1 and [CG], for
instance, \S5, eq. (5.76), (5.78)).  The constant $Q$ has been briefly
discussed in \S1 and we take it to be the minimal such that the
properties of peristence of the tori hold (\eg $Q=10$ if $l>2$ and $Q=1$
if $l=2$, see \S1).
\*

{\it If $l=2$ both bounds in \equ(8.2) are uniform in $J\ge J_0$ and
one can take $J\to+\io$; furthermore $\t$ can be taken $\t=0$.  If
$l>2$ the second bound in \equ(8.2) is still uniform in $J\ge J_0$, but
the first may fail to be uniform in the values $J_j\ge J_0$.
Perturbation theory in $\m$ is well defined to all orders, but it might
be non uniformly (as the $J_j\to\io$, \ie as the twist rate goes to
$0$) convergent even at small $\m$;} see, however, the conjecture at
the end of appendix A1.  \*

The bound \equ(8.2) is of great interest in the case $l=2$ as it is
sufficient to determine the exact asymptotic behaviour as $\h\to0$ of
the splitting, (hence in the periodically forced pendulum, by taking
$J=+\io$, a permissible choice). {\it Note that, if $l=2$, it is
$\e_h\=\e_1$}.
\*

We proceed to explain the strategy of our bounds.  The first inequality
in \equ(8.2) is just \equ(2.6).  Note also that since we always suppose
that $f$ is a trigonometric polynomial of degree $N$, it is actually
$\D_{\nn,j}^h=0$ if $|\nn|>Nh$.

To explain the appearance of the "small" factor $\e_h$ in \equ(8.2), let
us consider the most general resummation tree, $\th$, of order $h$ with
all its labels (see item 3) in part A) of \S6), representing one
contribution to the homoclinic splitting $\D^{h}_{\nn j}$.

Let $r$ be the root of $\th$; let $v_0$ be the first node after the
root; let $V_f$ be the set of top free nodes, \ie the set of top
nodes still outside the fruits (if any) of $\th$; and recall that $v_0$
cannot be surrounded by a fruit. Let $\tilde V_f$ be the set of top
nodes contained inside a fruit ($\tilde V_f$ might be empty): \ie
$\tilde V_f$ is the set of fruit seeds.

Recall that in general $h_{\l_v}$ denote the order of the subtree
following $v$: the fruits contribute to the order a quantity given by
the order label $h_{\l_v}$ leading to them.  We shall call
$v_1,\ldots,v_{m_v}$ the $m_v$ nodes following $v$, $v_0\le v\le
V_f$, and $v'$ will be the node preceding $v$; $m_v=0$ if $v\in V_f$.

With the above notations it is easy to ``write'' the sum over $\s$ of
the values of the contributions to $\b^{h,\g}_{\nn,j}$ from the
resummation trees $\th$ of order $h$ and total mode $\nn$ and root
label $j\=j_{v_0}\ge l$:
$$\eqalignno{
\b^{h
}_{\nn,j}(\th) =& {1\over
m!}\oint\fra{dR_{v_0}}{2\p i R_{v_0}} \igb_{-\io}^{\io} e^{-\s_{v_0}g
R_{v_0}\t_{v_0}} w^{\g_{v_0}}_{j_{v_0}}(\t_{v_0})\,d\,g\t_{v_0}\cdot\cr
&\cdot\prod_{v_0<v\le V_f} \oint\fra{d R_v}{2\p i R_v}\Big[
{1\over2}\igb_{+\io}^{\t_{v'}}d\,g\t_v+ {1\over2}\igb_{-\io}^{\t_{v'}}
d\,g\t_v\Big]e^{-\s_{v}g R_{v}\t_v} \cdot&\eq(8.3)\cr
&\cdot\prod_{v_0\le v\le V_f}\Big(({-i(\n_v)_{\tilde \jmath_v}\over2})
\ \ c_{\n_v}\ \ e^{i(n_v\f^0(\t_v)+\nn_v\cdot\oo\t_v)}
\prod_{s=0}^{l-1}(i\n_{vs})^{m^s_v}\Big)\cdot\cr &\cdot \prod_{v_0<v\le
V_f}\Big(w_{j_{v}}(\t_{v'},\t_v)\Big) \cdot\Big(\prod_{v\in \tilde V_f}
x^{\g_v}_{j_{\l_v}}(\t_{v'}) \b^{h_v,\g_v}_{\nn_{\l_v}
j_{\l_v}}\Big)\cr}$$
where (see \equ(5.12), \equ(3.4), \equ(5.13), \equ(5.14), \equ(6.6),
\equ(6.9), \equ(6.10)), and the dimensionless coefficients $c_{\n_v}$
are given by: $c_{\n_v}\=(J g^2)^{-1}f_{\n_v}\d_v\d_{j\ne0,l}$\- $+$\-
$(J_0 g^2)^{-1}$\- $ f_\n\d_v\d_{j=0,l}
-(1-\d_v)$ where $\d_{j=0,l}$ is $1$ if $j=0,l$ and $0$ otherwise and a
similar, complementary, meaning is given to $\d_{j\ne 0,l}$ (recall
also that if $\d_v=0$ then $\nn_v=\V0$) and:
$$\eqalignno{ w_{j_v}(t,\t)\=&\cases{w_{00}(t) \bar w_{0l}(\t) -\bar
w_{0l}(t) w_{00}(\t) , & $v>v_0\ , j_v=l$\cr g(t-\t), & $v>v_0\ ,
j_v>l$\cr}\cr x^\g_{j}(\t)\=&
\cases{
\s \bar w_{0l}(\t),& if $j=l$ and $\g=1$\cr
\s g\t,               & if $j>l$ and $\g=1$\cr
w_{00}(\t), & if $j=l$ and $\g=2$\cr
       1,          & if $j>l$ and $\g=2$\cr
}
\qquad w^\g_j=\cases{x^1_j& if $\g=2$\cr
x^2_j& if $\g=1$\cr}&\eq(8.4)\cr}$$
where the dimensionless matrix element $\bar w_{0l}$ was defined after
\equ(6.6); $\g=\g_v$ is $1$ or $2$, respectively, if the fruit
encircling $v$ is dry or ripe; $m$ is the total number of branches
(root included); the integers $m_v^s$ decompose $m_v$ and count the
number of branches emerging from $v$ and carrying the labels
$s=0,\ldots,l-1$; $\l_v,\nn_{\l_v},j_{\l_v}$ in the product over
$\tilde V_f$ are, respectively, the branch entering the fruit around
$v$, the momentum carried by $\l_v$ and its $j$-label.  \*

\\{\it Remark:} the $\t_v$ integrals in \equ(8.3) are, in general, not
convergent for $R_v$ small: therefore the improper integral symbols
$\igb$ are used and they have to be thought as the analytic
continuation in $R_v$ from $\Re R_v>0$ and large.  \*

If we do not perform the $R_v$ integrals in the above \equ(8.3), we see
that the result of the improper integrals is a holomorphic function of
$R_v$ which can be continued to the region of positive and large $\Re
R_v$.  At such large $R_v$ the integrands are well decaying as $\Re
\t_v\to\pm\io$.%

Suppose, furthermore, that the tree contains only ripe fruits and
represents a contribution to the splitting, (\ie $\g_{v_0}=1$), (the
case of no fruits being in our notations a special case of only ripe
fruits).  Then the integrals are holomorphic functions of the $\t_v$ in
the strip $g|\Im\t_v|<\fra\p2-d$, except for the factors $e^{-\s_v g
R_v\t_v}$ and provided the $(\g_v)_{v\in \tilde
V_f}$, are all equal to $2$, and provided $\g_{v_0}=1$.%

The residues have to be evaluated starting with those relative to the
top free nodes (and in arbitrary order) and continuing hierarchically,
(\ie always evaluating the residues relative to the highest remaining
nodes, in arbitrary order).%

Note that the evaluation of contributions from trees with dry fruits as
well as from trees with $\g_{v_0}=2$ (\ie to values of ripe fruits)
involve, instead, integrands with singularities at $\Re t=0$ see
\equ(6.9).  It is a fact, unfortunately, that the interplay between
$\g=1$ and $\g=2$ might look, at first, confusing.%

If the tree has $\g_{v_0}=1$ and $\g_v=2$ for all $v\in \tilde V_f$,
then its value \equ(8.3), \ie its contribution to $\b^{h,1}_{\nn,j}$,
is such that we shift the integrations over the $g\t_v$ to the branches
with fixed imaginary part $\x=\pm(\fra\p2-d)$ choosing the sign as we
like, by using the shift of contour formulae \equ(3.15).

This implies that, if $\nn_0$ is the total free momentum of our tree, we
can compute \equ(8.3) as:
$$\eqalignno{ \D^{h}_{\nn j}(\th) =&{e^{-\x g^{-1}|\oo\cdot\nn_0|} \over
m!}\oint\fra{dR_{v_0}}{2\p i R_{v_0}}\igb_{-\io}^{\io} d\,g\t_{v_0}
e^{-\s_{v_0} R_{v_0}(g\t_{v_0}+i\x)}w^1_j(\t_{v_0}+ig^{-1}\x)\cdot\cr
&\cdot\prod_{v_0<v\le V_f} \oint\fra{dR_{v}}{2\p i R_v}\Big[
{1\over2}\igb_{+\io}^{\t_{v'}} d\t_v+{1\over2}\igb_{-\io}^{\t_{v'}}
d\,g\t_v\Big]e^{-\s_{v} R_{v}(g\t_v+i\x)}\cdot&\eq(8.5)\cr
&\cdot\prod_{v_0\le v\le V_f}\Big(({-i(\n_v)_{\tilde \jmath_v}\over2})
\ \ c_{\n_v}\ \ e^{i(n_v\f^0(\t_v+ig^{-1}\x)+\nn_v\cdot\oo\t_v)}
\prod_{s=0}^{l-1}(i\n_{vs})^{m^s_v}\Big)\cdot\cr &\cdot \prod_{v_0\le
v\le V_f}\Big(w_{j_{v}}(\t_{v'}+ig^{-1}\x,\t_v+ig^{-1}\x)\Big)
\cdot\Big(\prod_{v\in \tilde V_f} x^{2}_{j_{\l_v}}(\t_{v'}+ig^{-1}\x)
\b^{h_v,2}_{\nn_{\l_v} j_{\l_v}}\Big)\cr}$$
if all the $\g_v,\,v\in V_f$, are equal to $2$ and if $\g_{v_0}=1$
(otherwise the $x^{\g_v}$ or the $w_j^{\g_{v_0}}$ are not holomorphic in
the strip and the argument does not apply).  The factor $e^{-\x
g^{-1}|\oo\cdot\nn_0|}$ is extracted, after shifting the contour up or
down according to the sign of $\oo\cdot\nn_0$, from the factors
$e^{i\oo\cdot\nn(\t_v+ig^{-1}\x)}$.

The factor $e^{-\x g^{-1}|\oo\cdot\nn_0|}$ can be bounded by $\e_h$,
provided $\nn_0\ne\V0$: the latter case $\nn_0=\V0$ can indeed be; but
we have proved that the cancellations I) show that it gives in fact
exactly $0$ contribution to the total splitting, after summing its
value for all the fruitless trees.  Thus the presence of the factor
$\e_h$ is explained, to all orders, in the contributions coming from
the (sums) of tree values of trees with only ripe fruits and
$\nn_0\ne\V0$ or from trees with no fruit at all and any $\nn_0$ (equal
to $\V0$ or not).

The case of trees with only ripe fruits and $\V0$ free momentum is
also giving a zero contribution to the splitting if the root label is
$j>l$, by the cancellations II) and IV) of \S6.  It remains the case in which
the root label is $j=l$: this case, however, is covered by the energy
cancellation of \S6, once we understand why the small factor $\e_h$
appears in the bound of the splitting to order $h$ in the cases $j>l$.

But the bound \equ(8.2) holds by direct calculation for $h=1$;
therefore one assumes it inductively for $h<k$.  It is clear that to
order $k$ the trees with at least one dry fruit will have a value
containing as a factor the value of the splitting to some lower order,
in turn bounded by a quantity containing a small factor $\e_h\le\e_k$,
by the inductive assumption.  Hence we just have to worry about the
trees with only ripe fruits (no fruit at all being a special case): but
they are the ones for which the above analyticity argument is correct.
Hence the factor $\e_k$ is present in the bounds of the contributions
to the splitting from the trees of order $k$ and either with total free
momentum $\ne\V0$ or with no fruits or with at least one dry fruit.

The  contribution to the splitting from the trees with ripe fruits only
and $\V0$ total free momentum also gives, by the above reasons, $\V0$
contribution to the splitting for the $\su$ components which are
therefore, to order $k$ bounded proportionally to $\e_k$: hence also the
$+$,\ie the $j=1$--component of the splitting is bounded
proportionally to ${\e_k}$.

It remains to estimate the integrals: but this can be done rather easily
from the analysis of integrals like \equ(8.3) or \equ(8.5) performed
with the aim of finding bounds which do not exhibit the small factors
$\e_h$ (which we already know have to be there).
In fact such integrals can be essentially exactly computed.

A full discussion of the straightforward but lenghty analysis
is reported, for completeness, in appendix A.

We proceed to show that the bounds \equ(8.2) ({\it i.e.} ultimately
the proved canellations) and the proved cancellations $\V\D^h_{\V0\su}=0$
imply that the splitting is smaller than any power.

By \equ(8.2), the splitting can be bounded, for any multiindex $\V a$,
by:
$$|\dpr_\aa^{\V a} \V\D_\su(\V 0)|\le D J_0g\sum_{h=1}^\io
\sum_{0<|\nn|\le Nh} |\m|^h |\nn|^{|\V a|}\min\{(b\h^Q)^{-h},
B_h \e_h(d)\}\Eq(8.6)$$
having denoted $B_h=B^{h-1}d^{-\b h}(h-1)!^p$.  Note that, if $N$
is the trigonometric degree of the polynomial $f$ in \equ(1.1), the
sums over $\nn$ can be suppressed by multiplying the $h$-th term by
the mode counting factor $(2N+1)^{h(l-1)+|\V a|}$ (\ie the maximum
number of non zero Fourier components times the maximum of $|\nn|^{\V
a}$).

{}From this bound it follows that $|\dpr_\aa^{\V a}\V\D_\su|$ is
smaller than any power in $\h$ (see \equ(1.2)), a result that, as
mentioned above after \equ(1.8), also follows from the theory of normal
forms ([Nei]).  In fact we can split the sum over $h$ in \equ(8.6) into
a finite sum, $\sum_{1\le h\le h_0}(\cdot)$ and a ``remainder",
$\sum_{h> h_0}(\cdot)$; then, if $|\m|<\fra12b\h^Q$ and $\h$ is small
we find:
$$\sum_{h>h_0}(\cdot) \le D \,J_0g\sum_{h> h_0} (\fra{|\m|\bar
C}{b\h^Q})^h\le 2D\,J_0g\, (\fra{\bar C|\m|}{b \h^Q})^{h_0}\Eq(8.7)$$
where $\bar C= (2N+1)^{(l-1)+|\V a|}$ takes into account the summation
over the $(2N+1)^{l-1}$ modes for which $\V\D^h_\nn$ does not vanish;
and:
$$
\sum_{h=1}^{h_0} (\cdot)\le  D J_0 g h_0|\m|\bar C^{h_0} d^{-\b h_0}
B^{h_0-1}(h_0-1)!^p \e_{h_0}(d)\Eq(8.8)$$
Thus if $\m=\h^{Q+s}$, $d=\sqrt\h$, and $s\ge1$ we see that fixing
$h_0=r/s$, for any $r>1$, the $|\dpr_\aa^{\V a} \V\D_\su|$ is bounded
by a ($r$--dependent) constant times $\h^r$ (as in such a case
\equ(8.8) is just a remainder, exponentially small in $\h^{-1/2}$).

Note that when $l=2$, the splitting $\d$ introduced in \equ(1.8) is just
$\dpr_\a \D_\su$ and $|\V a|=1$ in \equ(8.6).

{\it In the case $l=2$ ($\o_0>0$) we can get more precise asymptotic
estimates: namely we can prove the well known exponential decay of the
splitting, as $e^{-\p\o_0/2g\sqrt{\h}}$, for $\h\to0$ when $|\m|<
O(\h^{Q})$ with ${Q}$ large enough}, see the theorem 2 in \S1, and below.

To check the latter claim on the asymptotic value of the splitting
we simply remark that \equ(8.2) imply the convergence of the series for
the splitting for $|\m|< B^{-1}\h^{2(N_0+1)}$ (recall that $d=\sqrt\h$.
Thus the second and higher order terms can be simply bounded by:
$$\e_1\, J_0 g \,D\,\sum_{h=2}^\io \Big(\fra{
B|\m|}{\h^{N_0+1}}\Big)^2,\qquad
{\rm if}\quad |\m|<\fra1{2B} \h^{-2(N_0+1)}\Eq(8.9)$$
because, if $l=2$, it is $\e_h\=\e_1$.

Hence we compute explicitly the first order term.  We find (computing
from \equ(4.5)), with the above notations, that the first order term
$\d_1\=\dpr_\a\D_\su^1(0)$ is given by:
$$\sum_{{|n|\le N}\atop{|m|\le
N_0}} n^2 f_{mn} \ig_{-\io}^\io[\sin \o nt \sin m \f^0 - \cos \o n t
(\cos m \f^0 -1)]dt \Eq(8.10)$$
For all $J\le+\io$. Note that $J=+\io$ is, if $l=2$, possible.

A simple calculation shows that the leading order (easily studied in
terms of the elementarily expressible auxiliary functions:
$\ig_{-\io}^\io e^{i\o n t} [ e^{i m \f^0(t)}-1]dt$) is given, as
$\o\=\o_0/\sqrt{\h}$\- $\to\io$, by the terms in \equ(8.10) with
$m=\pm1$, $n=\pm N_0$ and it is:
$$ \h^{-N_0+\2} A_*
e^{-\p\o_0/2g\sqrt{\h}}\ ,\qquad A_*\=-g^{-1} \big({\o_0\over
g}\big)^{2N_0-1} (f_{N_0, 1}+f_{N_0, -1}) {4\p (-1)^{N_0}\over
(2N_0-1)!} \Eq(8.11)$$
provided $A_*\ne0$.

Thus we see that the first order dominates if:
$$\fra{|\m|}{\h^{N_0-\fra12}}>2\fra{J_0g}{A_*}
\,D\,\fra{B^2|\m|^2}{\h^{4N_0+4}}\Eq(8.12)$$
which means $|\m|<\fra{A_*}{2 J_0 g D}\h^{3N_0+\fra92}$ and it is
implied by $|\m|<\h^Q$ with $Q=3N_0+\fra92$ and $\h$ small enough.%

\*
{\bf Comments:}%
\*
1) A theory for the latter case, in a somewhat different set up%
\footnote{${}^7$}{\nota In [DS], [Ge], [HMS] a forced pendulum $\ddot\f+\h
\sin\f=\m\h\sin t$ is considered.  After suitable rescalings, such
equation arises as Hamilton equation for the hamiltonian
${A\over\sqrt{\h}}+{I^2\over 2} -\cos\f-\m\f\sin\a$, to which our
techniques seems to be adaptable.  The adaptation is needed because the
latter hamiltonian is not periodic in $\f$, a property used explicitly
in our analysis.  In [HMS], [Ge] the exponential smallness of the
splitting is obtained under the assumption that $|\m|\le$ const.
$\h^q$ with $q$ given, respectively, by $4$ and $5/2$; in [DS] $a$ is any
positive number.  In [La], [GLT] the splitting of separatices for the
so--called standard map is considered.  Our technique seems to be
adaptable to cover also such a case and would confirm the exponentially
small value of the separatrix splitting found in [La] and [GLT] (see
formula (6) in [GLT]).},
can also be found in [DS], [Ge], [GLT], [HMS], [La], based on arguments
not relying directly on the cancellations between the contributions to
the coefficients of the perturbation series.

2) Note that a value $p\ne0$ in \equ(8.1) would not spoil the asymptotic
formula \equ(1.9), if $l=2$: it is easy to check that it would simply
make it valid for $Q>N_0+\fra92+p$.

3)The above estimates can be obviously adapted to prove that the
quantities $\s^h_{\nn,j}$, obtained by looking at the definition
\equ(6.6) of $\b^{h,1}_{\nn,j}$ but replacing the $\s$ by $1$ inside
the integrals, also verify the bounds \equ(8.2), with different
constants.  Note, in fact, that the replacement of $\s$ by $1$ makes
the integrands, in the contribution to $\V\s^h_{\nn,j}$ coming from
trees with only ripe fruits, analytic thereby making it possible to
imitate the above shift of contour argument.

In [CG] the notion of homoclinic scattering, and of {\it
scattering phase shift function} $\V\s(\aa)$, was introduced and it
could be checked that the $\s^h_{\nn,j}$ have the interpretation of
Fourier transform of the scattering phase shift function $\V\s(\aa)$:
we do not give a formal proof of the latter identification.

\vskip1.truecm
\penalty-200

%\ciao

\fiat
%\footline={\rlap{\hbox{\copy200}\ $\st[\number\pageno]$}\hss\tenrm
%\foglioa\hss}
\vskip1.truecm

\penalty-200

{\bf Appendix A1: Estimates. A conjecture.}
\vskip0.5truecm\numsec=1\numfor=1\pgn=1

\penalty10000
Our purpose is to study the integral \equ(8.3) and \equ(8.5).  The
estimates are straightforward as we are willing to tolerate for $l>2$,
bounds proportional to a power, (any), of the factorial of the order
$k$ and to an inverse power, (any), of $d^k$.  The only really non
trivial point being the understanding of the cancellations in \S6
which, by the shift of contour argument of \S8, provide the small
factor $\e_k$.  We provide the estimates only for completeness, as the
results that we are deriving have often been presented in an incomplete
form.

The reader will realize that the following "estimates" are in fact an
essentially exact calculation of the integrals involved. Such a feat is
made possible by the simplicity of the model 2) in \equ(1.1).

Let $\th$ be a resummation tree with $m=m_f+m_0$ nodes, $m_f$ of which
are endnodes $v$ contained in some fruit ($m_f\ge0$).  We first consider
the \equ(8.3), \equ(8.5) with $\b^{h,\g}_{\nn,j}\=1$ and with the $m!$
replaced by $1$.

Such case contains the case of fruitless trees: once understood one
obtains the general case quite easily; see below.

To simplify the algebra we split \equ(8.3) and \equ(8.5) as a sum of
many ($c^m$ for some large $c$, see below) terms.  The splitting is
rather trivial and it is performed with the aim of being left only with
integrals of functions which are products of single argument functions.
Recall that $d=\fra\p2\pm\x$, see \equ(8.5).

1) each $w(\t,\t')$ is split into two addends, by using the expression
\equ(8.4), \ie $\bar w_{0l}(\t)$ $w_{00}(\t')$ $-$ $w_{00}(\t)$
$\bar w_{0l}(\t')$ is regarded as a difference of two terms, and so is
$\t-\t'$.  This splits each of \equ(8.3), \equ(8.5) into up to $2^m$
terms (in the estimates we sometimes exceed to get simpler expressions:
here we bound $m_0$ by $m$, for instance).  Hence we produce up to $2^m$
terms.

2) each $\bar w_{0l} (t)=2(gt) \,(\cosh gt)^{-1}-2\sinh gt$ is split
into its two composing addends. Hence we produce, for each of the
preceding ones, up to $2^m$ more terms.

3) in the case of \equ(8.5) we split the $g\t_v+i\x$ coming from
the action angle elements of the wronskians, or from $\bar w_{0l}$, into
the two addends composing them, thereby producing up to $2^m$ more terms
for each (see 2)) preceding ones.

4) we split $2^{-1}\sum_\r$ into two addends getting a factor $2^{-m_0}$
and up to $2^{m_0}\le2^m$ more terms (for each of the preceding), and
$2^{-m_0}$ will compensate in the bounds the $2^{m_0}$ that we get by
extracting the factor $2$ appearing in 2) above. Note however that it is
not important to keep track of powers of $m$: we are doing the counting
just to help the reader checking what we are doing.

The integral is thus split as a sum of up to $2^{5m}$ terms each of
which has the following form ($\o_v=\oo\cdot\nn_v$):
$$\prod_v\oint \fra{dR_v}{2\p i R_v}\prod_{v_0\le v\in\th}
\igb_{\r_{v'}\io}^{\t_{v'}}
d\,g\t_v\,\left[e^{-R_{v}\s_{v}i\x}\right]\,
e^{-R_{v}\s_{v}g\t_{v}}
(g\t_v)^{n_v}
e^{i\o_v\t_v}\,
\prod_{j=1}^{m_v+1} y^v_j(\t_v)\Eqa(A1.1)$$
where the terms in square brackets are present only if we are
condidering \equ(8.5) and the $y^v_j(\t)$ are elements of a {\it finite}
set of functions: namely, in the case of \equ(8.3), $e^{in\f(\t)}$
with $n=0,\pm1,$ $\ldots,$ $\pm N_0$ if $N_0$ is the
trigonometric degree of $f$ in \equ(1.1), and $(\cosh gt)^{-1}$, $\sinh
gt$, see \equ(2.15).  Here $\r_v=\pm$ {\it and it is not the sign of
$\t_v$}, but it is an independent variable.  The factor $2^{-m_0}$,
found in step 5) above, is not included in \equ(A1.1) and it will be
brought back later, (see \equ(A1.2) below).  The product over $v$ is
over the $m_0$ "free nodes" of $\th$, \ie which correspond to actual
integration operations. And $\t'_{v_0}\=0$.

Likewise, in the case of \equ(8.5), the $y^v_i(\t)$ are to be found among
the $e^{i n\f(\t+ig^{-1}\x)}$ with $n=0,\pm1,$ $\ldots,\pm N_0$, and
$(\cosh (g\t+i\x))^{-1}$, $\sinh (gt+i\x)$, and, as said above, the
terms in square brackets have to considered present.

Furthermore $n_v\le m_v+1$ is the number of factors $\t_v$ that are
collected from the wronskians in performing the above operations. But
$\sum_v n_v\le m$, because each node $v$ can contain at most one factor
$\t_v$ coming from the wronskians (recall that only $\bar w_{0l}(\t)$
and $w_{ij}$ with $0<i<l,l<j<2l-1$ can contain $\t$ explicitly).

The functions $y_j$ are holomorphic for $|\Im gt|<\fra\p2 -\x$ or, if
thought as functions of $x=e^{-\s gt}$, $\s=\sign t$, are holomorphic
for $|x|<1$ {\it and} outside the cone with half opening $\x$ centered at
the origin and symmetric about the imaginary axis (the case $\x=0$
refers to \equ(8.3)).  Furthermore they are in the class $\hat\MM$
introduced in \S2,\S3: see \equ(2.15) and \equ(3.3).

The contour integrals involving the $R_v$ are over a small
circle, around the origin, and they really denote the evaluation of the
appropriate residues, see \S8.

Note, as this will be quite relevant below, that the $y$ functions, as
functions of $x$ at fixed $\s$ are holomorphic near $x=\pm1$, as they
only have polar singularities at $x=\pm i$ or at $\pm e^{i\x}$

We reconstruct \equ(8.3), from the integrals \equ(A1.1), as a linear
combination of the above intergals, (up to $2^{4m}$, as one can check
by keeping track of the above accounted proliferation of the terms),
times suitable factors, all bounded by:
$$\tst F^{m_0}\Big(\fra\p2 \fra{N}2\Big)^{m_0} N^{m_0}\Eqa(A1.2)$$
and $N$ is the maximum degree of the trigonometric polynomial $f$ in
\equ(1.1) and $F\ge1$ is a bound on the Fourier components of $(J_0
g^2)^{-1}f$.  The factors bounded by powers of $\x$ ($\le{m_0}$)
coming from the
binomial expansions met in the splittings considered in 3) have been
bounded by $\fra\p2$ to simplify the notation.

5) We consider one of the nodes $\bar v$ following $v_0$ and split the
integration over $\t_{\bar v}$, \ie
$\igb_{\r_{\bar v}\io}^{\t_{v_0}}
d\t_{\bar v}$, as a sum of $\igb_{\r_{\bar v}\io}^0-\igb_{\r_{
v_0}\io}^0+\igb_{\r_{v_0}\io}^{\t_{v_0}}$, leaving for later
consideration the first two choices.  And we repeat the procedure,
generating up to $3^m$ terms: all of them left for future consideration
except the one which has all the $\t_v$ variables integrated in the
interval $[0,\r_{v_0}\io)$, \ie all having the same sign.  We shall
return to the terms left out for future consideration in item 8) below.

We are thus left with an integral like:
$$I_m=\prod_{v\ge v_0}\ig_{-\io}^{\t_v'} d\t_v [e^{R_v i\x}] e^{R_v\t_v g}
e^{i\o_v\t_v} (g\t_v)^{n_v} \prod_{j=1}^{m_v+1} y^v_j(\t_v)\Eqa(A1.3)$$
where the terms in square brackets are present only in the case
\equ(8.5) and $\r_{v_0}\io$ has been supposed, to fix the ideas,
to be $-\io$.  As before the term in square brackets is
present only in the case of \equ(8.5).

Of course one might think that we are undoing all that was painfully
done in \S6. And in fact this is essentially the case; the work of \S6
was performed only to exhibit the cancellations.

We shall let $d$ be $\fra\p4$ in the case \equ(8.3), while in the case
\equ(8.5) it will be actually $d$ (to unify the notation).
If we now show that \equ(A1.3) can be bounded by:
$$|I_m|\le I^0_m \= B_1^m [d^{-\b m}] \max_{0<|\nn|<
Nm}\big(g^{-1}|\oo\cdot\nn|\big)^{-2m}\Eqa(A1.4)$$
we will have shown that all the up to $3^m$ terms generated by the above
decoposition of the above integral are bounded by the same quantity. In
fact the terms left above "for later consideration" are manifestly
products of quantities bounded by $I^0_{m_1}\cdot\ldots\cdot I^0_{m_p}$ with
$\sum m_i=m$.

6) We write the first integral in \equ(A1.3) as
$\igb_{-\io}^{-g^{-1}} d\t_{v_0} +\ig_{-g^{-1}}^0 d\t_{v_0}$: this gives
us two terms. The first of which is:
$$
\ig_{-\io}^{-g^{-1}}\,d\,g\t_{v_0}\prod_{v>v_0}\ig_{-\io}^{\t_{v'}}\cdot
\Big( \prod_{v\ge v_0}
[e^{R_v i\x}] e^{R_v\t_v g}
e^{i\o_v\t_v} (g\t_v)^{n_v} \prod_{j=1}^{m_v+1} y^v_j(\t_v)\big)
\Eqa(A1.5)$$
We shall come back to the second integral in item 7) below.

The functions $y(\t)$ can be expanded into a series:
$$y^v_j(\t)=\fra1x
\sum_{p=0}^\io y^v_{j,p} x^p,\quad x=e^{-\s g\t}\Eqa(A1.6)$$
and some of the $y_{j,0}$ may vanish, but not necessarily all
(because of the possible choice: $y(\t)=\sinh g\t$).

If each of the $y_j$ is expanded as in \equ(A1.6), then each of
the \equ(A1.1) is broken into a sum over labels $\{k^v_j\}$, with
$v\in\th$, and $j=1,\ldots,m_v+1$, with convenient weights, namely:
$$\prod_{v\in\th}\prod_{j=1}^{m_v+1} y^v_{j, k^v_j}\Eqa(A1.7)$$
of integrals like:
$$\left[\prod_v e^{i\x R_v}\fra{\dpr^{n_v}}{\dpr E_v^{n_v}}\right]
\prod_{v_0\le v\in\th}\ig_{-\io}^{\t_{v'}}
e^{+R_{v}g\t_{v}} [e^{i R_v\x}]\,e^{(i\o_v+E_v)\t_v} x_v^{k_v}\,d\,g\t_{v}
\Eqa(A1.8)$$
where $v'$ denotes the node immediately preceding $v$, and the
auxiliary paramenters $E_v$ have to be put equal to $0$ after
differentiation while the $R_v$ will have to be integrated over the
above small contour around the origin.  The $[e^{i\x R_v}]$ factors
are present only in the case of \equ(8.5).

The integrals \equ(A1.8) are performed hierarchically. This means that we
first integrate with respect to the $\t_v$'s with $v$ being a top
node, in arbitrary order.

To perform such integration we use \equ(3.8):
$$\igb_{-\io}^t x^K e^{(i\O+E)\t}\,d\,g \t=\fra{x^K e^{(i\O +E)t}}{K+
g^{-1}(i\O+E)}\Eqa(A1.9)$$
valid if the denominator does not vanish, (needless to say), and if $t<0$.

After integrating over the $\t_v$ variable corresponding to a top free
node $v$, arbitrarily fixed, we compute the $E_v$ derivatives and the
residues at $R_v=0$, and then we shall repeat the procedure after
deleting the top free node $v$ and the branches outgoing from it.

The $E_v$ derivatives can act either on the numerators or on the
denominators of \equ(A1.9).  The $n_v$ differentiations with respect to
$E_v$ give results, if $v'$ denotes the node immediately preceding $v$,
like:
$$n!\,\fra{x^{k_v} e^{g R_v \t_{v'}} e^{i\x
R_v} e^{i\o_v\t_v}}{(k_v+R_v+ig^{-1}
\o_v)^{n+1}}\, (g\t_{v'})^{n'}\Eqa(A1.10)$$
with $n+n'=n_v$ and the terms like \equ(A1.10) add up
with coefficients $\pm1$ to give the result of the derivative of the
$\t_v$ integral that we are considering. The number of addends is not
greater than $2^{n_v}$ (because we are taking derivatives of a product
of at most two factors, when $\l_v=1$). Here
$\o_v=\oo\cdot\nn_v$.

The residue of $R^{-1}_v$ times \equ(A1.10) at $R_v=0$ is simply
\equ(A1.10) itself evaluated at $R_v=0$ if $k_v$ or $g^{-1}\o_v$ are
different from $0$ (because at $R_v=0$ the denominator in \equ(A1.10)
does not vanish being bounded below by either $|k_v|\ge1$ or by
$|g^{-1}\o_v|$; and otherwise it is:
$$\fra{n!}{(n+1)!} (g\t_{v'})^{n'}\big(g
\t_{v'}+i\x)^{n+1}\Eqa(A1.11)$$
still with $n+n'=n_v$ and, as above, the terms with $i\x$ are
present only in the bound of \equ(8.5).  Developing the binomial in
\equ(A1.11) we can say that the result of the derivative and residue
evaluation is a sum of at most $2^{n_v+1}$ terms, per each of the
already found $3^m2^{n_v+1}$, which have the form:
$$G\,\tilde n!\, x^{\prime \tilde k_v}_v e^{i\tilde\o_v\t_{v'}}
(g\t_{v'})^{\bar n}, \qquad x'=e^{-\s_{v'}\t_{v'}g}\Eqa(A1.12)$$
with $\tilde n+\bar n\le n_v+1$ and $\tilde\o_v=\oo\cdot\nn_v$; the
latter in general will be, when considering the other integrals
correponding to inner nodes, the sum of the modes $\nn_w$ with $w\ge v$.
The coefficients $G$ are bounded by the maximum between $1$
(corresponding to the a bound on the denominator in \equ(A1.10) when
$k_v\ne0$), or $(\min_{0<|\nn|\le N} |\oo\cdot\nn|g^{-1})^{-n_v-1}$
(corresponding to $k_v=0,\o_v\ne0$) or $(\p/2)^{n_v+1}$ (corresponding
to $k_v=\o_v=0$, $n=n_v,n'=0$, using \equ(A1.11) and recalling that
$|\x|<\fra\p2$).

Continuing the procedure we integrate successively over the $\t$
variables of all the top free nodes (in arbitrary order), each time
deleting the considered node and its outgoing branches.

Therefore after all the integrations have been performed, all
derivatives and residues computed, we shall have expressed the result
of the evaluation of \equ(A1.3) (taking into account the listed
proliferation of terms described after \equ(A1.4), ($3^m$), after
\equ(A1.10), ($\prod_v 2^{n_v}$), and before \equ(A1.12), ($\prod_v
2^{n_v+1}$)), as a sum of up to $3^m 2^{3m}$ terms: {\it we use here
that $\sum n_v\le m$ as there can be only up to one factor $\t_v$ per
node}.  Each of which is bounded by:
$$\eqalign{
 &e^{-\sum_{v,j} k^k_j}c_1^m\cdot\,\max_{0<|\nn|\le Nm}
 (g^{-1}|\oo_0\cdot\nn|)^{-2m} \quad{\rm if}\quad l>2\cr
&e^{-\sum_{v,j} k^k_j}c_1^m\cdot 1\kern3.truecm {\rm if}\quad l=2
\cr}\Eqa(A1.13)$$
because the factors $x_v$ are $\le e^{-1}$ in the integration interval,
and by \equ(A1.9), eventually, they are evaluated at $x_v=e^{-1}$;
here $c_1$ is a suitable constant and in the cases $l>2$ we supposed
$g/|\oo_0|>1$ and neglect the (favourable) $\sqrt\h$ in
$\oo=\oo_0/\sqrt\h$, to simplify the algebra (which is not restrictive
unless one cares about numerically good estimates, which is not our
desire here).  In the case $l=2$, on the contrary, we use explicitly
that $\min_{\nn\ne\V0} |\oo\cdot\nn|=\o_0/\sqrt\h$, very special for
this "non resonant" case.

The reason why the minimum has to be considered only for $|\nn|\le Nm$
is that at each integration the oscillating exponent will have the form
$\oo\cdot\nn$ with $\nn$ being sum of the, at most $m$,
modes present in $f$ in \equ(1.1) (and in
fact of $m_0\le m$).

Collecting all the above terms and estimates and using $\sum_v n_v\le
m$ we have a bound on \equ(A1.8):
$$\cases{
\quad\bar\NN_l^m\,{(Cm^\t)^{2m}}\le \NN_l^m m!^{2\t}& if $l\ge2$\cr
\quad\NN_2^m& if $l=2$\cr}\Eqa(A1.14)$$
for some suitable constants $\NN_l$, and if $C,\t$ denote respectively
the diophantine constant and the diophantine exponent of $\oo_0$, see
\equ(1.3).

The latter bound should be multiplied by the absolute values
of the weights \equ(A1.7), by \equ(A1.2) and by the "multiplicity''
(produced in the course of the analysis 1)\%5)) $2^{5m}$
in order to produce, after summing over the $k$ labels in
\equ(A1.7), a bound on \equ(8.3) and on \equ(8.5) with the
$\b^{.,.}_{.,.}=1$.

The weights are Taylor coefficients of a few (\ie a finite
number) of functions of $x$, with radius of convergence $|x|=1$.

Therefore they can be bounded by a common bound $M_\l$ on the maxima
of such functions in
any disk of radius $\l<1$ times $\l^{-k}$. The integrals over the $\t$'s
run over the interval $(-\io,-g^{-1})$ so that the $x_v$ are always
$\le e^{-1}$. Hence, for $\l=2^{-1}$, we get
convergent bounds because of the exponential factors in \equ(A1.13).
Clearly this was the reason for the splitting of the integrals over the
$\t_v$ into the part with $\t_v<-g^{-1}$ and the part with
$\t_v\in[-g^{-1},0]$.

Note that there are at most $(2N_0+3)$ different functions $y$ and at
most $3m$ of them appear: and of these there will be up to $m$
trigonometric functions of $n\f_0(\t)$ (in the case of \equ(8.3)) or of
$n\f_0(\t+i\x)$ (in the case of \equ(8.5)) and up to $2m$ functions
coming from the wronskians.  The first admit a bound proportional to
$d^{-2N_0}$ as the $\cos\f_0(\t)$ and $\sin\f_0(\t)$ carry a polar
singularity of second order on the unit circle at distance of order $d$
(\ie $\x$) from the real axis. The wronskians are bounded proportionally
to $d^{-1}$ as they carry, at worst, a simple pole on the unit circle.

Hence we see that a bound on the the products of $y_j$ that can be met
in the integrals \equ(8.5) has the form $M_*^m d^{-2(N_0+1)m}$ for some
constant $M_*$. In the case of \equ(8.3) we can take the bound to have
the same form with the same $M_*$ and $d=\fra\p4$, say.

7) We can therefore switch to considering the ``left out part''
$\ig_{-g^{-1}}^0 d\t_{v_0}$.  Let $v_1$ be one of the nodes following
$v_0$.  We break the integral $\igb_{-\io}^{\t_{v_0}} d\t_{v_1}$ as
$\igb_{-\io}^{-g^{-1}} d\t_{v_1}+\ig_{-g^{-1}}^0 d\t_{v_1}$.  If we
make the first choice and if $ m_1$ is the number of nodes following
$v_0$ in the direction $v_1$ the first integral can be bounded, by the
previous argument, by $I^0_{m_1}$ and we are left with the problem of
bounding $I^0_{m_1}\cdot\ig_{-g^{-1}}^0d\t_{v_0}\prod_{v\ni
\th_{v_0\bar v}}\igb_{-\io}^{\t_v}\ldots$.

We repeat the procedure hierarchically and we are eventually left with
up to $2^m$ integrals like $I^0_{m_1}\ldots$\-$ I^0_{m_p}$\-
$\prod_{v\in\tilde
\th} \ig_{-g^{-1}}^{\t_{v'}}d\t_v\ldots$ to bound, where $\tilde \th$
is a subtree of $\th$ with $\tilde m$ nodes, and $\tilde m+\sum m_j=m$.

The last integral is manifestly bounded by the maximum of the integrand,
which is $ \tilde B^m d^{-2 (N_0+1)\tilde m} $ for a suitable $\tilde B$,
because $\ig_{-g^{-1}}^0 d g\t=1$, we see that there is a constant $B_2$
such that each of the the up to $3^m2^{3m}$ terms we generated is
bounded by $B_2^m [d^{-2(N_0+1)m}]$.%
\*
8) If one selects for consideration any other of the $3^m-1$ integrals
left aside in the analysis of item 5), see lines preceding \equ(A1.3),
one is left with essentially the same problem discussed in items 5)\%7)
above. In fact such choices involve factorized integrals, each of which
has exactly the same form as the one studied in 5)\%7) above.
\*
9) {\it We conclude that the final bound on the $\underline{sum}$ over
the mode values $\nn$ and over all the other mode labels (which is a
sum over up up to to $(2N+1)^{lm}$ terms) of the absolute values of
\equ(8.3) or \equ(8.4) with all the $\b^{.,.}_{.,.}=1$, is}:
$$\eqalign{ &c_3^m M_*^m\, m!^{2\t},
\qquad\hbox{in\ the\ case\ of\ \equ(8.3)}\cr e^{-|\x\oo\cdot\nn_0|}
d^{-2(N_0+1)m}&c_3^m M_*^m\, m!^{2\t}, \qquad\hbox{in\ the\ case\ of\
\equ(8.5)}\cr}\Eqa(A1.15)$$
with $\x=\fra\p2-d$ and $c_3$ conveniently fixed; and we shall use the
\equ(A1.15) to get a recursive bound on the coefficients
$\b^{.,.}_{.,.}$.  {\it Furthermore if $l=2$ the diophantine exponent
$\t$ can be taken $\t=0$.}

The \equ(A1.15) can be used to play the role plaid by the bound on the
fruitless trees in the KAM proof in \S7 in terms of $D_0,B_0$: in our
case the corresponding quantities are given by $\e_1 M_*^2c_3^2
d^{-4h(N_0+1)}(2h)^{4\t}$ and $M_*^2c_3^2 d^{-4h(N_0+1)}(2h)^{4\t}$
for trees with order up to $h$.  The extra power of $2$ is lost
here because $m\le 2\t$ (recall that the order $h$ of a tree is in general
different from the number of nodes because some nodes may carry an
order label $\d_v=0$, and $h\le m< 2h$).\acapo
And one has simply to make an argument parallel to that presented in
the paragraph containing \equ(7.4) and \equ(7.5).  The cancellations
discussed in \S6 show that, among the subtrees that are generated by
"magnifying the fruit seeds", there has to be at least one which will
provide a small factor $e^{-|\x\oo\cdot\nn_0|}$ for some $\nn$ with
$|\nn|\le Nh$.  We leave the details to the reader, to avoid
repetitions of the bounds discussed in \S7.

The final estimate of $D,B$ is that $\b=4(N_0+1)$, $p=4\t$ for the same
reason which shows that in \equ(7.5) one can take $D,B$ proportional to
$D_0,B_0$.
\*
{\bf Remark: } A more careful analysis of the higher orders would show
that this can be improved to $|\m|< \h^Q$ with $Q>N_0+\fra92$, simply
because the origin of the $4N_0 h$ was a "poor" bound on the number of
factors $f_\n e^{i(n\f+\nn\cdot\aa)}$ present in the expressions of the
splitting to order $h$ and each of which contributes a $d^{-2N_0}$,
(hence eventually a $(\sqrt\h)^{-2 N_0}$): the number of such factors,
in deriving \equ(8.2), by $2h$.  But to order $h$ there are, obviously,
exactly $h$ such factors, and we can therefore essentially replace
$4N_0$ by $2 N_0$ (changing the various constants).%
\*
If $J=+\io$ the above analysis can be repeated: we realize that several
simplifications take place. For instance $\b^{h,\g}_{\nn,j}\=0$ if
$j=1,\ldots,l-1$ and in fact $\bar D$ can be replaced by $D_j$ with
$D_j=\bar D$ if $j=0$ and $D_j=\fra{J_0}J$ if $j>0$. It follows also
that the trees with at least one inner branch with $j_\l=1$ give a
vanishing contribution (corresponding to the fact that $X^h_\giu\=\V0$
as the angles $\aa$ are isochronous, \ie $\dot\aa=\oo$ even if
$\m\ne0$). The splitting being $Jg \b^{h,1}_{\nn,j}$ we see that the
second bound in \equ(8.2) holds uniformly in the size of $J\ge J_0$,
\ie the constants $D,B,\b,p$ do not depend on $J$. This is again as in
the KAM theory of \S7.

The first bound in \equ(8.2) is more delicate, if $J=+\io$, as it is not
"purely algebraic", resting on the KAM theory: if $l=2$ one does not
really need the KAM theory as mentioned in \S1 and also the first bound
holds.  If $l>2$ the KAM theory does not apply, in general, and only the
second estimate \equ(8.2) is proved by the above arguments.  This
implies in particular the (not surprising) fact that perturbation theory
for the whiskers and their splitting is well defined to all orders; but
convergence is not guaranteed (\ie the first of \equ(8.2) may fail) and
one cannot guarantee even the persistence of the invariant tori.

To summarize \equ(8.2) holds uniformly in $J\ge J_0$ if $l=2$; if $l>2$
only the second of \equ(8.2) holds uniformly in $J\ge J_0$, and it is
not even sufficient to yield the convergence of the formal perturbation
theory for the tori and the splitting. If $l>2$ the first of \equ(8.2)
holds at fixed $J<+\io$: as needed and claimed in \S8.
\*

\0{\it A conjecture.}
\*

Nevertheless we think that even if $l>2$, and for all $J\le+\io$, but
with $J\ge J_0$, the $\t$ can be set equal to $0$: in other words {\it
we conjecture that the above theory can be completely freed from its
dependence on a KAM type of proof of existence of the invariant tori
and whiskers (like the quoted [CG]), so that the two theories are
essentially equivalent and independent even if $l>2$}.  We believe that
the essential has been done already in the present paper. One
would have to improve a little the cancellations analysis of \S6.
And the form of \equ(1.1) will have to play an essential role (as it
already did above).  The main feature of \equ(1.1) to keep, in order to
be able to believe the above conjecture, is the absence of action
variables in the perturbation $f$.
\*
\vglue0.5truecm

\penalty-200

{\bf Appendix A2: Symmetry of rootless trees.}

\penalty10000

\vskip0.5truecm\numsec=2\numfor=1\pgn=1

We first prove \equ(6.12).  Integrating the second of \equ(6.11) over
$R_1,R_2$ on circles with radii $r_1,r_2$ with $0<r_2<r_1$ and $r_1$
small enough we get a l.h.s.  equal to the l.h.s.  of the relation
\equ(6.12); while the r.h.s.  of \equ(6.12) is obtained by taking
$0<r_1<r_2$, and $r_2$ small enough.  Hence we want to deform the radius
$r_1$ to become larger than $r_2$.  This can be done if no singularities
in $R_1,R_2$ are met near the product of two circles with radius $r_2$.
The evaluation of the double integral in \equ(6.11) leads to a function
which can have some singular terms when $R_1=\pm R_2$: it is easy to
see, representing $F,G$ as a sum of finitely many terms like \equ(3.2)
plus very fastly decreasing functions at $\pm\io$, that the singular
terms have the form $(R_1\pm R_2)^{-n}$, $n>0$.  Therefore after
multiplying them by $R_1^{-1}R_2^{-1}$ we see that they have equal
double residue at $0$, whether we take first the residue in $R_1$ or in
$R_2$ (and the result is zero in both cases).  So that \equ(6.12)
follows.

Given a tree $\th_0$ (with all its labels) we can consider the
object obtained by deleting the root branch (with its labels).  The
object $\bar \th$ thus obtained will be a {\it rootless tree} $\bar\th$.

The same rootless tree can be obtained from several trees, as by
deleting the root branch we forget the action label it beared. And we
also forget which node was the first node.

We can find the trees which generate the same rootless tree $\bar\th$ by
attaching a root branch to any of the free nodes $v\in\bar\th$.  Some
restrictions may apply to the labels that can appear on the new root
branch.  In fact from the rules of tree labeling of \S5, and fron the
\equ(2.14), \equ(4.10) it appears that if the type $\d_v$ of the node is
$1$ no restrictions apply; but if $\d_v=0$ the action label of the new
branch has to be $0$.

A rootless tree with a distinguished node will be a pair $(\bar\th,v)$
of a rootless tree and one of its nodes.

It is convenient to define the value $V(\bar\th,v)$ of a rootless tree
with a distinguished node: we shall use the following prescription.

1) Imagine to attach a root branch at the node $v$ with any possible
action label.

2) Evaluate the tree value with the rules of \S5, but the node function
introduced in \equ(5.12) relative to the distinguished node will be
modified by deleting the factor $-\fra{i}2(\n_v)_{j_v}$ and replacing it
by $1$.

3) The $\EE^T_v$ operation is {\it always} $\II$ (even when the attached
root branch action label is $j_v=0$, which would require, for a normal
tree, the interpretation $\EE^T_v(\cdot)=\II(w_{00}\cdot)$).

By construction $V(\bar\th,v)$ does not depend on the label $j_v$ that
we attach to the new root to perform the above calculation. In
principle, however, it does depend on the distinguished node $v$. The
main consequence of the symmetry \equ(6.12) is that {\it in fact it is
$v$ independent}. Thus we can denote it $V(\bar \th)$, calling it the
value of the rootless tree $\bar\th$.

To prove the above independence we write the just defined value
$V(\bar\th,v)$ as:
$$\igb_{-\io}^{+\io} d\t\,\Big(\prod_{p=1}^r(i\n_v)_{j_p}\Big) e^{i
n_v\f_0(\t)+(\aa+\oo\t)\cdot\nn_v}\prod_{p=1}^r \tilde X^{h_p}_{j_p}(\t)
\Eqa(A2.1)$$
if $r$ is the number of branches arriving in $v$ and $h_p$ is the order
of the $p$-th branch (defined as the sum of the type labels of the nodes
that can be reached leaving $v$ along the $p$-th branch; the $j_p$ is
the angle label of the $p$-th branch. The $w(\t,\t')$ is either
$(\t-\t')$ if $j_1>0$ or $w_{00}(\t)w_{0l}(\t')-w_{0l}(\t)w_{00}(\t')$
if $j_1=0$ (apart from some proportionality constant fixing the
dimensions). The $\tilde X^{h_p}_{j_p}(\t)$ is the result of the
integrations over the time labels of the nodes of $\bar\th$ which
follow the $p$-th branch emerging from $v$ (in the order generated on
$\bar\th$ by the insertion of a root branch).

Let $v'$ be one of the $r$ nodes linked by a branch of $\bar\th$ to $v$.
We suppose that the branch is the one corresponding to $p=1$. Then, by
the definition of the $\RR$--tree evaluation we see that:
$$X^{h_1}_{j_1}(\t)=\sum_\r\2\igb_{\r\io}^\t w(\t,\t')
d\t'\,\Big(\prod_{q=1}^s(i\n_{v'})_{j_q'}\Big)
e^{i n_v\f_0(\t')+(\aa+\oo\t')\cdot\nn_v}\prod_{q=2}^r
\tilde X^{h_q'}_{j'_q}(\t')\Eqa(A2.2)$$
if $s$ is the number of branches arriving in $v'$ and $h_q$ is the order
of the $q$-th branch (defined as the sum of the type labels of the nodes
that can be reached leaving $v'$ along the $q$-th branch. The $j'_q$ is
the angle label of the $q$-th branch.

Substituting \equ(A2.2) into \equ(A2.1) we get an expression like the
l.h.s. of \equ(6.12): then it is immediate to check that the expression
to which it is equal, by applying \equ(6.12), is in fact
$V(\bar\th,v')$.

Finally one remarks that the value of a tree $\th_0$ with root label
$l+j_{v_0}$ with $j=j_{v_0}>0$ is expressed in terms of the
corresponding rootless tree $\bar\th$ value as $-\2(i\n_{v_0})_j
V(\bar\th)$.  This would not be true for $j=0$ because the evaluation of
the tree value would require using the operation $\II(w_{00}\cdot)$
instead of $\II$.

If $\th_0,\th_1,\ldots,\th_u$ are the trees that can be associated with
a given rootless tree $\bar\th$ and which have a root branch action
label $l+j$ with $j>0$, we see that the sum of their values is:
$$-\fra{i}2\sum_k (\n_{v_k})_j V(\bar\th)=-\fra{i}2
(\nn_{0})_jV(\bar\th)\Eqa(A2.3)$$
where $\nn_0$ is the total free mode of the trees (which is the same for
all).

Hence we see that if the total free mode vanishes the corresponding
trees give a vanishing contribution to the homoclinic splitting.  And
the cancellation takes place separately among the trees that have the
same rootless tree.  In particular we can say that the family of all the
trees with only ripe fruits and $\V0$ total free mode gives a zero
contribution to the action splitting for $j>0$, (of course the same
could be said of the family of the trees with only dry fruits, or with a
prefixed number of dry and ripe fruits and of free nodes).

\vskip1.truecm
{\bf Appendix A3: Analysis of the approximate cancellations.}
\vskip0.5truecm
\numsec=3\numfor=1\pgn=1
We have seen in \S7 the basic mechanism, {\it given for the purpose of
illustration}, showing that, restricting the sum in \equ(6.23) to a sum
over the trees such that:
\item{I) } $\nn_f(v)\ne\V0$ \ if\ $v$ is any node (seeds included).
\item{II) } $\nn_f(v)\ne\nn_f(v')$\ for all pairs of comparable nodes
$v',v$, (not necessarily next to each other in the tree order, however),
with $v'\ge v_0$.
\*
\0leads to a convergent series. The trees not verifying II) are called
{\it resonant} by Eliasson. And in this section we deal with them. Which
is the hardest part of the problem and the most original contribution by
Eliasson to the field.

I shall consider only the case of trees without fruits, as the
reduction, in \S7, to this case was not based on assumption II).

However {\it there are} resonant trees.  The key remark is that they
cancel {\it almost exactly}.  The reason is very simple: imagine to
detach from a tree $\th$ the subtree $\th_2$ with first node $v$.
Then attach it to all the {\it remaining} nodes $w\ge v', w\in
\th/\th_2$.  We obtain a family of trees whose contributions to
$h^{(k)}$ differ because:
\item{1)} some of the branches above $v'$
have changed total momentum by the amount $\nn(v)$: this means that some
of the denominators $(\oo\cdot\nn(w))^{-2}$ have become
$(\oo\cdot\nn(v)\pm\e)^{-2}$ if $\e\=\oo_0\cdot\nn(v)$; and:
\item{2)} because there is one of the node factors which changes by
taking successively the values $\n_{wj}$, $j$ being the branch label of
the branch leading to $v$, and $w\in\th/\th_2$ is the node to which such
branch is reattached.
\*

Hence if $\oo\cdot\nn=\e=0$ we would build in this {\it resummation}
a quantity proportional to: $\sum \nn_w= \nn(v)-\nn(v')$ which is zero,
because $\nn(v)=\nn(v')$ means that the sum of the $\nn_w$'s vanishes.
Since $\oo\cdot\nn=\e\ne0$ we can expect to see a sum of order $\e^2$,
if we sum as well on a overall change of sign of the $\nn_w$ values
(which sum up to $\V0$).

But this can be true only if $\e\ll \oo\cdot\nn'$, for any branch momentum
$\nn'$ of a branch in $\th/\th_2$.  If the latter property is not true
this means that $\oo\cdot\nn'$ is small and that there are many nodes
in $\th/\th_2$ of order of the amount needed to create a momentum with
small divisors of order $\e$.

Examining carefully the proof of Brjiuno's lemma one sees that such
extreme case would be essentially also treatable. Therefore the problem
is to show that the two regimes just envisaged (and their "combinations")
do exhaust all possibilities.

Such problems are very common in renormalization theory and are called
"overlapping divergences". Their systematic analysis is made through the
renormalization group methods. We argue here that Elliasson's method can
be interpreted in the same way.

The above introduced trees will play the role of {\it Feynman
diagrams}; and they will be plagued by overlapping divergences. They
will therefore be collected into another family of graphs,
that we shall call {\it trees}, on which the bounds are easy.
The $(\oo\cdot\nn)^{-2}$ are the {\it propagators}, in our analogy.

We fix an {\it scaling} parameter $\g$, which we take $\g=2$
for consistency with \equ(7.1), and we also define $\oo\=C_0\oo_0$: it
is an adimensional frequency.  Then we say that a propagator
$(\oo\cdot\nn)^{-2}$ is {\it on scale $n$} if $2^{n-1}<|\oo\cdot\nn|\le
2^n$, for $n\le0$, and we set $n=1$ if $1<|\oo\cdot\nn|$.

Proceeding as in quantum field theory, see [G3], given a tree $\th$
we can attach a {\it scale label} to each branch $v'v$ in \equ(6.23) (with
$v'$ being the node preceding $v$): it is
equal to $n$ if $n$ is the scale of the branch propagator.
Note that the labels thus attached to a tree are uniquely
determined by the tree: they will have only the function of
helping to visualize the orders of magnitude of the various tree
branches.

Looking at such labels we identify the connected clusters $T$ of
nodes that are linked by a continuous path of branches with the same
scale label $n_T$ or a higher one.  We shall say that {\it the cluster
$T$ has scale $n_T$}.

Among the clusters we consider the ones with the property that there is
only one tree branch entering them and only one exiting and both carry
the same momentum. Here we use that the tree branches carry an arrow
pointing to the root: this gives a meaning to the words ``incoming'' and
``outgoing''.

If $V$ is one such cluster we denote $\l_V$ the incoming branch: the branch
scale $n=n_{\l_V}$ is smaller than the smallest scale $n'=n_V$ of the
branches inside $V$.  We call $w_1$ the node into which the branch $\l_V$
ends, inside $V$.  {\it We say that such a $V$ is a {\it resonance} if
the number of branches contained in $V$ is $\le E\,2^{-n\e}$}, where
$n=n_{\l_V}$, and $E,\e$ are defined by:
$E\=2^{-3\e}N^{-1},\,\e=\t^{-1}$.  We shall say that $n_{\l_V}$ is {\it
the resonance scale}.

Let us consider a tree $\th$ and its clusters.  We wish to estimate
the number $N_n$ of branches with scale $n\le0$ in it, assuming $N_n>0$.

Denoting $T$ a cluster of scale $n$ let $m_T$ be the number of
resonances of scale $n$ contained in $T$ (\ie with incoming branches of
scale $n$), we have the following inequality, valid for any tree
$\th$:
$$N_n\le\fra{3k}{E\,2^{-\e n}}+\sum_{T, \,n_T=n}(-1+m_T)\Eqa(A3.1)$$
with $E=N^{-1}2^{-3\e},\e=\t^{-1}$. This is a version of Brjuno's
lemma: a proof is in appendix A4.

Consider a tree $\th^1$ we define the family $\FF(\th^1)$ generated
by $\th^1$ as follows. Given a resonance $V$ of $\th^1$ we detach the
part of $\th^1$ above $\l_V$ and attach it successively to the points
$w\in\tilde V$, where $\tilde V$ is the set of nodes of $V$
(including the endpoint $w_1$ of $\l_V$ contained in $V$) outside the
resonances contained in $V$. Note that all the branches $\l$ in $\tilde V$
have the same scale $n_\l=n_V$.

For each resonance $V$ of $\th^1$ we shall call $M_V$ the number of
nodes in $\tilde V$.  To the just defined set of trees we add the
trees obtained by reversing simoultaneously the signs of the node
modes $\nn_w$, for $w\in \tilde V$: the change of sign is performed
independently for the various resonant clusters.  This defines a family
of $\prod 2M_V$ trees that we call $\FF(\th_1)$. The number
$\prod 2M_V$ will be bounded by $\exp\sum2M_V\le e^{2k}$.

It is important to note that the definition of resonance is such that
the above operation (of shift of the node to which the branch entering
$V$ is attached) does not change too much the scales of the tree
branches inside the resonances: the reason is simply that inside a
resonance of scale $n$ the number of branches is not very large being
$\le\lis N_n\=E\,2^{-n\e}$.

Let $\l$ be a branch, in a cluster $T$, contained inside the resonances
$V=V_1\subset V_2\subset\ldots$ of scales $n=n_1>n_2>\ldots$; then the
shifting of the branches $\l_{V_i}$ can cause at most a change in the size
of the propagator of $\l$ by at most $2^{n_1}+2^{n_2}+\ldots< 2^{n+1}$.

Since the number of branches inside $V$ is smaller than $\lis N_n$ the
quantity $\oo\cdot\nn_\l$ of $\l$ has the form
$\oo\cdot\nn^0_\l+\s_\l\oo\cdot\nn_{\l_V}$ if $\nn^0_\l$ is the momentum
of the branch $\l$ "inside the resonance $V$", \ie it is the sum of all
the node modes of the nodes preceding $\l$ in the sense of the
branch arrows, but contained in $V$; and $\s_\l=0,\pm1$.

Therefore not only $|\oo\cdot\nn^0_\l|\ge 2^{n+3}$ (because $\nn^0_\l$
is a sum  of $\le \lis N_n$ node modes, so that $|\nn^0_\l|\le N\lis
N_n$) but $\oo\cdot\nn^0_\l$ is "in the middle" of the diadic interval
containing it and by \equ(7.1) does not get out of it if we add a quantity
bounded by $2^{n+1}$ (like $\s_\l\oo\cdot\nn_{\l_V}$). Hence no branch
changes scale as $\th$ varies in $\FF(\th^1)$, if $\oo_0$ verifies \equ(7.1).

{\it This implies, by the strong diophantine hypothesis on $\oo_0$,
\equ(7.1), that the resonant clusters of the trees in $\FF(\th^1)$ all
contain the same sets of branches, and the same branches go in or out of each
resonance (although they are attached to generally distinct nodes
inside the resonances: the identity of the branches is here defined by the
number label that each of them carries in $\th^1$).  Furthermore the
resonance scales and the scales of the resonant clusters, and of all the
branches, do not change.}

Let $\th^2$ be a tree not in $\FF(\th^1)$ and construct
$\FF(\th^2)$, \etc.  We define a collection
$\{\FF(\th^i)\}_{i=1,2,\ldots}$ of pairwise disjoint families of
trees.  We shall sum all the contributions to $\V h^{(k)}$ coming
from the individual members of each family.  This is the {\it
Eliasson's resummation}.

We call $\e_V$ the quantity $\oo\cdot\nn_{\l_V}$ associated with the
resonance $V$. If $\l$ is a branch with both extremes in $\tilde V$ we
can imagine to write the quantity $\oo\cdot\nn_\l$ as
$\oo\cdot\nn^0_\l+\s_V\e_V$, with $\s_V=0,\pm1$. Since
$|\oo\cdot\nn_\l|> 2^{n_V-1}$ we see that the product of the propagators
is holomorphic in $\e_V$ for $|\e_V|<2^{n_V-3}$. While $\e_V$ varies in
such complex disk the quantity $|\oo\cdot\nn_\l|$ does not become
smaller than $2^{n_V-1}-2^{n_V-3}\ge2^{n_V-2}$. Note the main point
here: the quantity $2^{n_V-3}$ will usually be $\gg 2^{n_{\l_V}-1}$
which is the value $\e_V$ actually can reach in every tree in
$\FF(\th^1)$; this can be exploited in applying the maximum priciple, as
done below.

It follows that, calling $n_\l$ the scale of the branch $\l$ in $\th^1$,
each of the $\prod 2 M_V\le e^{2k}$ products of propagators
of the members of the family $\FF(\th^1)$ can be bounded above by
$\prod_\l\,2^{-2(n_\l-2)}=2^{4k}\prod_\l\,2^{-2n_\l}$, if regarded as a
function of the quantities $\e_V=\oo\cdot\nn_{\l_V}$, for $|\e_V|\le
\,2^{n_V-3}$, associated with the resonant clusters $V$.  This even
holds if the $\e_V$ are regarded as independent complex parameters.

By construction it is clear that the sum of the $\prod 2M_V\le e^{2k}$
terms, giving the contribution to $\V h^{(k)}$ from the trees in
$\FF(\th^1)$, vanishes to second order in the $\e_V$ parameters (by the
approximate cancellation discussed above).  Hence by the maximum
principle, and recalling that each of the scalar products in \equ(6.23) can
be bounded by $N^2$, we can bound the contribution from the family
$\FF(\th^1)$ by:
$$\left[\fra1{k!} N\Big(\fra{f_0 C_0^2 N^2}{J_0}\Big)^k 2^{4k} e^{2k}
\prod_{n\le0}2^{-2nN_n}\right]\left[\prod_{n\le0}\prod_{T,\,n_T=n}
\prod_{i=1}^{m_T}\,2^{2(n-n_{i}+3)}\right]\Eqa(A3.2)$$
where:
\acapo
1) $N_n$ is the number of propagators of scale $n$ in $\th^1$ ($n=1$
does not appear as $|\oo\cdot\nn|\ge1$ in such cases),\acapo
2) the first square bracket is the bound on the product of
individual elements in the family $\FF(\th^1)$ times the bound $e^{2k}$
on their number,
\acapo
3) The second term is the part coming from the maximum principle, applied
to bound the resummations, and is explained as follows.
\acapo
i) the dependence on the variables $\e_{V_i}\=\e_i$ relative to
resonances $V_i\subset T$ with scale $n_{\l_V}=n$ is holomorphic for for
$|\e_i|<\,2^{ n_i-3}$ if $n_i\=n_{V_i}$, provided $n_i>n+3$ (see above).
\acapo
ii) the resummation says that the dependence on the $\e_i$'s has a
second order zero in each.  Hence the maximum principle tells us that
we can improve the bound given by the first factor in \equ(A3.2) by the
product of factors $(|\e_i|\,2^{-n_i+3})^2$ if $n_i>n+3$.  If $ n_i\le
n+3$ we cannot gain anything: but since the contribution to the bound
from such terms in \equ(A3.2) is $>1$ we can leave them in it to simplify
the notation, (of course this means that the gain factor can be
important only when $\ll1$).

Hence substituting \equ(A3.1) into \equ(A3.2) we see that the $m_T$ is taken
away by the first factor in $\,2^{2n}2^{-2n_{i}}$, while the remaining
$\,2^{-2n_i}$ are compensated by the $-1$ before the $+m_T$ in \equ(A3.1)
taken from the factors with $T=V_i$,  (note that there are always enough
$-1$'s).

Hence the product \equ(A3.2) is bounded by:
$$\fra1{k!}N\,(C_0^2J_0^{-1}f_0 N^2)^k e^{2k}2^{4k}2^{6k}
\prod_n\,2^{-4 n k E^{-1}\,2^{\e n}}\le \fra1{k!}N\, B_0^k\Eqa(A3.3)$$
with: $B_0=2^{10}e^2 C_0^2 f_0 N^2 \exp [N\,
(2^{2+2\t^{-1}}\log2\big)\sum_{p=1}^\io p 2^{-p\t^{-1}}]$.

To sum over the trees we note that, fixed $\th$ the collection of
clusters is fixed.  Therefore we only have to multiply \equ(A3.3) by the
number of tree shapes for $\th$, ($\le 2^{2k}k!$), by the number of
ways of attaching mode labels, ($\le (3N)^{lk}$), so that we can bound
$|h^{(k)}_{\nn j}|$ by an exponential of $k$ and \equ(1.4) follows.
\*
%%%%%%%%%%%%%%%%%%%%%%%%%%%%%%%%%%%%%%%%%%%%%%%*********
\0{\it Remarks.}
\*
The strong diophantine condition is quite unpleasant as it seems to put
an extra requirement on $\oo_0$: I think that in fact such condition is
not necessary: I explain why in [G3].  The basic reason is that one is
not forced to introduce the scales $2^n$ in an exact geometric growth:
an approximate one, in which $2^n$ is replaced by $\g_n$ and $1\le
\g_n2^{-n}\le2$ would be enough. This gives much more freedom in
fulfilling the \equ(7.1), given $\oo_0$ verifying \equ(1.3) only. Such a
squence can be shown to exist always if $\oo_0$ verifies \equ(1.3),
possibly replacing $C_0$ by a larger constant (to fulfill the analogue
of \equ(7.1)).

\vskip1.truecm
{\bf Appendix A4: Resonant Siegel-Brjuno bound.}
\vskip0.5truecm
\numsec=3\numfor=1\pgn=1

Calling $N^*_n$ the number of non resonant lines carrying a scale label
$\le n$. We shall prove first that $N^*_n\le 2k (E 2^{-\e n})^{-1}-1$ if
$N^*_n>0$.

If $\th$ has the root line with scale $>n$ then calling
$\th_1,\th_2,\ldots,\th_m$ the subdiagrams of $\th$ emerging from the
first vertex of $\th$ and with $k_j>E\,2^{-\e n}$ lines, it is
$N^*(\th)=N^*(\th_1)+\ldots+N^*(\th_m)$ and the statement is inductively
implied from its validity for $k'<k$ provided it is true that
$N^*(\th)=0$ if $k<E2^{-\e n}$, which is is certainly the case if $E$ is
chosen as in \equ(A3.1).\footnote{${}^*$}{\nota Note that if $k\le
E\,2^{-n\e}$ it is, for all momenta $\nn$ of the lines, $|\nn|\le N E
\,2^{-n\e}$, \ie $|\oo\cdot\nn|\ge(NE\,2^{-n\e})^{-\t}=2^3\,2^{n}$ so
that there are {\it no} clusters $T$ with $n_T=n$ and $N^*=0$. The
choice $E=N^{-1}2^{-3\e}$ is convenient: but this, as well as the whole
lemma, remains true if $3$ is replaced by any number larger than $1$.
The choice of $3$ is made only to simplify some of the arguments based
on the resonance concept.}

In the other case it is $N^*_n\le 1+\sum_{i=1}^mN^*(\th_i)$, and if
$m=0$ the statement is trivial, or if $m\ge2$ the statement is again
inductively implied by its validity for $k'<k$.

If $m=1$ we once more have a trivial case unless the order $k_1$ of
$\th_1$ is $k_1>k-\fra12 E\,2^{-n\e}$.  Finally, and this is the real
problem as the analysis of a few examples shows, we claim that in the
latter case the root line is either a resonance or it has scale $>n$.

Accepting the last statement it will be: $N^*(\th)=1+N^*(\th_1)=
1+N^*(\th'_1)+\ldots+N^*(\th'_{m'})$, with $\th'_j$ being the $m'$
subdiagrams emerging from the first node of $\th'_1$ with orders
$k'_j>E\,2^{-\e n}$: this is so because the root line of $\th_1$ will
not contribute its unit to $N^*(\th_1)$.  Going once more through the
analysis the only non trivial case is if $m'=1$ and in that case
$N(\th'_1)=N^*(\th"_1)+\ldots+N(\th"_{m"})$, \etc, until we reach a
trivial case or a diagram of order $\le k-\fra12 E\,2^{-n\e}$.

It remains to check that if $k_1>k-\fra12E\,2^{-n\e}$ then the root line of
$\th_1$ has scale $>n$, unless it is entering a resonance.

Suppose that the root line of $\th_1$ is not entering a resonance. Note
that $|\oo\cdot\nn(v_0)|\le\,2^n,|\oo\cdot\nn(v)|\le
\,2^n$, if $v_0,v_1$ are the first vertices of $\th$ and $\th_1$
respectively.  Hence $\d\=|(\oo\cdot(\nn(v_0)-\nn(v_1))|\le2\,2^n$ and
the diophantine assumption implies that $|\nn(v_0)-\nn(v_1)|>
(2\,2^n)^{-\t^{-1}}$, or $\nn(v_0)=\nn(v_1)$.  The latter case being
discarded as $k-k_1<\fra12E\,2^{-n\e}$ (and we are not considering the
resonances), it follows that $k-k_1<\fra12E\,2^{-n\e}$ is inconsistent:
it would in fact imply that $\nn(v_0)-\nn(v_1)$ is a sum of $k-k_1$
vertex modes and therefore $|\nn(v_0)-\nn(v_1)|< \fra12NE\,2^{-n\e}$
hence $\d>2^3\,2^n$ which is contradictory with the above opposite
inequality.

A similar, far easier, induction can be used to prove that if $N^*_n>0$
then the number $p$ of clusters of scale $n$ verifies the bound
$p<2 k \,(E2^{-\e n})^{-1}-1$. Thus \equ(11) is proved.

{\it Remark}: the above argument is a minor adaptation of Brjiuno's
proof of Siegel's theorem, as remarkably exposed by P\"oschel, [P].

\vskip1.truecm

{\bf Acknowledgements:} partial support from NSF grant \# DMR 89-18903,
from Rutgers University and Institut des Hautes Etudes Scientifiques
(IHES).  The work was also partly supported by Ministero della Ricerca
(fondi 40\%).  I am indebted to L.  Chierchia for many suggestions and
for critical readings of early versions; but mainly for showing me,
prior to publication, his "tree root identity" which is the key idea for
the homoclinic splitting theory, and without which the part of the paper
on the homoclinic splitting would have been left incomplete, and at best
with a strange looking conjecture.  It is with gratitude that I thank H.
Epstein and S. Miracle--sol\'e for criticism, suggestions and
encouragement; I am also indebted to J. Bricmont and M. Vittot formany
clarifying discussions.

\penalty-200

%\ciao
\vskip0.5truecm

\penalty-200

\numsec=0\numfor=1
{\bf References}

\vskip0.5truecm

\penalty10000

\item{[A] } Arnold, V.: {\it Instability of dynamical systems with several
degrees of freedom}, Sov. Mathematical Dokl., 5, 581-585, 1966.

\item{[A2] } Arnold, V.: {\it Proof of a A.N. Kolmogorov theorem on
conservation of conditionally periodic motions under small perturbations
of the hamiltonian function}, Uspeki Matematicheskii Nauk, 18, 13-- 40,
1963.

\item{[ACKR] } Amick C., Ching E.S.C., Kadanoff L.P., Rom--Kedar V.:
{\it Beyond All Orders: Singular Perturbations in a Mapping} J.
Nonlinear Sci. 2, 9--67, 1992.

\item{[BG] } Benettin, G., Gallavotti, G.: {\it Stability of motions near
resonances in quasi-integrable hamiltonian systems}, J. Statistical Physics,
44, 293-338, 1986.

\item{[BfG] } Benfatto, G., Gallavotti, G.: {\it Perturbation theory of
the Fermi surface in a quantum liquid. A general quasi particle
formalism and one dimensional systems}, Journal of Statistical Physics,
59, 541- 664, 1990.

\item{[CG] } Chierchia, L., Gallavotti, G.: {\it Drift and diffusion in
phase space}, in mp\_arc, \# 92-92.  {This paper
is deposited in the archive mp\_arc: to get a TeX version send an
empty E-mail message to {\tt mp\_arc@math.utexas.edu}: instructions will
be sent back.} In print in Annales de l'Institut H. Poincar\`e.

\item{[CZ] } Chierchia, L., Zehnder, E.: {\it Asymptotic expansions of
quasi-periodic motions}, Annali della Scuola Normale Superiore di Pisa,
Serie IV Vol XVI Fasc.2 (1989).

\item{[DS]} Delshams, A., Seara, M.T.:
{\it An asymptotic expression for the splitting of separatrices of rapidly
forced pendulum}, preprint 1991.

\item{[E] } Eliasson L. H.: {\it Absolutely convergent series expansions
for quasi--periodic motions}, report 2--88, Dept. of Math., University of
Stockholm, 1988.

\item{[FG] } Felder, G., Gallavotti, G.: {\it Perturbation theory and
non renormalizable scalar fields}, Communications in Mathematical
Physics, 102, 549-571, 1986.

\item{[G1] } Gallavotti, G.: {\it The elements of Mechanics}, Springer, 1983.

\item{[G2] } Gallavotti, G.: {\it Renormalization theory and ultraviolet
stability for scalar fields via renormalization group methods}, Reviews
in Modern Physics, 57, 471- 572, 1985. See also, Gallavotti, G.:
{\it Quasi integrable mechanical systems}, Les Houches, XLIII (1984),
vol. II, p. 539-- 624, Ed. K. Osterwalder, R. Stora, North Holland, 1986.

\item{[Ge] } Gelfreich, V.: {\it Separatrices splitting for the rapidly
forced pendulum} preprint 1992.

\item{[GLT] } Gelfreich, V.  G., Lazutkin, V.F., Tabanov, M.B.:{\it
Exponentially small splitting in Hamiltonian systems}, Chaos, 1 (2),
1991.

\item{[H] } Harary, F., Palmer, E.: {\it Graphical enumeration},
Academic Press, 1973, New York.

\item{[HMS] } Holmes, P., Marsden, J., Scheurle,J: {\it Exponentially Small
Splittings of Separatrices in KAM Theory and Degenerate Bifurcations},
Preprint, 1989.

\item{[K] } Kolmogorov, N.: {\it On the preservation of conditionally
periodic motions}, Doklady Akademia Nauk SSSR, 96, 527-- 530, 1954. See
also: Benettin, G., Galgani, L., Giorgilli, A., Strelcyn, J.M.:
{\it A proof of Kolmogorov theorem on invariant tori using canonical
transormations defined by the Lie method}, Nuovo Cimento, 79B, 201--
223, 1984.

\item{[La] } Lazutkin, V.F.: {\it Separatrices splitting for standard and
semistandard mappings}, Pre\-pr\-int, 1989.

\item{[M] } Moser, J.: {\it On invariant curves of an area preserving
mapping of the annulus}, Nach\-rich\-ten Akademie Wiss. G\"ottingen, 11,
1-- 20, 1962.

\item{[N] } Nekhorossev, N.: {An exponential estimate of the time of
stability of nearly integrable hamiltonian systems}, Russian Mathematical
Surveys, 32, 1-65, 1975.

\item{[Nei] } Neihstad, A.I.: {\it The separation of motions in systems
with rapidly rotating phase}, Prikladnaja Matematika i Mekhanika {\bf
48}, 133--139, 1984 (Translation in Journal of Applied Mathematics and
Mechanics).

\item{[P] } P\"oschel, J.: {\it Invariant manifolds of complex analytic
mappings}, Les Houches, XLIII (1984), vol. II, p. 949-- 964, Ed. K.
Osterwalder, R. Stora, North Holland, 1986.

\item{[S] } Siegel, K.: {\it Iterations of analytic functions}, Annals
of Mathematics, {\bf 43}, 607-- 612, 1943.

\penalty10000

\item{[T] } Thirring, W.: {\it Course in Mathematical Physics}, vol. 1,
p. 133, Springer, Wien, 1983.

\item{[V]} Vittot, M.: {\it Lindstedt perturbation series in hamiltonian
mechanics: explicit formulation via a multidimensional Burmann--Lagrange
formula}, Preprint CNRS--Luminy 1992.

\ciao